%% file: main.tex
\documentclass[preprint2]{aastex62}

\newcommand{\fesclyc}{$f_{esc}^{LyC}$}
\newcommand{\fesclycrel}{$F_{\lambda LyC}/F_{\lambda 1100}$}
\newcommand{\fesclya}{$f_{esc}^{Ly\alpha}$}
\newcommand{\orat}{$O_{32}$}
\newcommand{\lya}{$\rm Ly\alpha$}
\newcommand{\sigsfr}{$\Sigma_{\rm SFR}$}
\newcommand{\ssfr}{${\rm sSFR}$}
\newcommand{\oh}{$\rm 12+\log_{10}( O/H )$}
\newcommand{\Izotov}{\citet{2016Natur.529..178I,2016MNRAS.461.3683I,2018MNRAS.474.4514I,2018MNRAS.478.4851I,2021arXiv210301514I}}
\newcommand{\pubsamp}{\citet{2016Natur.529..178I,2016MNRAS.461.3683I,2018MNRAS.474.4514I,2018MNRAS.478.4851I,2021arXiv210301514I,2019ApJ...885...57W}}

\graphicspath{{./plots/}}

\shorttitle{Low-z LyC Survey II: \fesclyc Diagnostics}
\shortauthors{Flury et al.}
\submitjournal{ApJ}
\received{}
\revised{}
\accepted{}

\begin{document}

\title{The Low-Redshift Lyman Continuum Survey II:\\New Insights into LyC Diagnostics}

\input{authors}

\begin{abstract}

{ { The Lyman continuum (LyC) cannot be observed at the epoch of reionization ($z\ga6$) due to intergalactic \ion{H}{1} absorption.} To identify Lyman continuum emitters (LCEs) and infer the fraction of escaping LyC, astronomers have developed various indirect diagnostics of LyC escape. Using measurements of the LyC from the Low-redshift Lyman Continuum Survey (LzLCS), we present the first statistical test of these diagnostics. While optical depth indicators based on \lya, such as peak velocity separation and equivalent width, perform well, we also find that other diagnostics, such as the [\ion{O}{3}]/[\ion{O}{2}] flux ratio and star formation rate surface density, predict whether a galaxy is a LCE. The relationship between these galaxy properties and the fraction of escaping LyC flux suggests that LyC escape depends strongly on \ion{H}{1} column density, ionization parameter, and stellar feedback. We find LCEs occupy a range of stellar masses, metallicities, star formation histories, and ionization parameters, { which may indicate episodic and/or different physical causes of LyC escape.}}

\end{abstract}


\section{Introduction}\label{sec:intro}

{ While the reionization of the Universe likely completed by a redshift of $z\sim6$ \citep[e.g.,][]{2001AJ....122.2850B,2020JCAP...09..005P,2020MNRAS.493.3194P,2020arXiv200913544Y}}, substantial contention persists concerning which galaxies are responsible. { Intense, concentrated star formation seems to be a necessary characteristic of reionizing galaxies because of the associated stellar feedback \citep[e.g.,][]{2001ApJ...558...56H,2002MNRAS.337.1299C,2011ApJ...730....5H,2017MNRAS.468.2176S,2020ApJ...892..109N}}. Many predictions suggest that dwarf galaxies are the primary Lyman continuum (LyC) photon donors because the weak gravitational potentials  of low mass galaxies exacerbate the clearing effects of feedback \citep[e.g.,][]{2010ApJ...710.1239R,2014MNRAS.442.2560W,2015MNRAS.451.2544P,2021arXiv210301514I}. Others indicate that more massive galaxies dominate reionization because stronger gravitational potentials better facilitate efficient star formation, resulting in a higher LyC photon budget \citep[e.g.,][]{2013MNRAS.428.2741W}. An increase in dust extinction with galaxy mass could reduce LyC escape at the highest masses, causing galaxies with masses in between these two regimes to dominate reionization \citep[e.g.,][]{2020arXiv200305945M}.

Central to this debate is the fraction of LyC photons that escape from a galaxy to reionize the intergalactic medium (IGM). This ``escape fraction'' \fesclyc\ relates the cosmic star formation rate density and LyC production efficiency to the cosmic ionization rate. Based on results from stacked spectra at $z\sim3$ \citep[e.g.,][]{2018ApJ...869..123S} and from cosmological simulations \citep[e.g.,][]{2012ApJ...758...93F}, galaxies that exhibit \fesclyc$\ga0.05$ are the most relevant to reionization. Due to the increase in intergalactic Lyman continuum absorption with redshift, direct measurement of the LyC, let alone \fesclyc, is statistically unlikely beyond $z\sim4$ \citep[e.g.,][]{2012ApJ...751...70V,2014MNRAS.445.1745W} to ascertain which galaxies are responsible for reionizing the universe.

Therefore, we must study lower redshift LyC-leaking galaxies to develop indirect diagnostics for \fesclyc. These diagnostics may also provide insight into the physical mechanism(s) and environmental conditions that contribute to leaking of LyC photons. In the following, we will explore some of the most promising diagnostics for \fesclyc, including the equivalent widths (EWs) of nebular emission lines, emission line flux ratios, the profile of the \lya\ emission line, the star formation rate surface density (\sigsfr), and the UV continuum { slope ($\beta$) and magnitude ($M_{1500}$)}.

Of these properties, nebular emission line EWs are one of the easiest to measure. {Because hydrogen Balmer lines originate from recombination across the entire extent of \ion{H}{2} regions, measuring a smaller EW in these lines may indicate lower optical depth: if gas does not extend out to the Str\"omgren radius, not all LyC photons will photoionize \ion{H}{1}, and thus, the nebula will emit fewer Balmer photons \citep[e.g.,][]{2013A&A...554A..38B}.} To date, few tests of this diagnostic have been conducted. Moreover, many local LyC emitters (LCEs) exhibiting high Balmer emission line EWs have been found \citep{2016MNRAS.461.3683I,2018MNRAS.474.4514I,2018MNRAS.478.4851I}. In any case, Balmer line EWs are sensitive not only to nebula size but also starburst age and star formation history \citep{2017ApJ...836...78Z,2018MNRAS.479..368B,2020arXiv200705519A}. Plagued by such degeneracies, the EW diagnostic may be best when paired with another indicator.

Similarly motivated by optical depth, the emission line flux ratio \orat$=$\newline [\ion{O}{3}]$\lambda5007$/[\ion{O}{2}]$\lambda\lambda3726,3729$ probes the relative sizes of the O$^{+2}$ and O$^{+}$ ionized zones. When the nebula is optically thin to LyC, the O$^{+}$ zone will be smaller relative to the O$^{+2}$ zone \citep{2013ApJ...766...91J,2014MNRAS.442..900N}. Albeit straightforward, \orat\ also depends strongly on ionization parameter to the extent that \orat\ is commonly used to measure it \citep[e.g.,][]{2004ApJ...617..240K,2015A&A...576A..83S,2018ApJ...868..117S}. It is unclear whether this dependence will obfuscate \fesclyc\ diagnostics. {For instance, \fesclyc\ itself may depend on ionization parameter if the incident stellar radiation field is responsible for a density-bounded scenario wherein gas does not persist out to the Str\"omgren radius.} \orat\ also does not probe the outer extent of the nebula, the region that is most sensitive to optical depth. Observationally, there is substantial scatter in the \fesclyc\ - \orat\ relation \citep[e.g.,][]{2018MNRAS.478.4851I}, making this diagnostic tenuous.

Complementing the \orat\ diagnostic, the\\ $[$\ion{O}{1}$]\lambda6300$ emission line traces neutral gas at the outer extent of a nebula because O$^0$ has an ionization potential similar to that of hydrogen ($13.62$ eV) and is coupled to hydrogen recombination by charge exchange reactions \citep{source:osterbrock2006}. Comparing the $[$\ion{O}{1}$]\lambda6300$ flux to the flux of another emission line can thus indicate the prevalence of neutral gas at the nebula boundary. \citet{2015A&A...576A..83S} consider the $O_{31}=$[\ion{O}{3}]$\lambda5007$/[\ion{O}{1}]$\lambda6300$ flux ratio as a diagnostic for LyC escape. They argue that, while sensitive to physical conditions and other ionization sources, $O_{31}$ could indicate a density-bounded nebula as it traces the size of the O$^0$ zone relative to the O$^{+2}$ zone. Similarly, the [\ion{O}{1}]/H$\beta$ ratio could indicate a deficiency in neutral gas relative to hydrogen recombining from \ion{H}{2} to \ion{H}{1} across the nebula, comparable to the effect described by \citet{2019ApJ...885...57W} for {[\ion{S}{2}]/H$\alpha$}. The [\ion{O}{1}] line is not without complications, though. In models where LyC optical depth and ionization parameter are not constant across the nebula \citep[the so-called ``picket fence" scenario, cf.][]{2001ApJ...558...56H}, \citet{2020A&A...644A..21R} predict that $O_{31}$ and [\ion{O}{1}]/H$\beta$ depend on conditions such as starburst age, ionization parameter, and LyC escape conditions. Shocks can also produce [\ion{O}{1}] emission, further complicating the [\ion{O}{1}] line as a diagnostic.
Even so, these [\ion{O}{1}] flux ratios have promise as indicators of \fesclyc. \citet{2019ApJ...885...96J} found that both flux ratios correlate with the \lya\ profile peak velocity separation, itself an indicator of \fesclyc\ \citep[e.g.,][]{2018MNRAS.478.4851I}.

The \lya\ recombination emission line is sensitive to the neutral \ion{H}{1} opacity of a galaxy \citep[e.g.,][]{2014A&A...563A..77B,2015A&A...578A...7V}. First, the shape of the \lya\ profile depends on both the amount of absorption and re-emission of \lya\ photons by \ion{H}{1} in the host galaxy \citep[essentially a ``scattering'' effect, e.g.,][]{2017A&A...597A..13V,2020A&A...639A..85G}. Second, the line-of-sight \ion{H}{1} column density affects the fraction of \lya\ photons that escape \citep[e.g.,][]{2015ApJ...809...19H,2017ApJ...844..171Y,2018MNRAS.478.4851I}. The discovery of triple-peaked \lya\ profiles associated with LCEs \citep[e.g.,][]{2019Sci...366..738R,2018MNRAS.478.4851I,2020MNRAS.491.1093V} suggests low optical depth cavities in the ISM facilitating both \lya\ and LyC photon escape \citep{2020arXiv200610041B}. While these results are promising, the \lya\ line can be sensitive to IGM extinction, metallicity, and stellar populations in the host galaxy \citep[e.g.,][]{2015ApJ...809...19H,2017ApJ...844..171Y}. Moreover, the increasing IGM neutral covering can significantly extinguish \lya\ at high redshifts \citep[e.g.,][]{2011ApJ...743..132P,2014ApJ...795...20S}, particularly the blue peak in the emission line profile \citep[e.g.,][]{2011ApJ...728...52L,2017ApJ...838....4Y,2021ApJ...908...36H}. Focusing on \lya\ emitters may additionally introduce a sampling bias since a non-negligible fraction of high-redshift Lyman break galaxies and some LCEs exhibit little to no \lya\ \citep[e.g.,][]{2009ApJ...695.1163V,2020ApJ...888..109J}. Nevertheless, at reionization, \lya\ falls in an ideal ground-based observing window \citep{2017ApJ...837...11B,2018MNRAS.477.5406Y}, making \lya\ more feasible a diagnostic of \fesclyc\ than rest-frame optical lines.

Star formation rate surface density, on the other hand, may serve simultaneously as a probe of the physical mechanism of LyC leakage and a proxy for \fesclyc. Feedback from star formation can blow bubbles/chimneys in the host galaxy's ISM \citep[e.g.,][]{2001ApJ...558...56H,2002MNRAS.337.1299C,2016MNRAS.458L..94S,2020ApJ...892..109N,2020A&A...639A..85G}, suggesting high \sigsfr\ will correspond to high \fesclyc. Previously detected LCEs also appear to be compact \citep[e.g.,][]{2018MNRAS.478.4851I,2018A&A...614A..11M}. Compactness indicates concentrated star formation and may correlate with LyC emission \citep[e.g.,][]{2018A&A...614A..11M}, although it is yet unclear whether this is a defining characteristic of all LCEs. High \sigsfr\ appears to correlate with weaker low-ioniztion metal absorption lines \citep[e.g.,][]{2015ApJ...810..104A}, which suggests a connection with low optical depth.

All the diagnostics outlined above are plausible but remain, as of yet, insufficiently explored due to small sample sizes. Future observations with {\it JWST} and other observatories require reliable indirect diagnostics to infer \fesclyc\ at reionization. The recent Low-Redshift Lyman Continuum Survey \citep[LzLCS;][hereafter Paper {\sc i}]{2022arXiv220111716F} presents an unprecedented opportunity to investigate the properties of nearby ($z\sim0.3$) LCEs to test each diagnostic. In this paper, we summarize the survey in \S\ref{sec:lzlcs} but refer the reader to Paper {\sc i} for details. We provide assessments of the success of indirect diagnostics for selecting LCEs and inferring \fesclyc\ (\S\ref{sec:diagnostics}). In \S\ref{sec:2d-diag}, we compare different parameters and consider the implications both for diagnostics and for populations of LCEs. Finally, we discuss our results in the contexts of different LyC escape scenarios, cosmological simulations, and high redshift surveys in \S\ref{sec:discussion}. As in Paper {\sc i}, we assume $H_0=70$ km s$^{-1}$\ Mpc$^{-1}$, $\Omega_m=0.3$, and $\Omega_\Lambda=0.7$.

\section{The LzLCS Sample}\label{sec:lzlcs}

In Paper {\sc i}, we outline the LzLCS sample selection, data processing, LyC measurements and ancillary measurements. { Here, we summarize the LzLCS but refer the reader to Paper {\sc i} for details. The LzLCS consists of \emph{HST}/COS G140L observations of 66 new LCE candidates in the nearby ($z\sim 0.3-0.4$) Universe.
The significance of the LzLCS over previous surveys is not just its size but also its scope: the LzLCS spans a much broader range in, among other properties, \orat, UV $\beta$, \sigsfr, stellar mass, metallicity, and burst age than prior studies.

From the COS spectra, we measure the LyC and \lya\ flux, the \lya\ EW, and the spectral slope of the attenuated starlight continuum. To assess the detection of the LyC, we define the probability $P(>N|B)$ given by the Poisson survival function that the observed signal arises by chance from background fluctuations \citep[see][]{2016ApJ...825..144W,2020arXiv201207876M}. In summary, we detect LyC emission from 35 galaxies at $>$97.725\% confidence ($P(>N|B)<0.02275$, $>2\sigma$ significance) with 13/35 of these galaxies having fairly-detected LyC flux ($P(>N|B)<0.00135$, $3-5\sigma$ significance) and 12/35 of these galaxies having well-detected LyC flux ($P(>N|B)<2.867\times10^{-7}$, $>5\sigma$ significance). Following \citet{2019ApJ...882..182C}, we also perform SED fitting using instantaneous burst model spectra from {\sc Starburst99} \citep{2010ApJS..189..309L} to infer the intrinsic and observed UV magnitudes at 1500 \AA, starlight attenuation, starburst age, and intrinsic LyC flux \citep[see][for details]{2022arXiv220111800S}. From the COS acquisition images, we determine the UV half-light radius from the background-subtracted empirical cumulative light distribution, which is consistent with results from S\'ersic profile fits.

Optical spectra from the SDSS provide additional information. We measure nebular emission line fluxes and EWs from these spectra. To correct the fluxes for reddening, we infer nebular extinction from Balmer lines assuming Case B recombination with \citet{1995MNRAS.272...41S} atomic data and the \citet{1989ApJ...345..245C} reddening law. From the extinction-corrected fluxes, we obtain diagnostic flux ratios like \orat. Since [\ion{O}{3}]$\lambda4363$ is detected in the majority of the galaxies in our sample, we determine gas-phase metallicities by the direct method, inferring the auroral line flux by the ff-relation \citep{2006MNRAS.367.1139P} when not available.

Combining the SDSS and COS measurements with {\it GALEX} observations, we infer additional properties of the sample. We obtain stellar masses by fitting aperture-matched photometry from SDSS and {\it GALEX} using {\sc Prospector} \citep{2017ApJ...837..170L,2019ascl.soft05025J}. From the Balmer-line star formation rate \citep{2012ARA&A..50..531K}, UV half-light radius, and stellar mass, we calculate the star formation rate surface density and specific star formation rate. From the H$\beta$ and Ly$
\alpha$ fluxes, we also determine the fraction of \lya\ photons \fesclya\ which have escaped the host galaxy.
}

{ As discussed in \S\ref{sec:intro}, the key objective of establishing LyC escape diagnostics is to determine which properties most strongly correlate with the LyC escape fraction \fesclyc. We measure \fesclyc\ by three different methods: (i) the empirical \fesclycrel\ ratio, (ii) using H$\beta$ to infer the intrinsic LyC from {\sc Starburst99} continuous starburst model spectra (\fesclyc$({\rm H\beta})$), and (iii) using spectral energy distribution (SED) fits to the COS FUV spectra to infer the intrinsic LyC (\fesclyc$({\rm UV})$). We summarize these metrics below.
\renewcommand{\theenumi}{\roman{enumi}}
\begin{enumerate}
    \item The \fesclycrel\ ratio is simply a ratio of flux density measurements and is therefore free of any assumptions. However, \fesclycrel\ is sensitive to a number of factors in addition to the LyC escape fraction, including dust attenuation, star formation rate, metallicity, and burst age, complicating the interpretation of this quantity as a metric of \fesclyc.
    \item The H$\beta$ approach introduced by \citet{2016MNRAS.461.3683I} uses the H$\beta$ flux, a measure of the total number of absorbed ionizing photons, the H$\beta$ EW, a proxy for burst age, and the gas phase metallicity to determine which {\sc Starburst99} model to use to predict the intrinsic LyC flux density corresponding to the measured LyC. The observed LyC flux density relative to the intrinsic LyC flux density then gives the \fesclyc. This method can be sensitive to the assumed star formation history, particularly for older burst ages or for galaxies with high stellar mass. In Paper {\sc i}, we find that a continuous star formation scenario provides the best agreement with the other two \fesclyc\ metrics.
    \item The UV approach relies on fitting the COS UV spectra over the rest-frame wavelength range of 925-1345 \AA. These SED fits consist of a uniform \citet{2016ApJ...828..107R} attenuation law applied to a weighted linear combination of single-burst {\sc Starburst99} spectra and accompanying nebular continua \citep[see][]{2019ApJ...882..182C,2022arXiv220111800S}. The best-fit SED allows us to predict the intrinsic LyC flux density, which we then use to obtain the \fesclyc. While the UV fits are still affected by assumptions about the star formation history and dust attenuation, the COS UV \fesclyc\ is less sensitive to systematics than the H$\beta$ \fesclyc, making this metric the most reliable of the three.
\end{enumerate}
We discuss the systematics and caveats of each \fesclyc\ measurement in detail in Paper {\sc i}. Since each metric has its caveats, we choose to consider all three in our assessment of \fesclyc\ diagnostics below.

In addition to the 66 LzLCS galaxies, we include 23 LCE candidates which have been observed by \emph{HST}/COS and previously published in other studies \citep{2016Natur.529..178I,2016MNRAS.461.3683I,2018MNRAS.474.4514I,2018MNRAS.478.4851I,2021arXiv210301514I,2019ApJ...885...57W}. To obtain LyC fluxes and \fesclyc\ values in a consistent manner, we re-process their {\it HST}/COS observations. Our significance estimates are comparable to the published values. However, our more stringent $P(>N|B)<0.02275$ detection criterion includes two fewer LCEs than previous assessments, yielding a total of 15 LyC detections out of 23 published observations. For consistency, we also re-measure and infer the diagnostic properties from SDSS and COS observations in the same manner as the LzLCS galaxies. A detailed comparison of these published LCE candidates with the LzLCS is presented in Paper {\sc i}.}

\section{Indirect $\MakeLowercase{f_{esc}^{LyC}}$ Diagnostics}\label{sec:diagnostics}

\begin{figure*}
    \centering
    \includegraphics[width=0.32\textwidth]{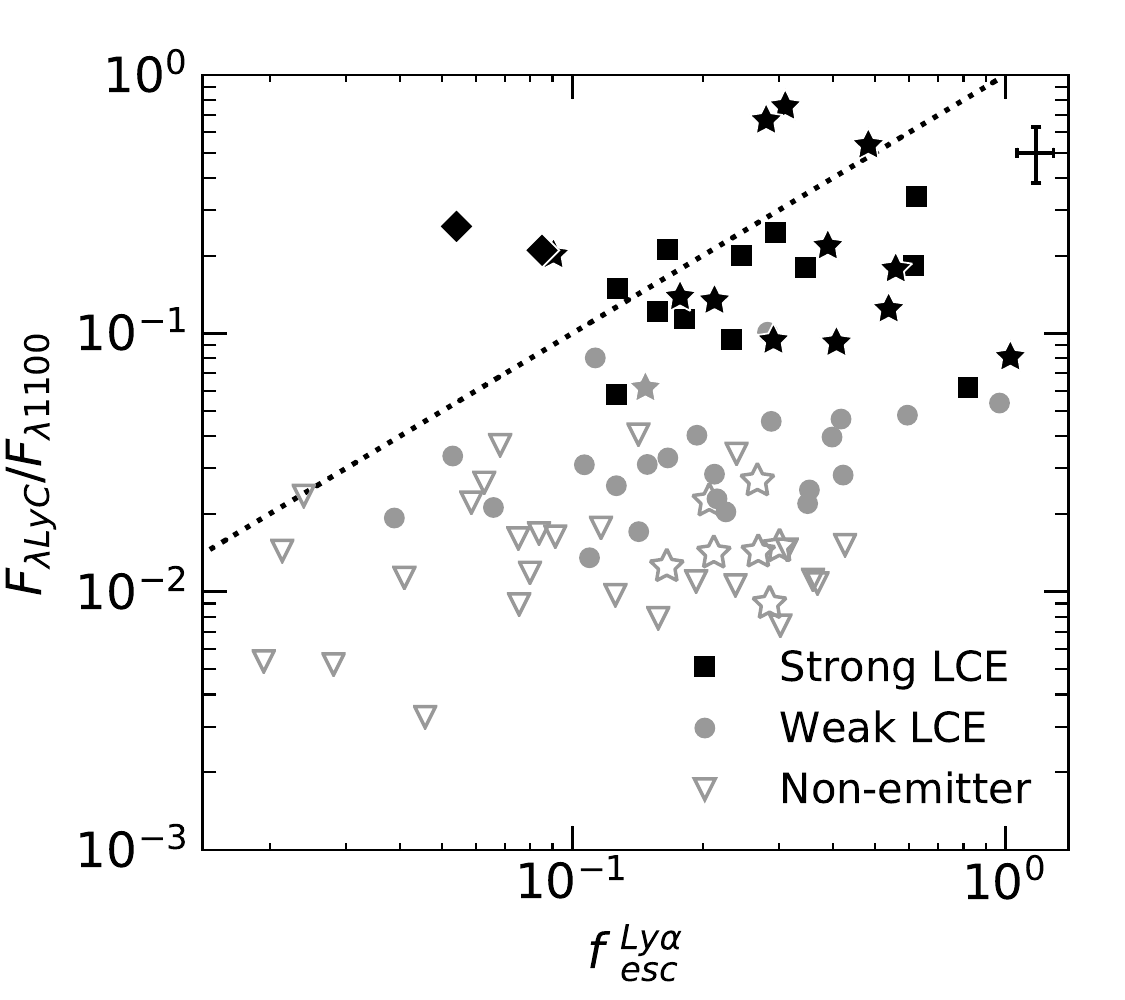}
    \includegraphics[width=0.32\textwidth]{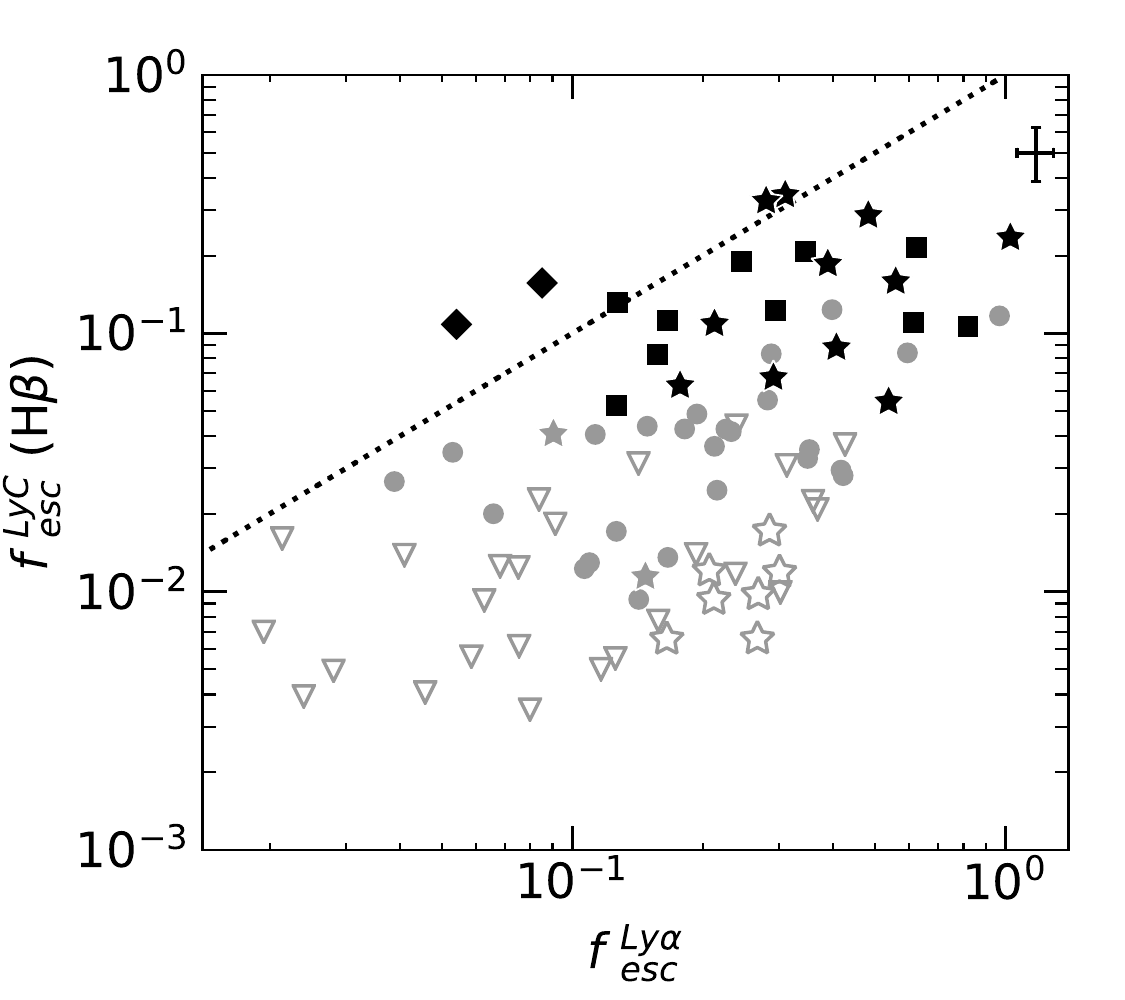}
    \includegraphics[width=0.32\textwidth]{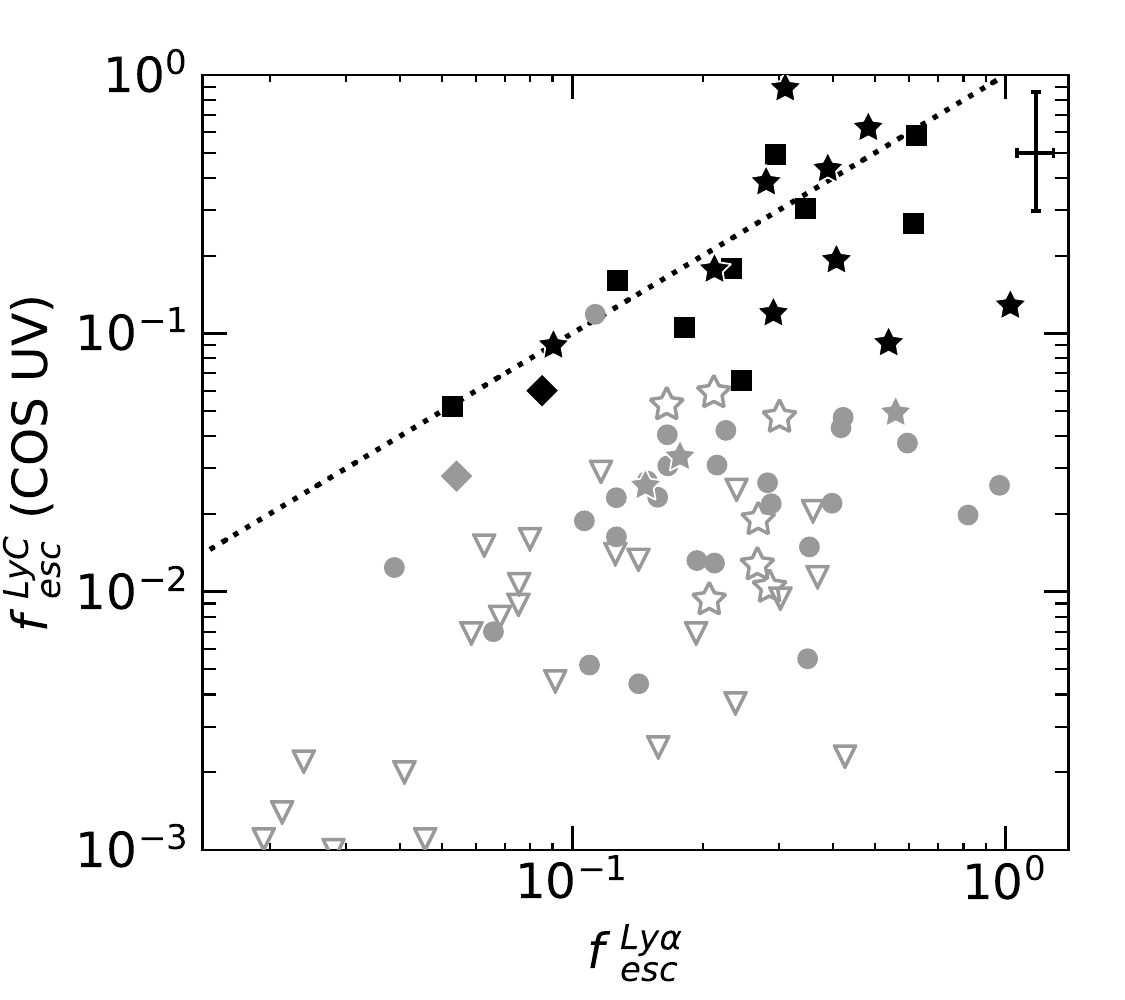}
    \caption{{Comparison of \fesclycrel\ (left), \fesclyc\ from H$\beta$ (center), and \fesclyc\ from fitting the COS UV spectrum (right) with \fesclya\ for strong LCEs ($>5\sigma$ LyC detection and \fesclyc$>0.05$, filled black), weak LCEs ($>2\sigma$ LyC detection but not strong, filled grey), and non-emitters ($<2\sigma$ LyC detection, open grey). For non-emitters, \fesclyc\ is shown for the $1\sigma$ upper limit on the LyC flux. LzLCS results are shown as squares (strong LCEs), circles (weak LCEs), and triangles (non-detections) while the \Izotov\ galaxies are shown as stars and the \citet{2019ApJ...885...57W} are shown as diamonds. Characteristic $1\sigma$ uncertainties are shown in the upper right.} Dotted line indicates 1:1 equivalence.}
    \label{fig:fesc_lya}
\end{figure*}

The LzLCS aims to test various \fesclyc\ diagnostics proposed in the literature as discussed in \S\ref{sec:intro}. To that end, we compare our measured \fesclyc\ against each parameter in Figures \ref{fig:fesc_lya}-\ref{fig:sigma_sfr}. { As discussed in \S\ref{sec:lzlcs}, we consider three different metrics of \fesclyc\ since each is susceptible to various systematic uncertainties. Another reason to use different \fesclyc\ indicators stems from the fact that one of the \fesclyc\ measurements is sometimes obtained from the same data as the proposed indirect diagnostics. Additional \fesclyc\ metrics allow us to assess whether an apparent correlation is real or simply introduced by using the same data in the abscissa and ordinate variables.}

{ To quantify possible correlations, we compute the Kendall $\tau$ rank correlation coefficient following the \citet{1996MNRAS.278..919A} prescription for censored data to account for upper limits on \fesclyc\ when $P(>N|B)>0.02275$ and any left- or right-censored indirect indicators of \fesclyc. For non-emitters, we use \fesclyc\ derived from the $1\sigma$ upper limit on the LyC flux for all three \fesclyc\ indicators. If we impose a $3\sigma$ detection requirement on the LyC instead of the adopted $2\sigma$ criterion, we find no significant change in the value of $\tau$.} We report $\tau$ for each of the three \fesclyc\ indicators (\fesclycrel, \fesclyc(H$\beta$), and \fesclyc(UV)) in Table \ref{tab:corrcoef} for the combined LzLCS and \pubsamp\ samples. We also report the probability $p$ that the measured $\tau$ is consistent with the null hypothesis that \fesclyc\ is not correlated with the indirect diagnostic. { We consider correlations to be significant if $p<1.350\times10^{-3}$ ($3\sigma$ confidence) and strong if $|\tau|>0.2$.}

\input{kendall_tau}

Below, we examine diagnostics motivated by LyC optical depth and by properties related to stellar feedback mechanisms facilitating LyC escape. { Many proxies exhibit an intrinsic, often significant correlation with \fesclyc\ but with substantial scatter in the data. This scatter often extends from a parameter-dependent upper bound on \fesclyc\ down to \fesclyc$=0$. These upper bounds in \fesclyc\ suggest that line of sight effects (e.g., orientation, covering fraction, etc.) may be responsible for obscuring otherwise clear trends, although relationships between various properties may also contribute to the scatter.}

{ To illustrate the combination of obscuring scatter and upper-bound effects, we refer to Figure \ref{fig:fesc_lya} comparing \fesclyc\ and \fesclya. For each of the LyC metrics, the highest \fesclyc\ values at any given \fesclya\ closely follow the dotted line indicating where \fesclyc=\fesclya. However, the lowest \fesclyc\ values at any given \fesclya\ are largely invariant with \fesclya, consistent with \fesclyc$\approx10^{-3}$ estimated in the non-detections. Thus, in this example, the identity line \fesclyc=\fesclya\ is an upper bound, or ``envelope'', to the distribution of \fesclyc values at a particular value of \fesclya. The envelope represents the intrinsic relationship between \fesclyc\ and \fesclya, a trend obscured by the scatter apparent in the observed distribution. Here, this upper bound could originate from a line of sight effect. When optically thin regions align with the aperture, \fesclyc=\fesclya. When optically thin regions are not aligned with the aperture, \fesclyc\ decreases significantly while resonant scattering of \lya\ photons into the line of sight maintains a high \fesclya. This upper bound effect appears in the \fesclya, [\ion{O}{1}], \orat, EW H$\beta$, $M_{1500}$, $\beta_{1200}$, $M_\star$, $r_{50}$, sSFR, and \oh\ diagnostics for at least one of the LyC escape metrics, although the scatter down to \fesclyc=0 tends to be more pronounced in \fesclycrel\ and H$\beta$ \fesclyc.}

{To highlight the intrinsic trends obscured by downward scatter in \fesclyc, we calculate the fraction of galaxies that are prodigious LCEs (LyC detected at $>5\sigma$ and either \fesclyc\ or \fesclycrel$>0.05$) in each diagnostic. This ``LCE fraction" indicates the affinity of strong LCEs for certain global properties. Such a preference could be used to select LCEs in future studies.

We define the LCE fraction as the ratio of strong LCEs ($P(>N|B) < 2.867\times10^{-7}$ for $5\sigma$ significance as in \S\ref{sec:lzlcs}, and \fesclyc\ indicator exceeds 5\%) to total galaxies in the combined sample in a given bin of a particular property. Bin counts are given as the sum of the probabilities for the strong LCE and total LCE samples, i.e., the Poisson binomial expected value for each bin. We define the final LCE detection fraction as the ratio of the strong LCE counts to that of the total counts in a given bin. We determine uncertainty in the LCE detection fraction using a Monte Carlo simulation of the ratio of Poisson binomial distributions using $10^4$ independent Bernoulli trials for each measurement to determine the distribution of counts in each bin. In cases of upper limits where no uncertainty is reported for a measurement (notably the [\ion{O}{1}] line), we follow the root-finding minimization method from \citet{1986ApJ...303..336G} to define confidence intervals on the detection fraction. The \fesclycrel\ flux ratio tends to agree more with \fesclyc(H$\beta$) than with \fesclyc(UV); however, \fesclyc(UV) is less dependent on assumptions about stellar populations and star formation history than \fesclyc(H$\beta$) and contains corrections for dust that are absent in \fesclycrel. Despite some disagreements in values of LCE fraction, \fesclyc, and \fesclycrel, the trend in LCE fraction does not change significantly depending on the \fesclyc\ indicator.}

\subsection{\lya}

\begin{figure*}
    \centering
    \includegraphics[width=0.32\linewidth]{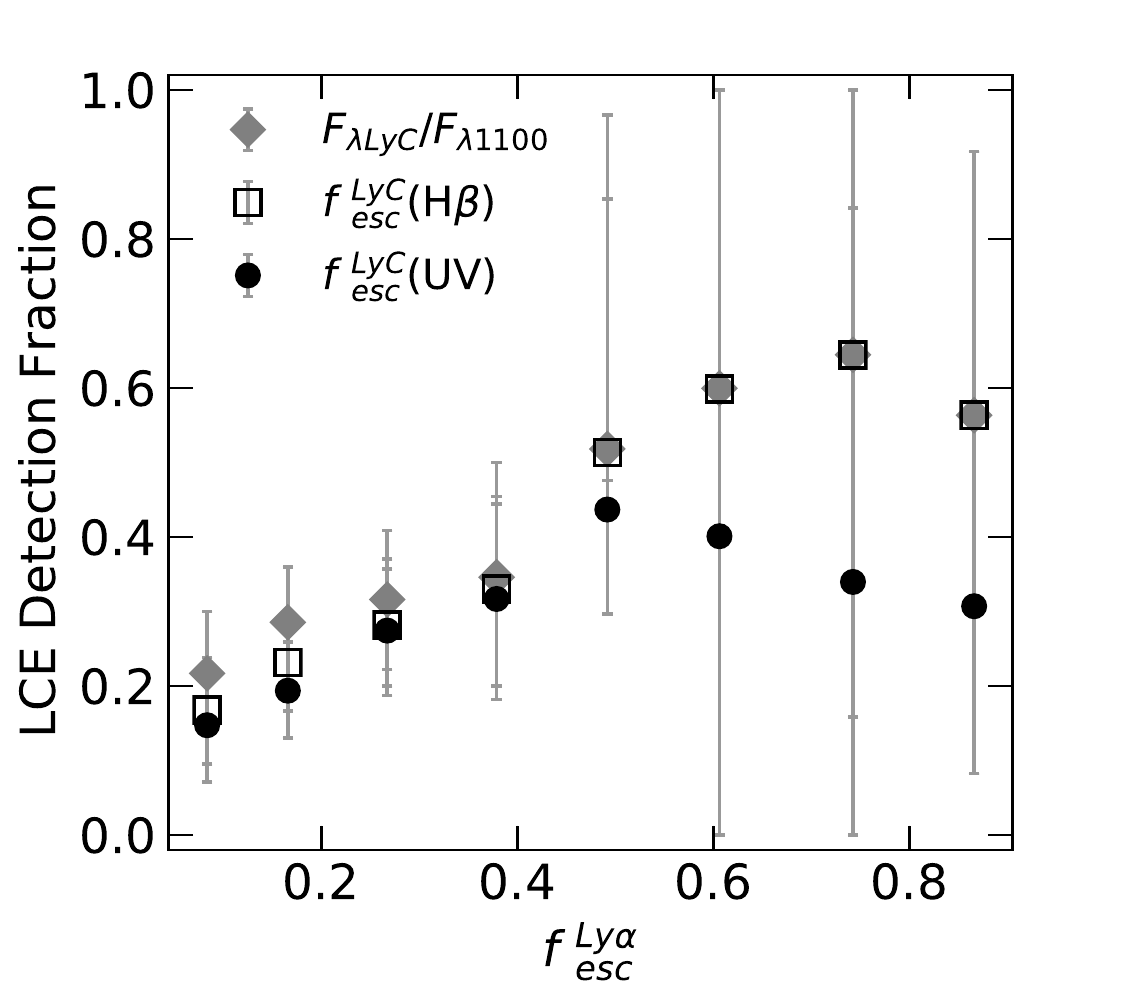}
    \includegraphics[width=0.32\linewidth]{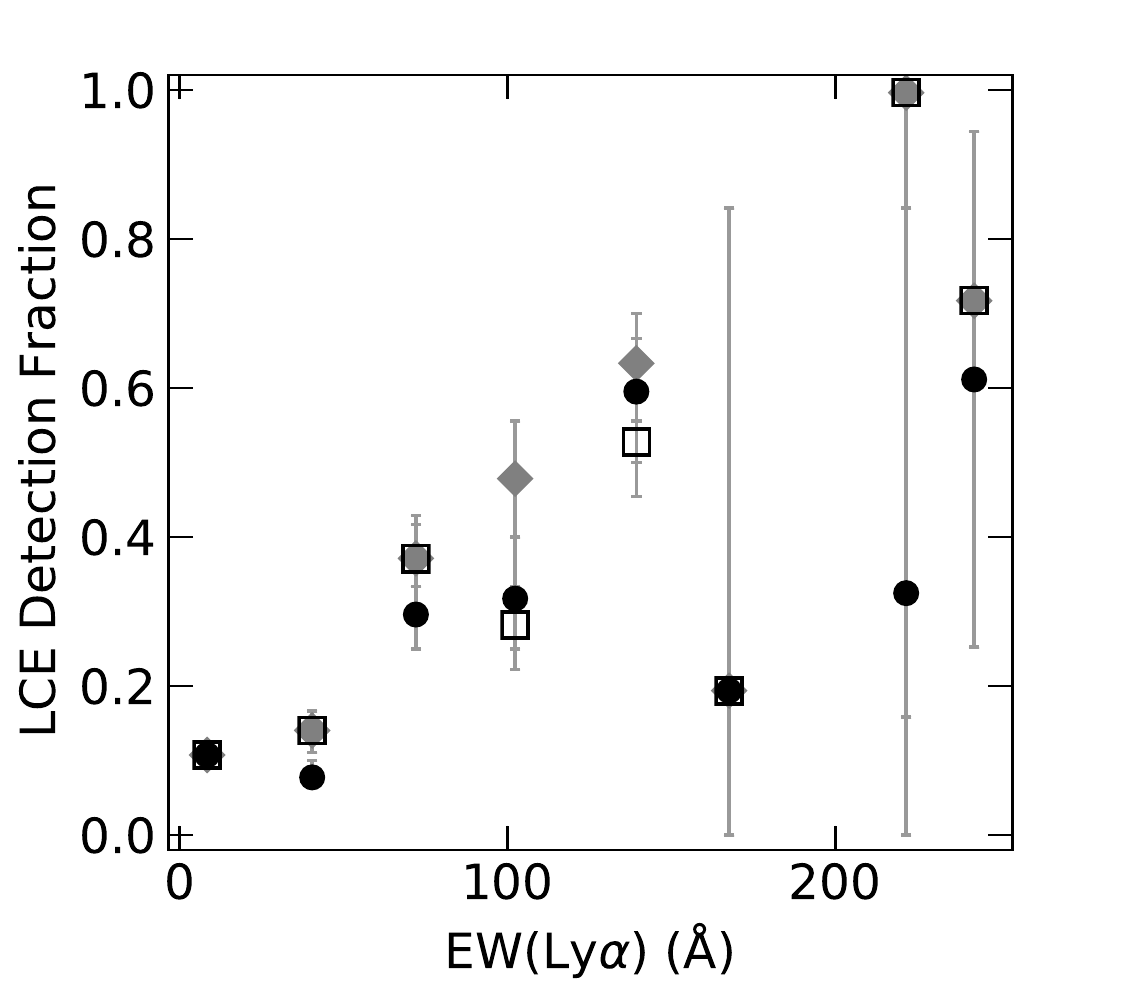}
    \includegraphics[width=0.32\linewidth]{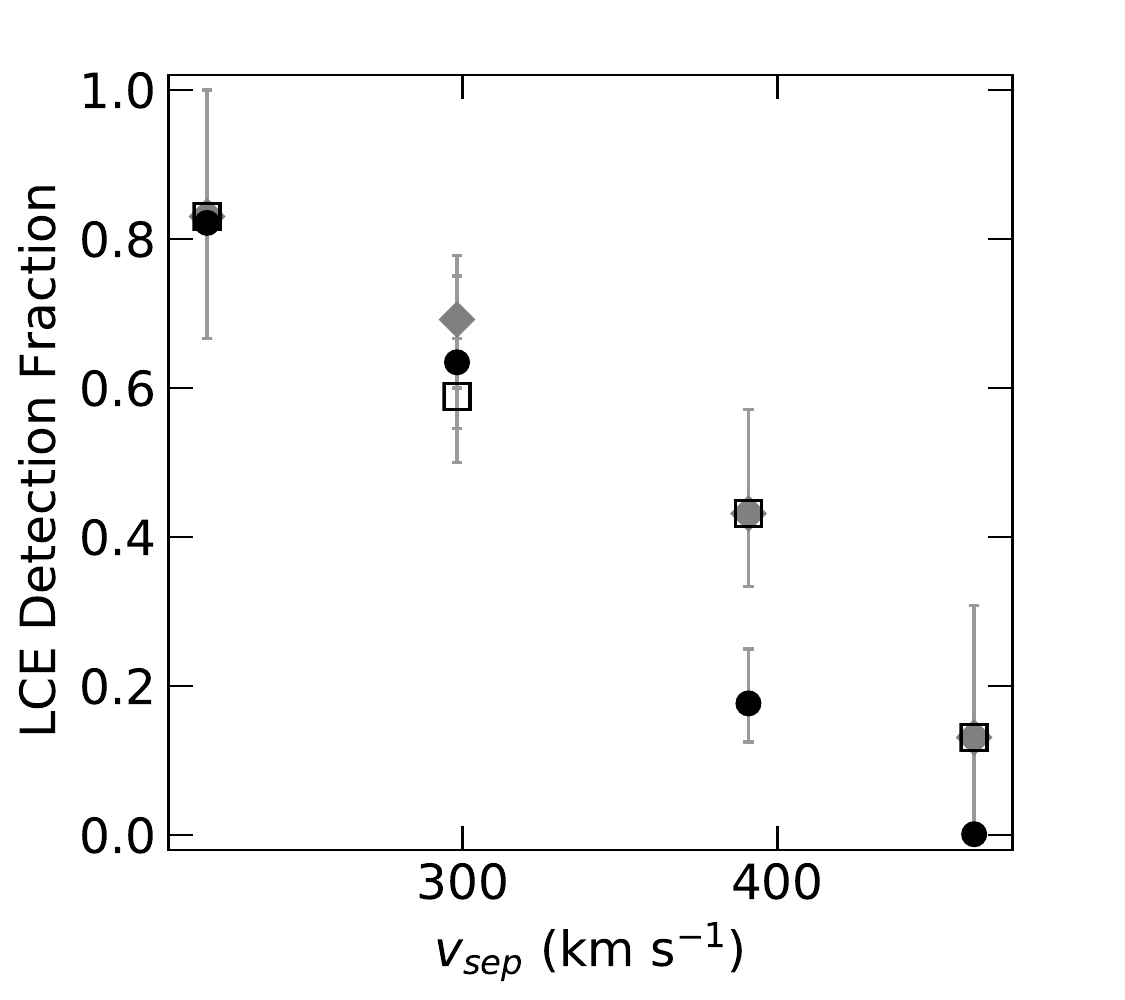}
    \caption{Fraction of galaxies in a given bin of \fesclya\ ({\it left}), EW(\lya) ({\it center}), or $v_{sep}$ ({\it right}) that are strong LCEs according to \fesclycrel\ (grey diamonds), \fesclyc(H$\beta$) (open squares), and \fesclyc(UV) (filled black circles) for the combined LzLCS and published sampels. \label{fig:lya_lce_frac}}
\end{figure*}

As both \lya\ and the LyC are sensitive to the line of sight, we anticipate the former to trace the LCE fraction rather well even as the emission line experiences higher optical depths than the continuum for the same column density. We demonstrate this relation is the case in Figure \ref{fig:lya_lce_frac}. The LCE fraction increases nearly monotonically with both \fesclya\ and the \lya\ EW. Deviations from this trend occur at high \fesclya\ due to two galaxies having abnormally high \lya/H$\beta$ ratios corresponding to values of \fesclya\ close to $1$.

{ The overwhelming success of all three \lya\ diagnostics in correlating with \fesclyc\ demonstrates that both \lya\ and LyC escape depend on the distribution of neutral hydrogen in a galaxy. While all three are promising candidates for \fesclyc\ indicators at high redshift, \lya\ EW is perhaps the easiest to measure as it requires neither rest-frame optical nor high resolution spectroscopy. The wealth of \lya\ EW measurements of LCEs at higher redshifts \citep[$z\sim3-4$, e.g.,][]{2018ApJ...869..123S,2018A&A...614A..11M,2019ApJ...878...87F,2021MNRAS.505.2447P} suggests these diagnostics will likely remain applicable for galaxies at the epoch of reionization. We discuss this possibility in detail in \S\ref{sec:disc_highz}.}

\subsubsection{\lya\ Escape Fraction}\label{sec:lyaesc}

{In Paper {\sc i}, we derived \fesclya\ from \lya\ and H$\beta$ emission line fluxes using Case B recombination coefficients determined by the measured electron temperatures and densities.} The \fesclyc exhibits a strong correlation with the fraction of escaping \lya\ photons. Consistent with previous studies \citep[e.g.,][]{2016ApJ...828...71D,2017A&A...597A..13V,2020MNRAS.491..468I}, we typically find that \fesclya$\ga$\fesclyc\ and that the envelope of \fesclyc\ increases with increasing \fesclya (Figure \ref{fig:fesc_lya}). This result illustrates that \lya\ has an advantage over the LyC in escaping a galaxy because it can resonantly scatter. However, \lya\ and LyC escape have a clear physical connection: our results in Figure \ref{fig:fesc_lya} suggest only \lya\ emitters (LAEs) can be strong LCEs. Indeed, from Figure \ref{fig:fesc_lya}, \fesclyc$=$\fesclya\ serves as an approximate envelope to the distribution of \fesclyc.

There remains a great deal of scatter in \fesclyc\ values for any given \fesclya, which is to be expected. Observed \lya\ flux depends on scattering effects from dust and gas along and outside the line of sight, making \fesclya\ sensitive to dust and gas content, ISM clumpiness, and gas dynamics \citep[e.g.,][]{2016ApJ...828...71D}. {While ISM content and geometry also affect \fesclyc, the scatter evident in Figure \ref{fig:fesc_lya} suggests the dependence of \fesclyc\ on these properties may differ from that of \fesclya.}

\subsubsection{\lya\ Equivalent Width}

\begin{figure*}
    \centering
    \includegraphics[width=0.32\textwidth]{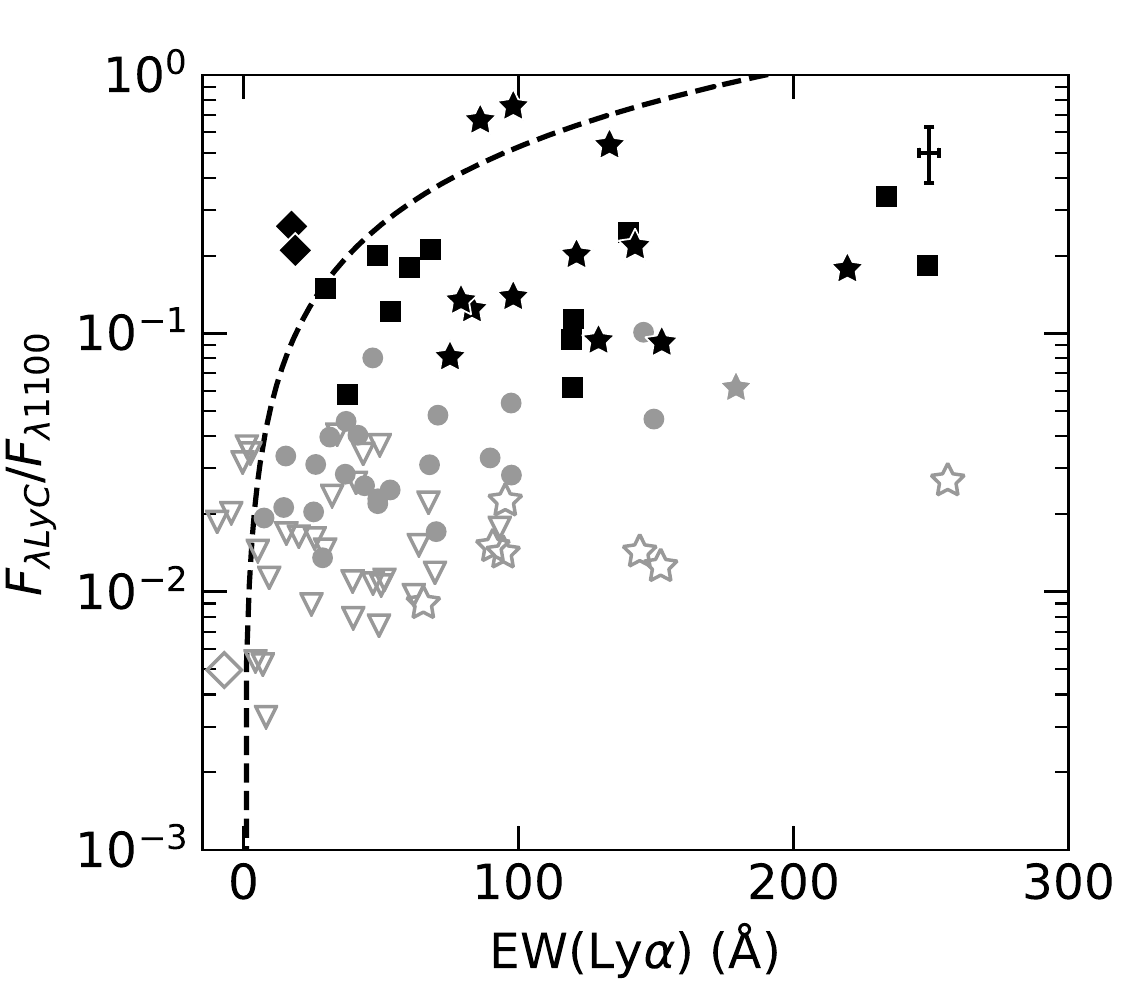}
    \includegraphics[width=0.32\textwidth]{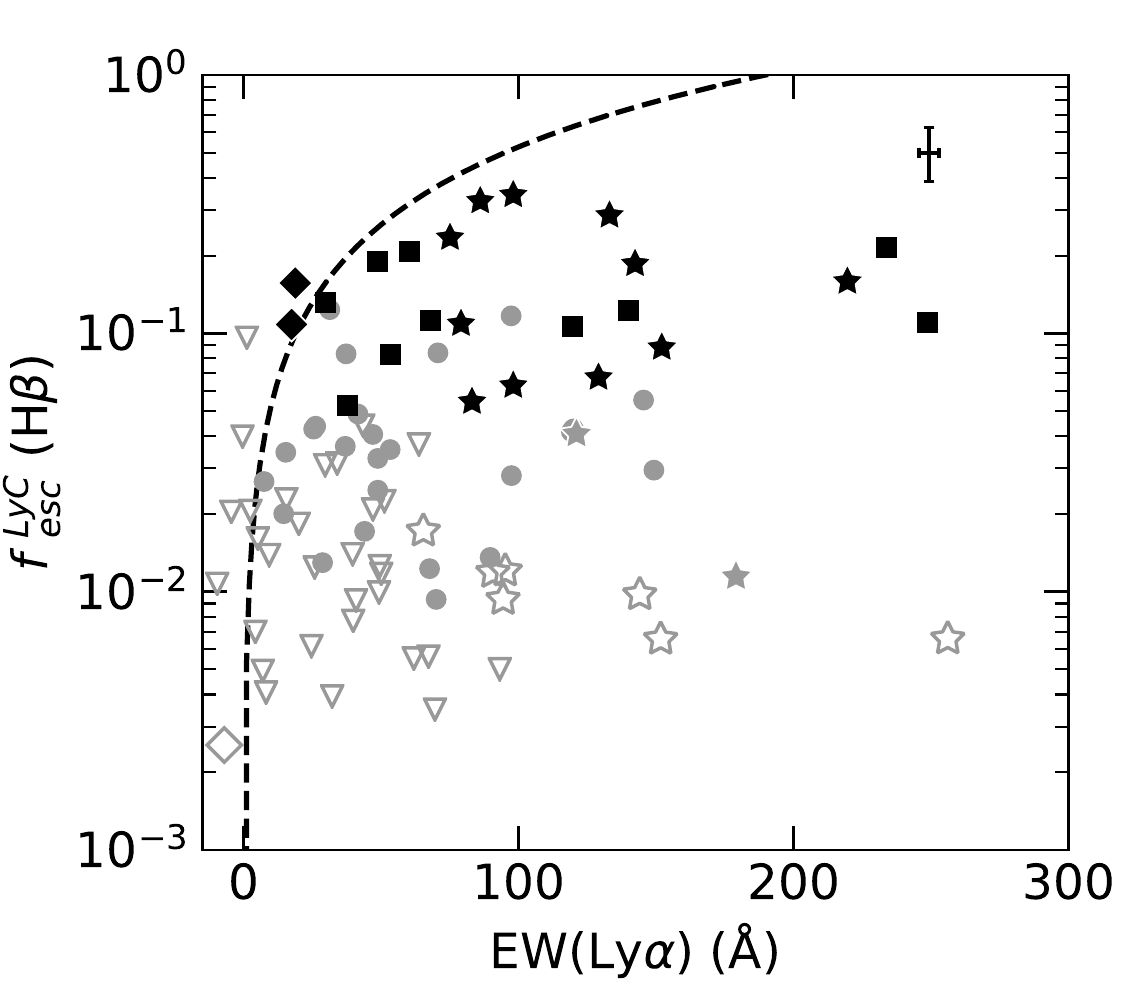}
    \includegraphics[width=0.32\textwidth]{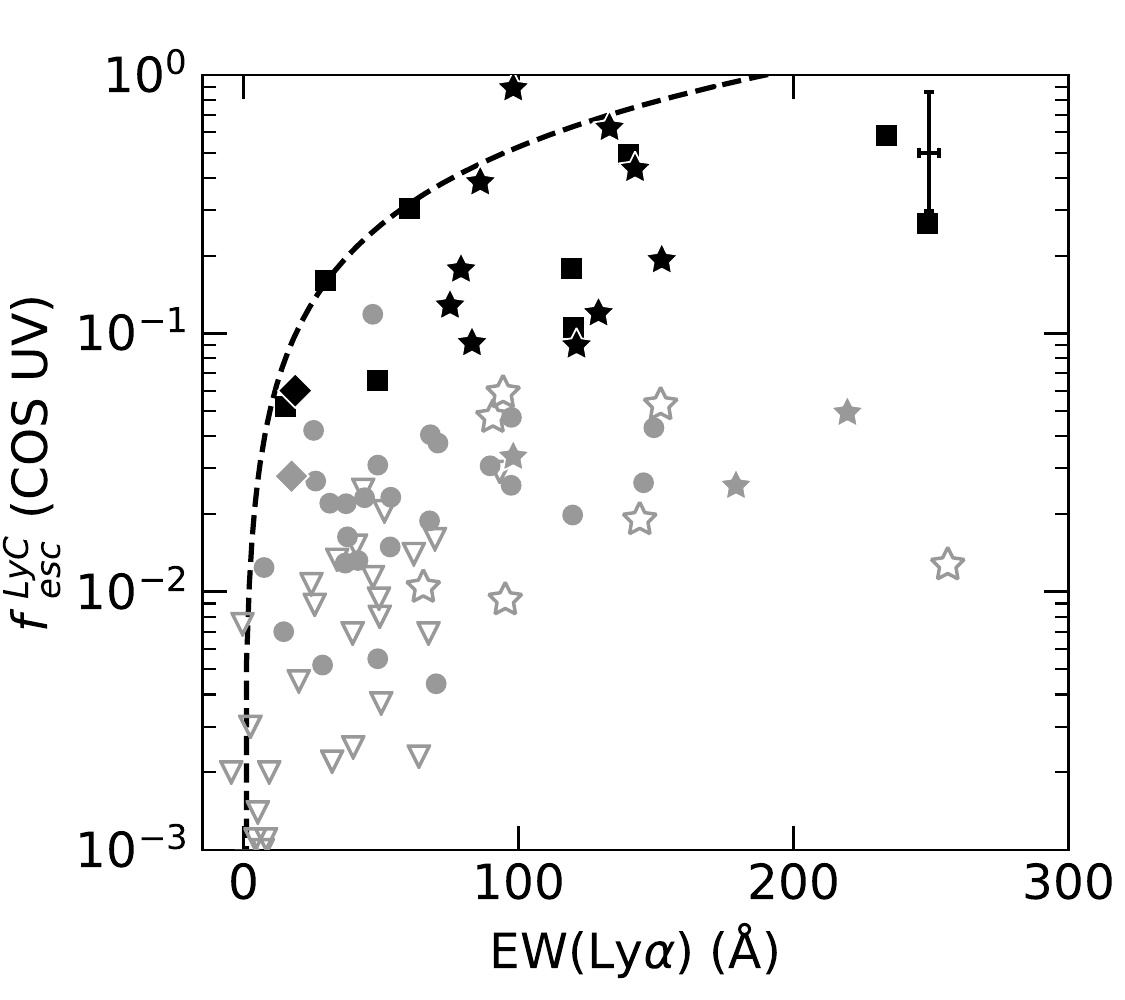}
    \caption{Same as Figure \ref{fig:fesc_lya} but for \lya\ EW. Dashed line is the relation from \citet{2021MNRAS.505.2447P} based on measurements of the LyC from stacks of spectra from galaxies at $z\sim3$.}
    \label{fig:ew_lya}
\end{figure*}

In place of \fesclya, we consider \lya\ EW as another possible indicator of \fesclyc. While less directly tracing optical depth than \fesclya, \lya\ EW is far easier to measure at high redshift because, unlike \fesclya, it does not require observing the Balmer lines. Furthermore, both \citet{2015ApJ...809...19H} and \citet{2017ApJ...844..171Y} demonstrate a strong correlation between \fesclya\ and \lya\ EW for Green Peas (GPs).

From Figure \ref{fig:ew_lya}, \fesclyc\ is bounded by an upper limit that increases with \lya\ EW. { We find that the relation between \lya\ EW and \fesclyc\ given by \citet{2021MNRAS.505.2447P} provides a reasonable description of this envelope in \fesclyc\ values up to $\sim150$ \AA. Interestingly, this trend extends beyond the 110 \AA\ maximum predicted by \citet{2018ApJ...869..123S}, perhaps due to their assumed continuous star formation history.

The correlation between \fesclyc\ and \lya\ EW is one of the stronger, more significant of the diagnostics we consider, with $\tau\approx0.3$ across all three \fesclyc\ indicators. Ten of the sixteen galaxies with high \lya\ EW are strong leakers, indicating a possible transition in \lya\ EW at 100 \AA. Galaxies with EWs above this value are far more likely to be strong LCEs than those below it (see also Figure \ref{fig:lya_lce_frac}). Further supporting this notion of a 100 \AA\ transition, LCEs with EWs below this value are not as prodigious emitters as above it, which may suggest LCEs with high EW contribute more significantly to the cosmic LyC budget. }

\lya\ is sensitive to both column density and \fesclyc, which affects the distribution of \fesclyc\ over the EWs. As the optical depth decreases, more \lya\ and LyC photons can escape. { The drop in column density will then begin to limit the intrinsic emission of \lya\ photons, causing \lya\ EW to decrease with increasing \fesclyc\ \citep{2014MNRAS.442..900N,2018ApJ...869..123S}.} This effect may explain why some strong LCEs persist at \lya\ EWs at or even below 100 \AA. { However, \lya\ EW also depends very strongly on the continuum flux and, by extension, the stellar population(s), making unclear how significantly \fesclyc\ affects the \lya\ EW. Nevertheless, \lya\ EW is still one of the most promising diagnostics for \fesclyc, performing comparably to \fesclya.}

\subsubsection{\lya\ Peak Velocity Separation}

\begin{figure*}
    \centering
    \includegraphics[width=0.32\textwidth]{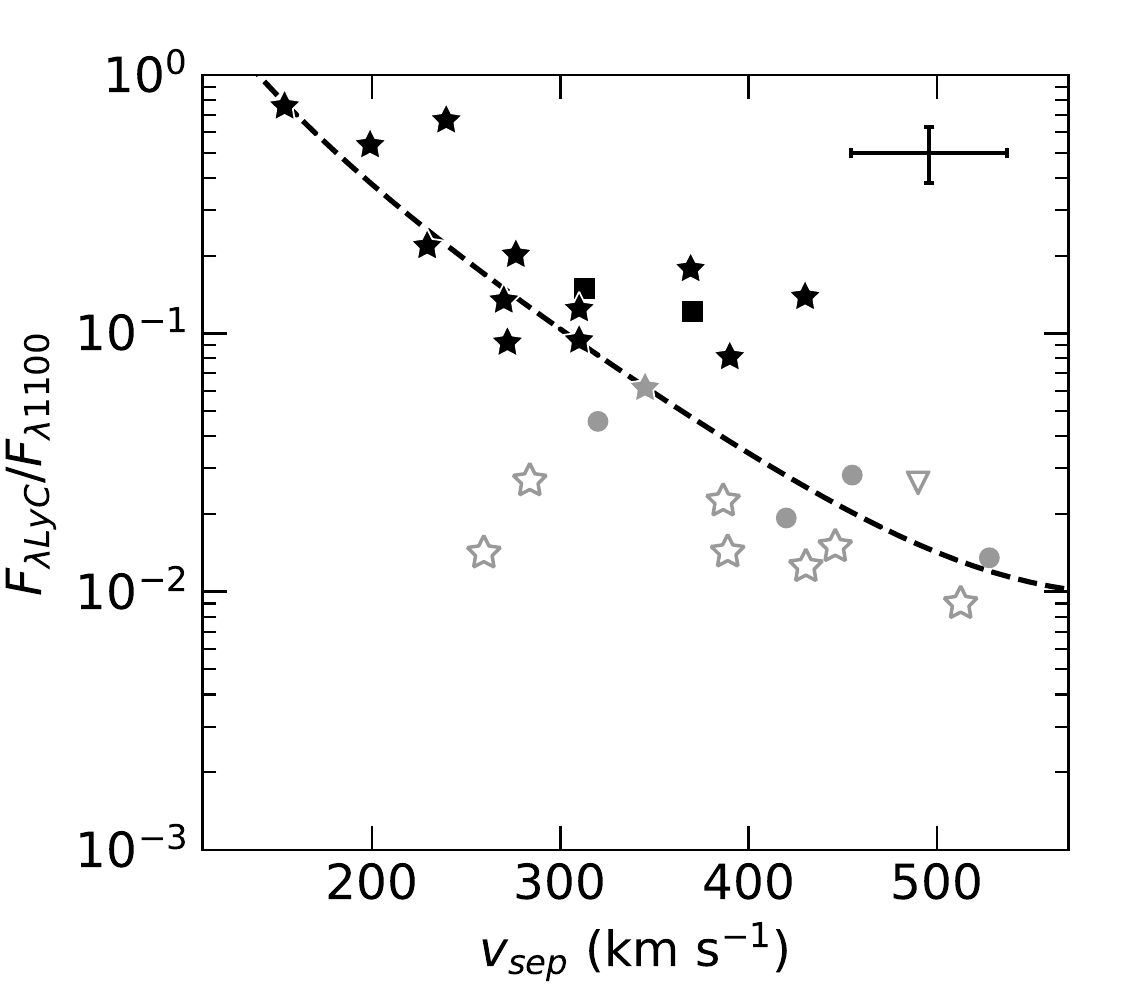}
    \includegraphics[width=0.32\textwidth]{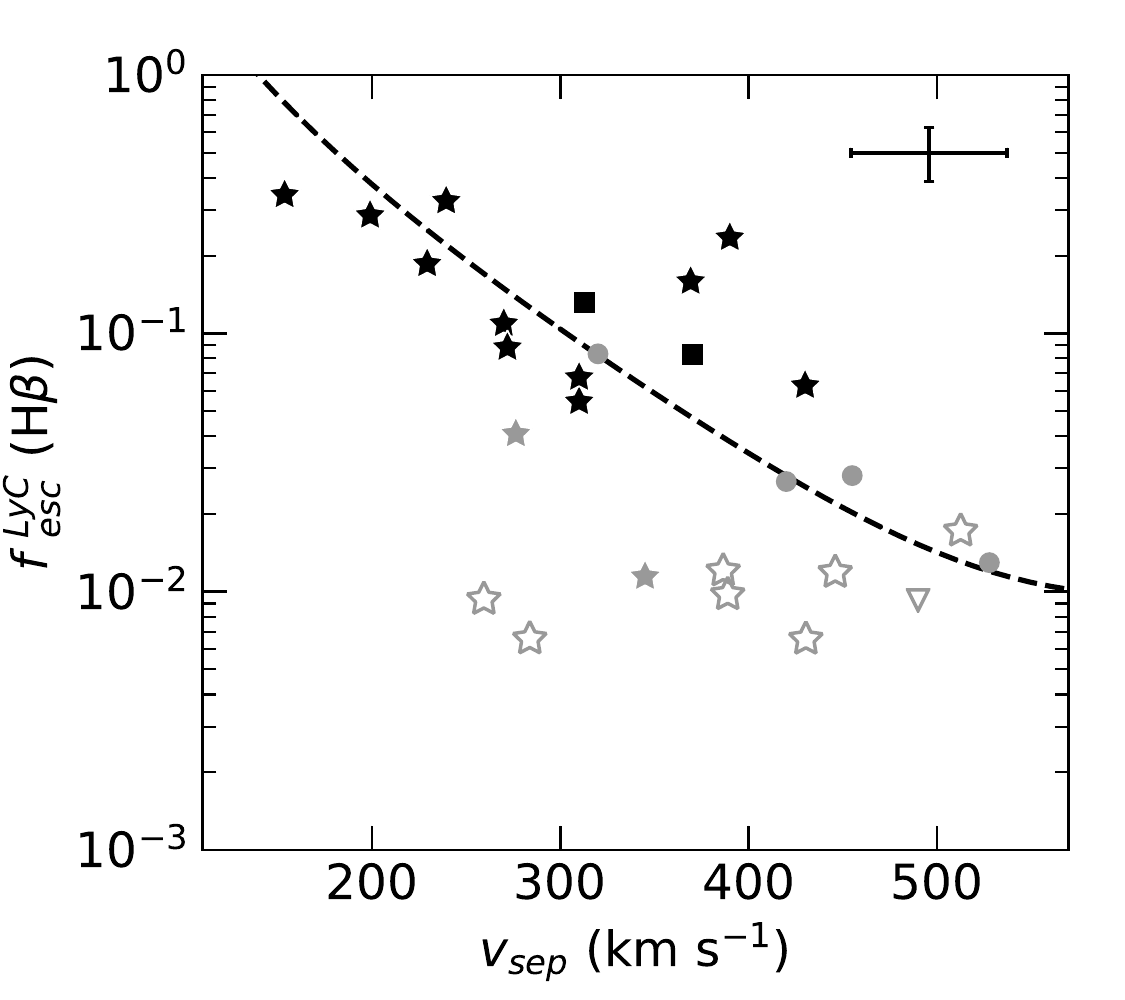}
    \includegraphics[width=0.32\textwidth]{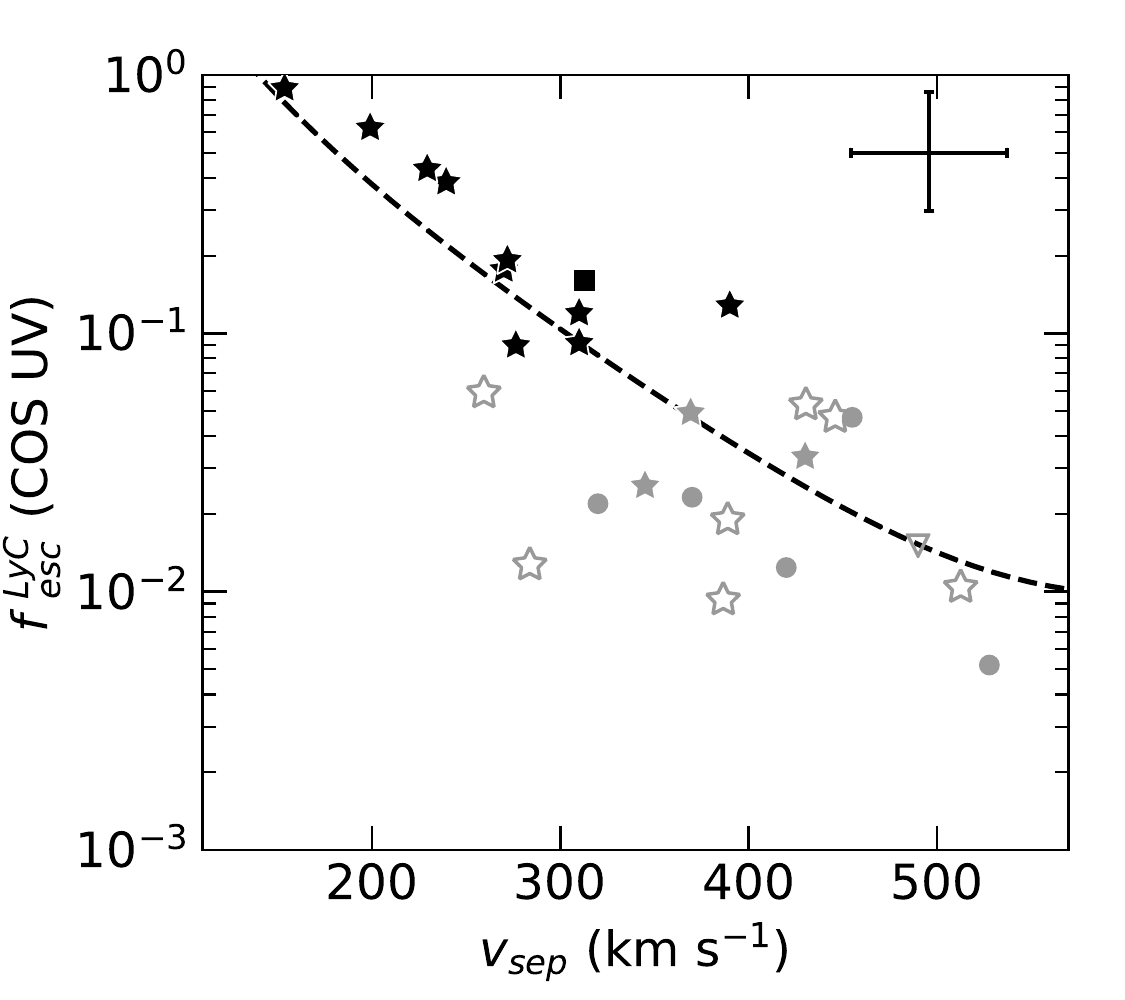}
    \caption{Same as Figure \ref{fig:fesc_lya} but for \lya\ peak velocity separation $v_{sep}$. Results include only 7 galaxies from the LzLCS plus the \Izotov\ samples. Dashed line is the description by \citet{2018MNRAS.478.4851I}.}
    \label{fig:v_sep}
\end{figure*}

{
\lya\ profiles are typically double-peaked with the velocity separation of the two peaks directly depending on the \ion{H}{1} column density \citep[e.g.,][]{2015A&A...578A...7V}. Previously, \citet{2018MNRAS.478.4851I,2021arXiv210301514I} demonstrated that the velocity separation $v_{sep}$ of \lya\ peaks strongly correlates with \fesclyc\ as a result of the strong dependence of both values on the \ion{H}{1} column. We compare \fesclyc\ with $v_{sep}$ for a subset of 7 LzLCS galaxies with G160M measurements \citep{2015ApJ...809...19H,2017ApJ...844..171Y} and a subset of 20 published LCEs \citep{2017A&A...597A..13V,2018MNRAS.478.4851I,2021arXiv210301514I} in Figure \ref{fig:v_sep}. We find a strong relationship between the values. Neither weak leakers nor non-detections persist below $v_{sep}\sim250$ km s$^{-1}$ regardless of \fesclyc\ metric, which is consistent with predictions by \citet{2015A&A...578A...7V}. This result favors the \ion{H}{1} column density interpretation that motivated this diagnostic.}

{
In Figure \ref{fig:v_sep}, we also compare \fesclyc\ and $v_{sep}$ to the relation from \citet{2018MNRAS.478.4851I} and find that their fit roughly describes the envelope of \fesclyc\ values. Moreover, we find $v_{sep}$ has one of the most pronounced correlations of any indirect \fesclyc\ indicator regardless of \fesclyc\ metric with $\tau\sim-0.4$.
However, additional measurements of $v_{sep}$ from \lya\ profiles are necessary to fully test this \fesclyc\ diagnostic.}

\subsection{$[$\ion{O}{1}$]\lambda6300$}

\begin{figure*}
    \centering
    \includegraphics[width=0.32\textwidth]{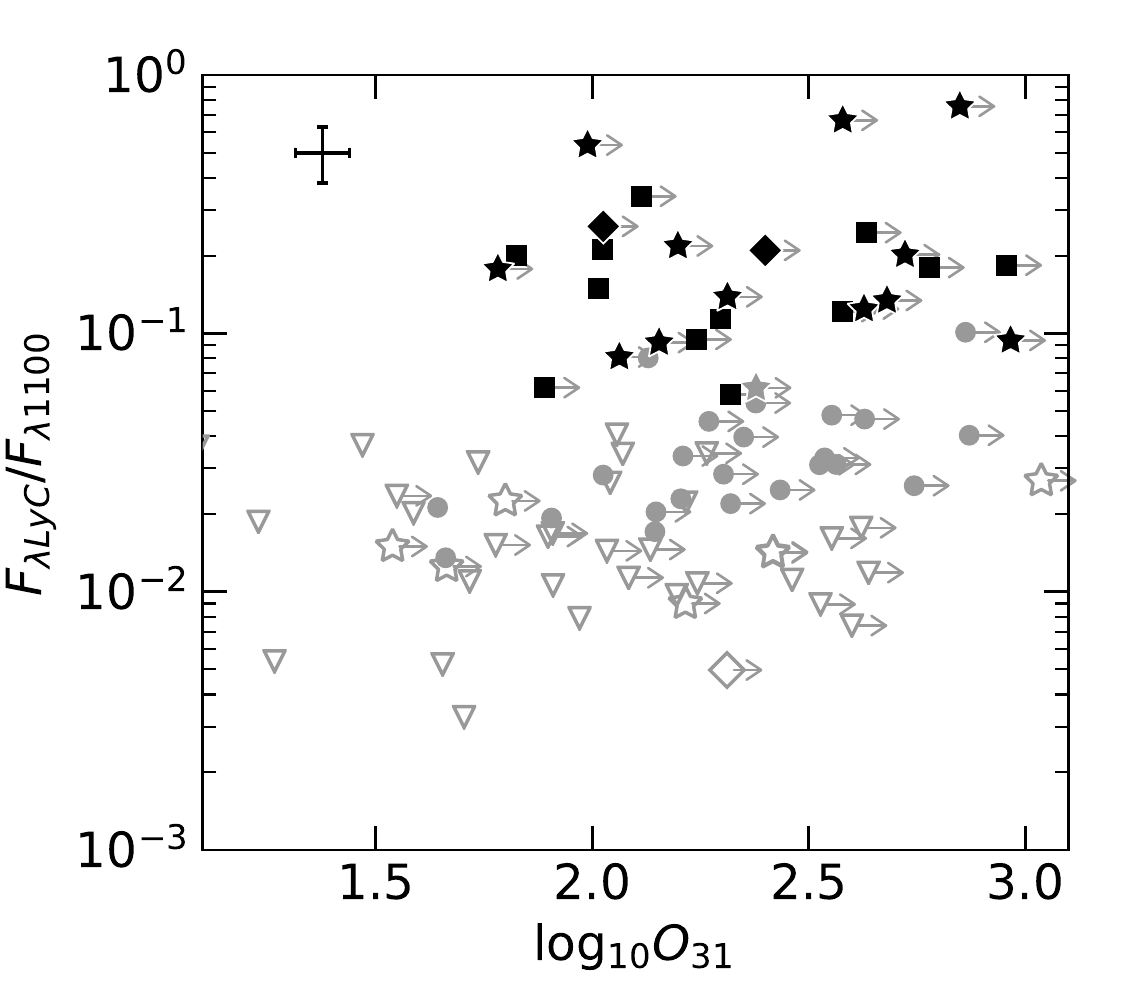}
    \includegraphics[width=0.32\textwidth]{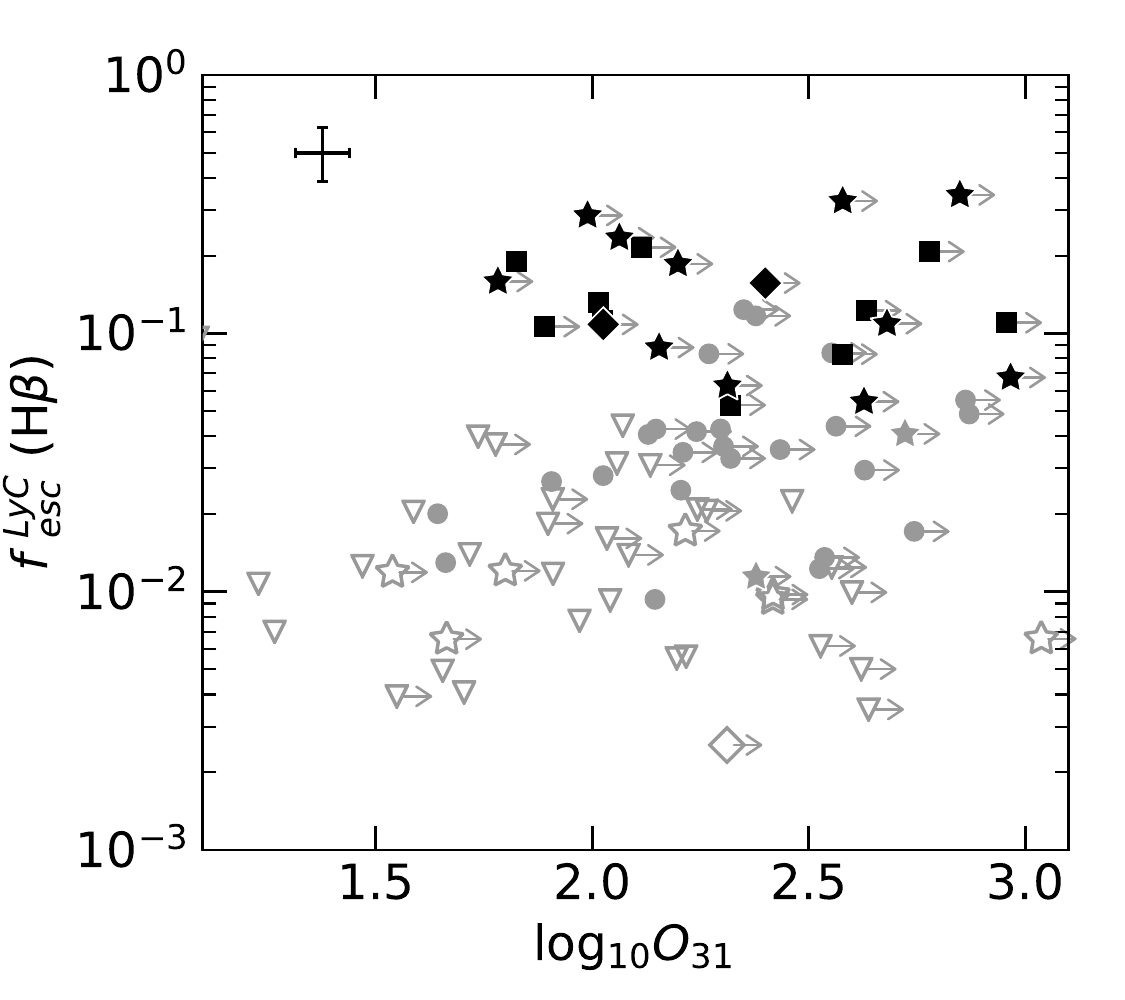}
    \includegraphics[width=0.32\textwidth]{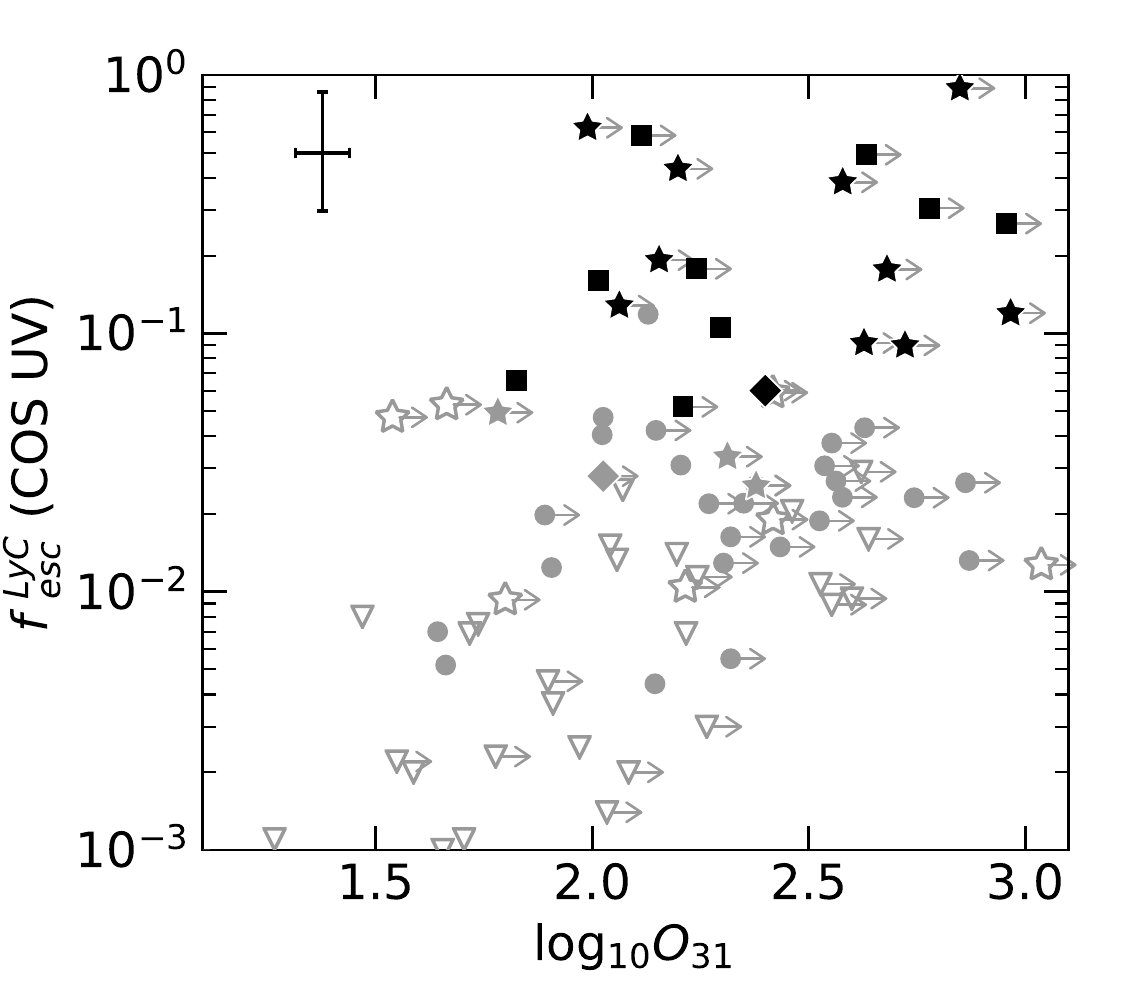}
    \includegraphics[width=0.32\textwidth]{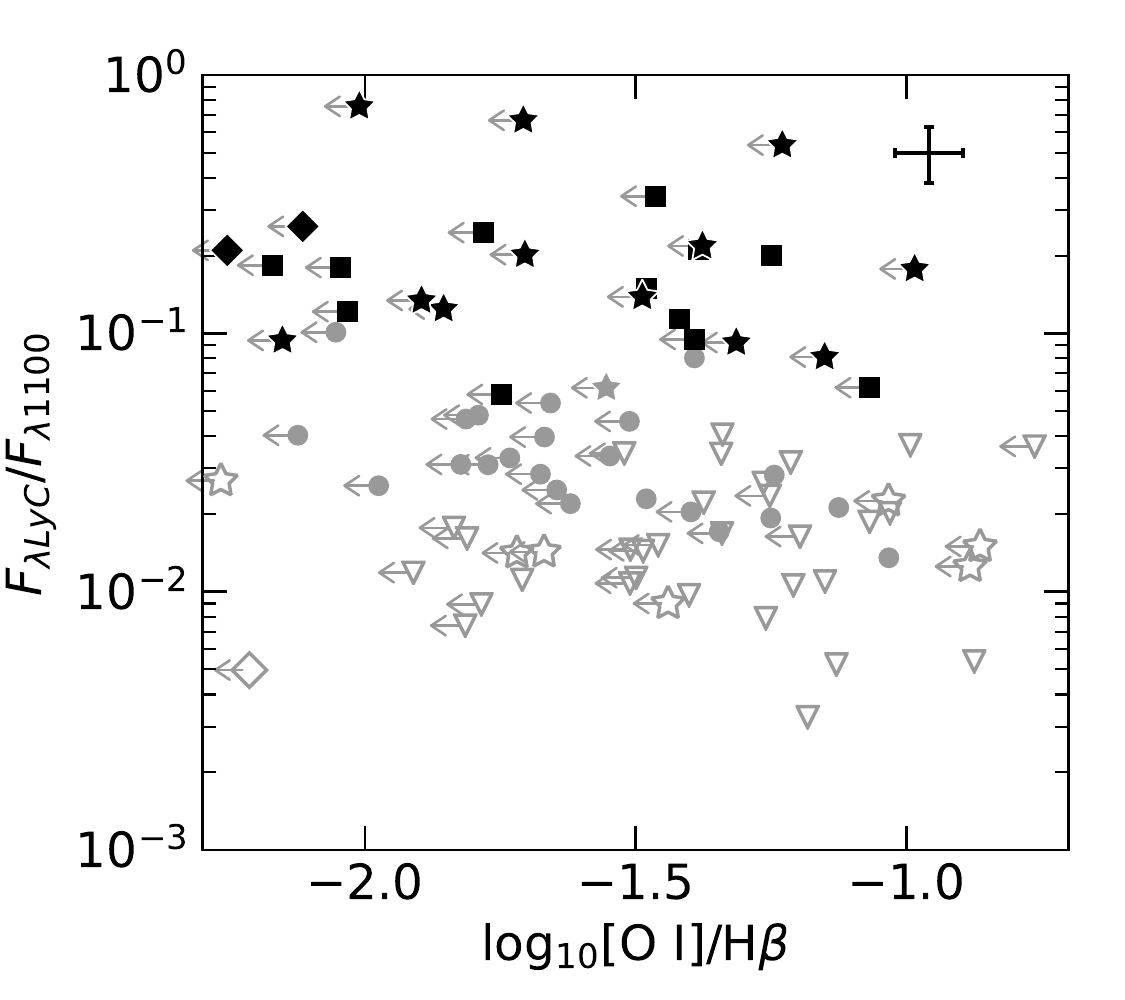}
    \includegraphics[width=0.32\textwidth]{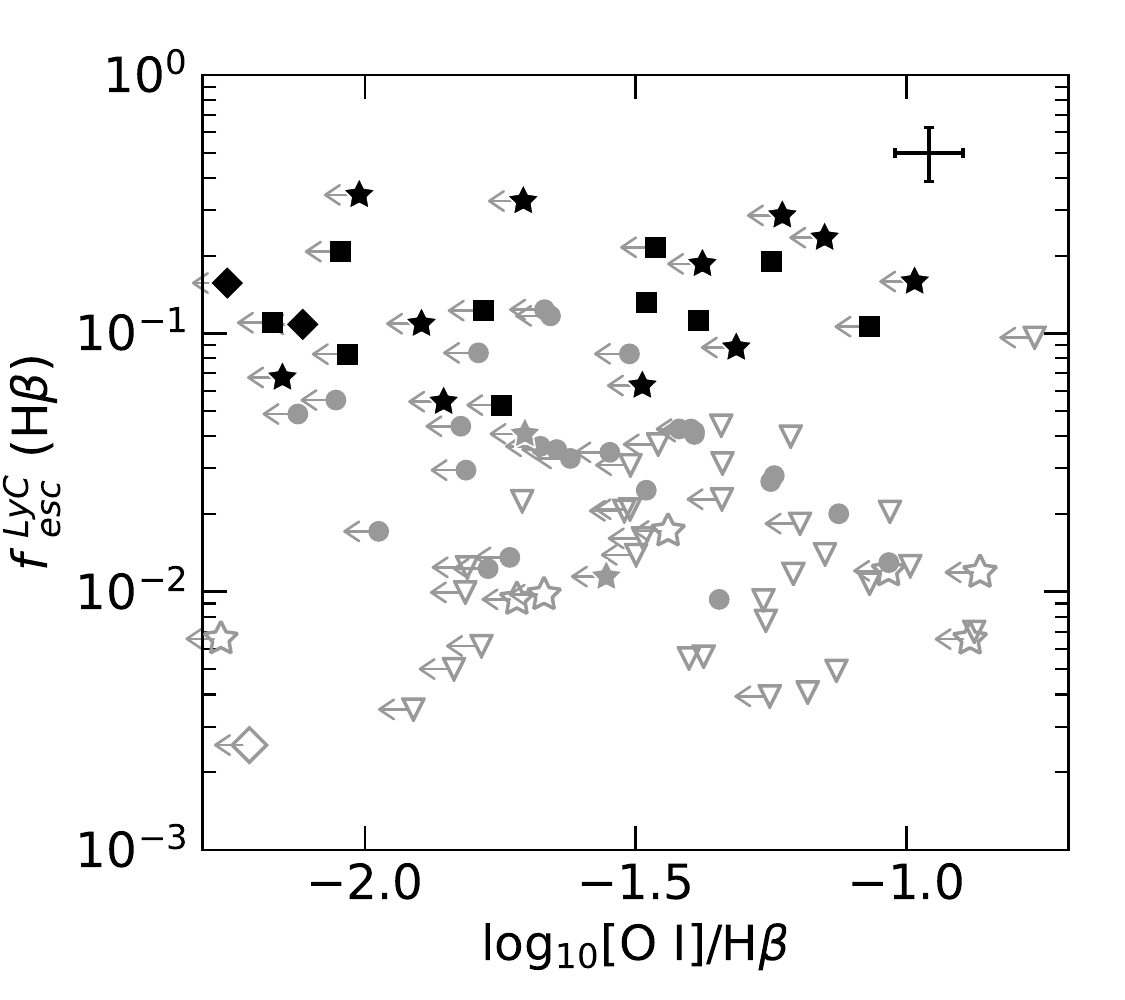}
    \includegraphics[width=0.32\textwidth]{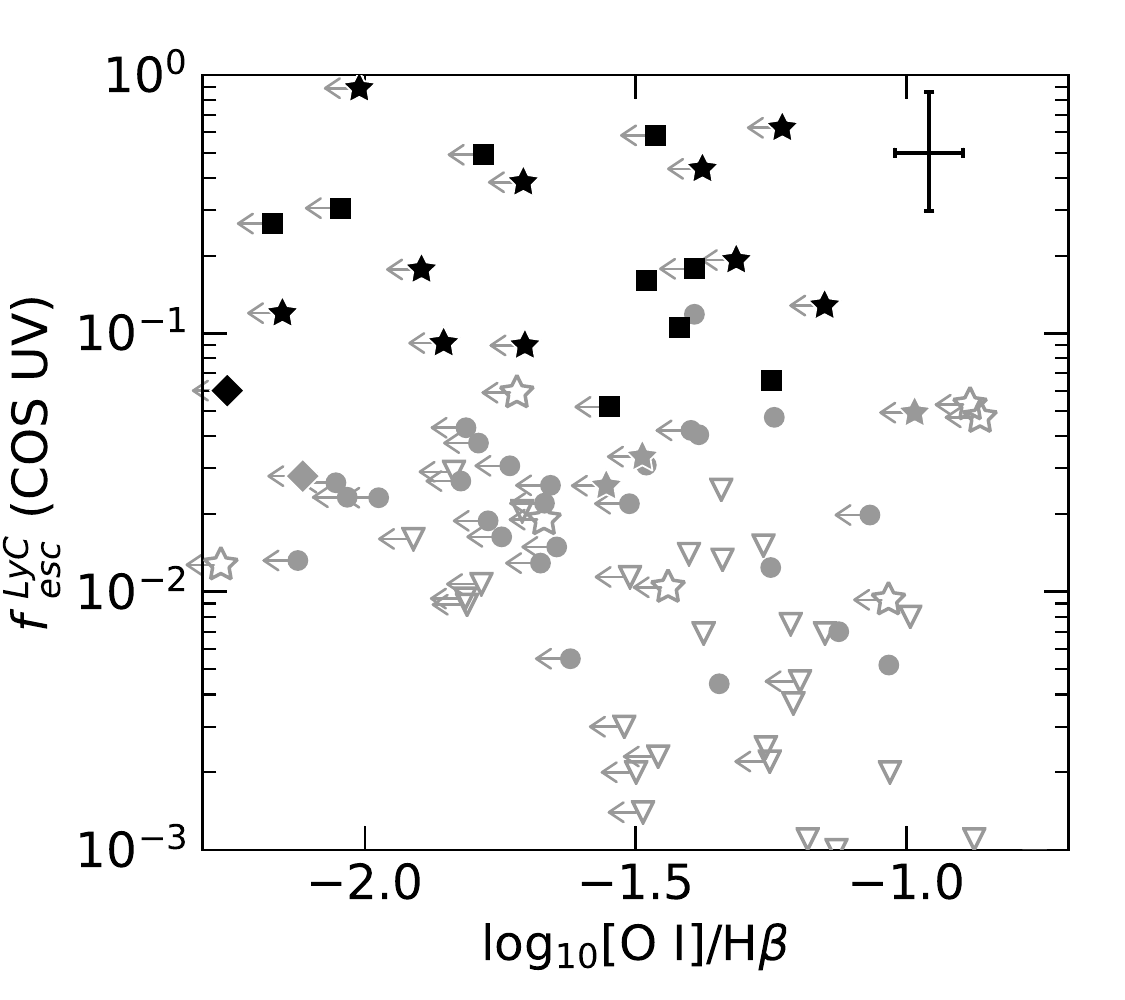}
    \caption{Same as Figure \ref{fig:fesc_lya} but for $O_{31}$ and [\ion{O}{1}]/H$\beta$.}
    \label{fig:o16300}
\end{figure*}

{ Few of the LCEs have detected [\ion{O}{1}]$\lambda6300$, with 39 of the 50 LCEs in the combined sample having only upper limits in the emission line flux, indicating that [\ion{O}{1}] may be particularly weak in LCEs \citep[see, however,][]{2019MNRAS.490..978P,2020A&A...644A..21R}.} The preponderance of [\ion{O}{1}] upper limits in LCEs over non-emitters prevents determination of the shape of the distribution of \fesclyc\ with respect to the [\ion{O}{1}] flux ratios, as evidence in Figure \ref{fig:o16300}. These upper limits may contribute to the lack of significant correlation (as shown in Table \ref{tab:corrcoef}, $p>0.09$ for all three \fesclyc\ metrics for both [\ion{O}{1}] flux ratios).

\begin{figure}[!ht]
    \centering
    \includegraphics[width=\columnwidth]{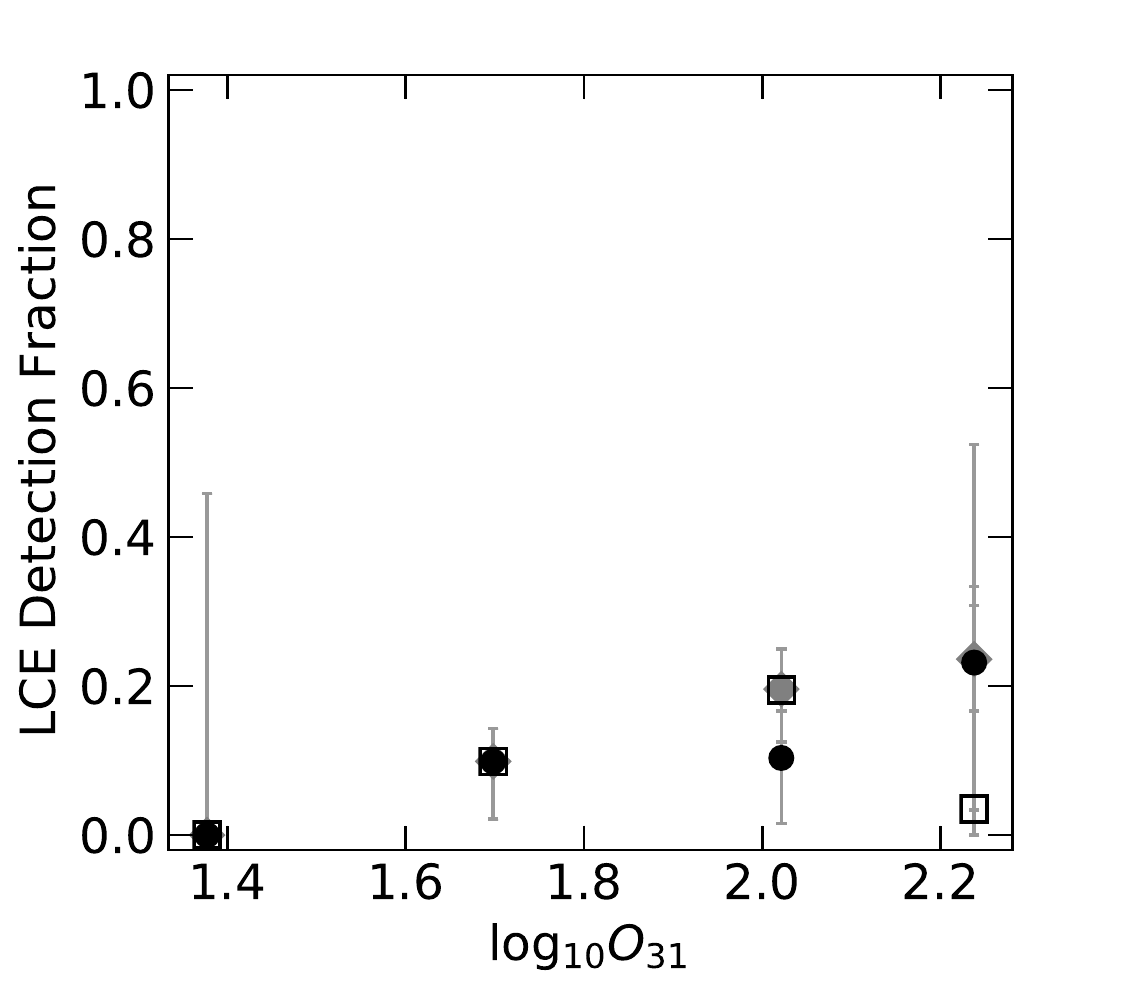}
    \includegraphics[width=\columnwidth]{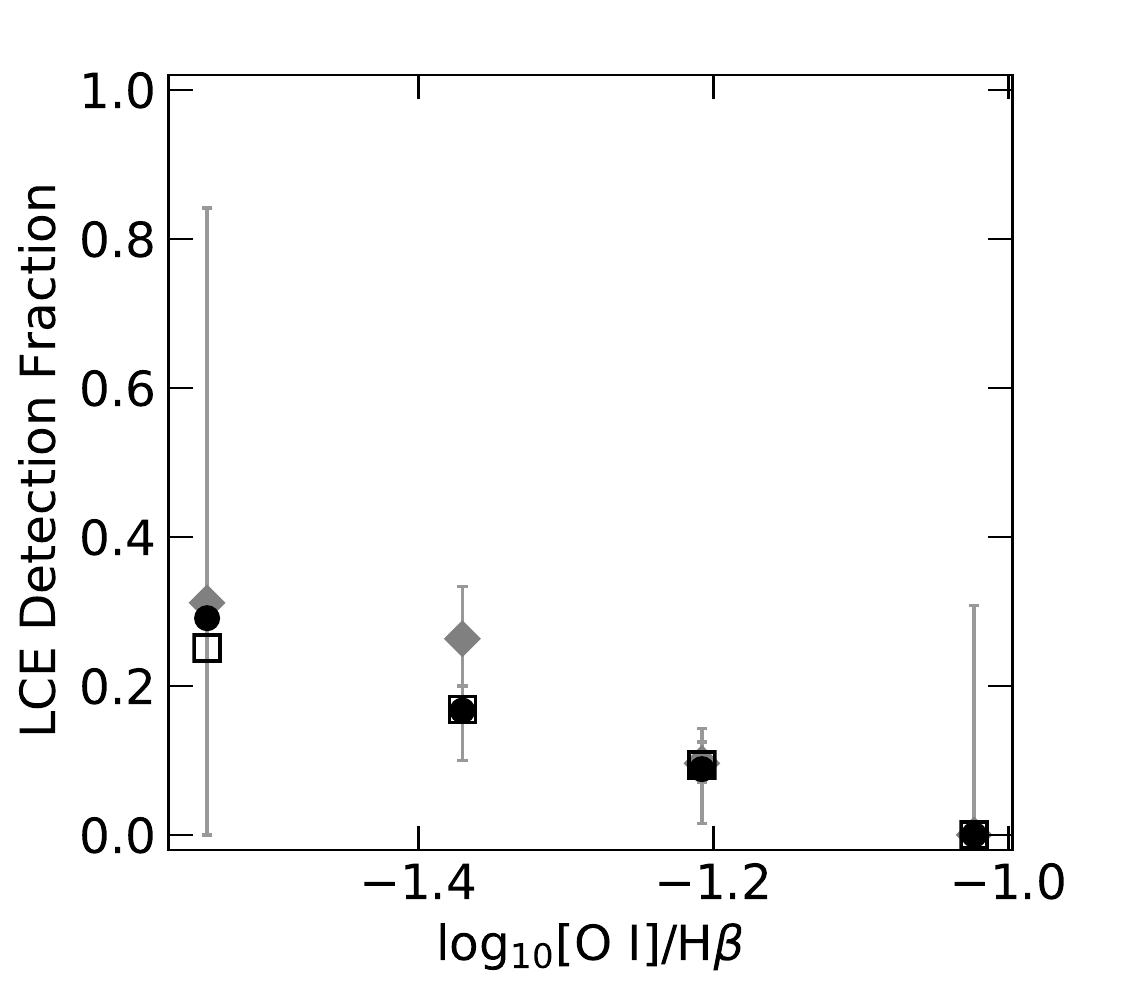}
    \caption{Same as Figure \ref{fig:lya_lce_frac} but for $O_{31}$ (left) and [\ion{O}{1}]/H$\beta$ (right).}
    \label{fig:oi6300_lce_frac}
\end{figure}

We compute the [\ion{O}{1}] LCE fraction using upper limits when no flux is detected. The LCE fraction distribution (Fig. \ref{fig:oi6300_lce_frac}) appears relatively monotonic over the [\ion{O}{1}] flux ratios $O_{31}$ and [\ion{O}{1}]/H$\beta$. As shown in Figure \ref{fig:oi6300_lce_frac}, there is little difference between the three LyC escape indicators. Above $\log_{10}O_{31}\sim1.8$ and below [\ion{O}{1}]/H$\beta\sim-1.3$, the LCE fraction is a roughly constant 30\%. The trends with both flux ratios suggest that LCE detection fraction increases with decreasing [\ion{O}{1}] flux.

\subsection{\orat}

\begin{figure*}
    \centering
    \includegraphics[width=0.32\textwidth]{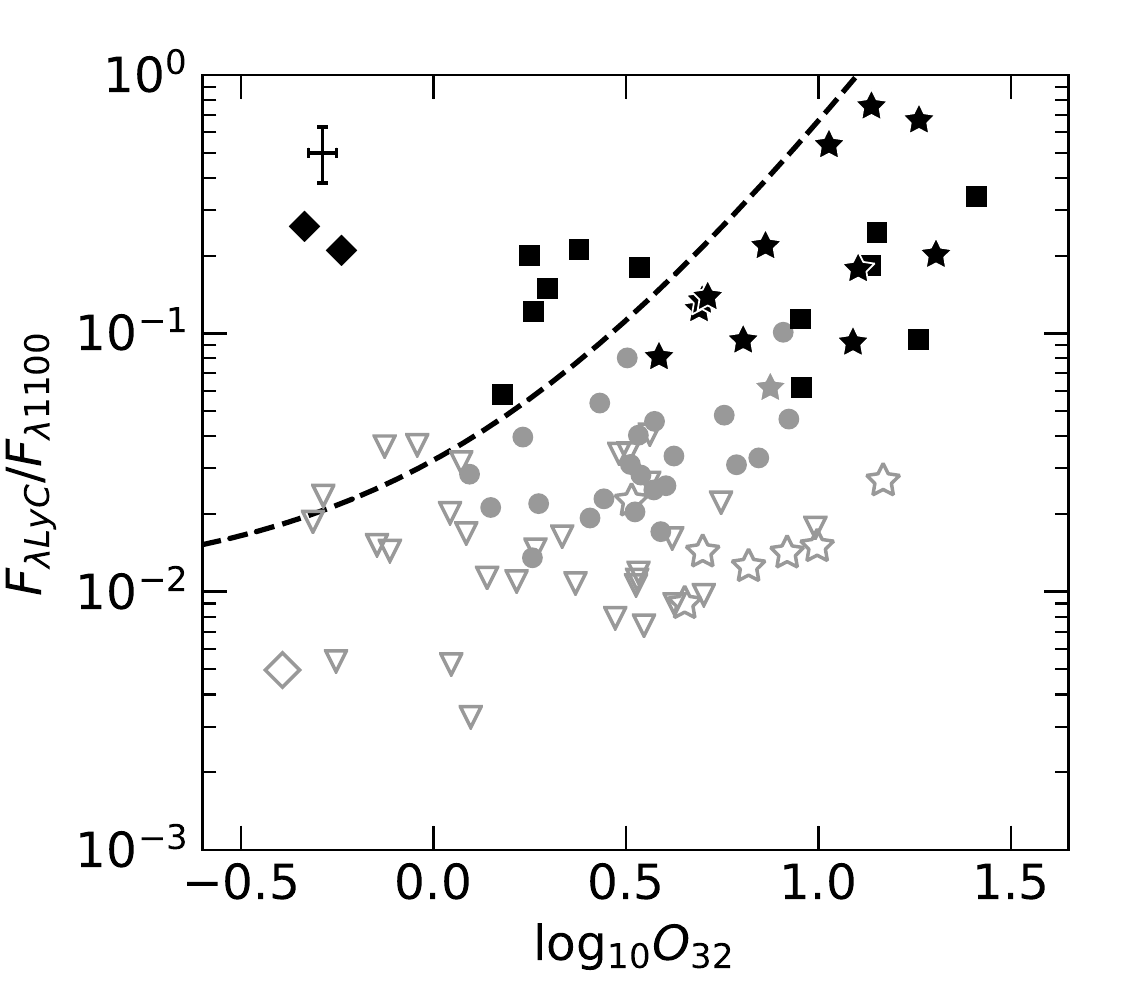}
    \includegraphics[width=0.32\textwidth]{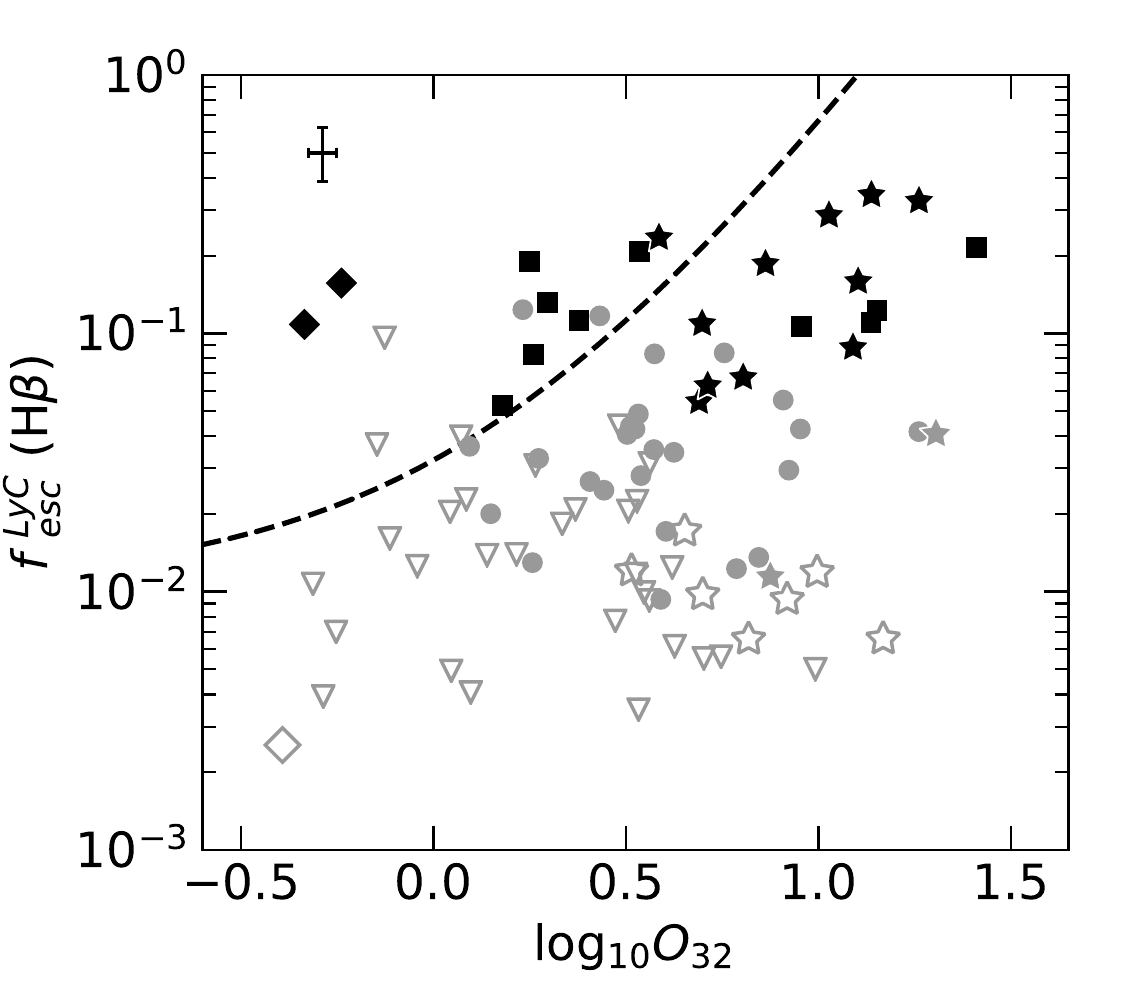}
    \includegraphics[width=0.32\textwidth]{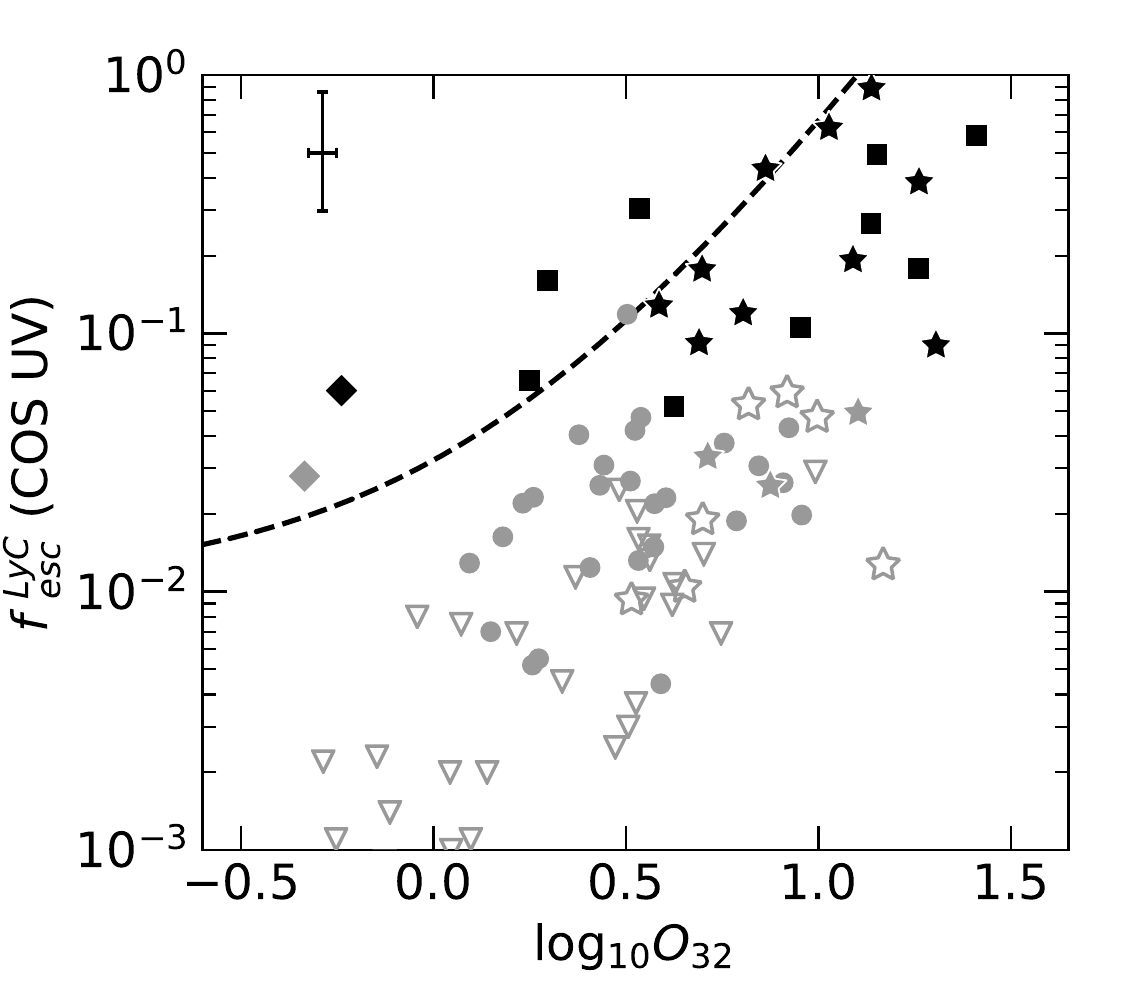}
    \caption{Same as Figure \ref{fig:fesc_lya} but for the extinction-corrected \orat\ flux ratio. Dashed line indicates the \citet{2018MNRAS.474.4514I} relation. Note that the number of strong LCEs changes depending on the \fesclyc\ metric.}
    \label{fig:o32}
\end{figure*}

\begin{figure}
    \centering
    \includegraphics[width=\columnwidth]{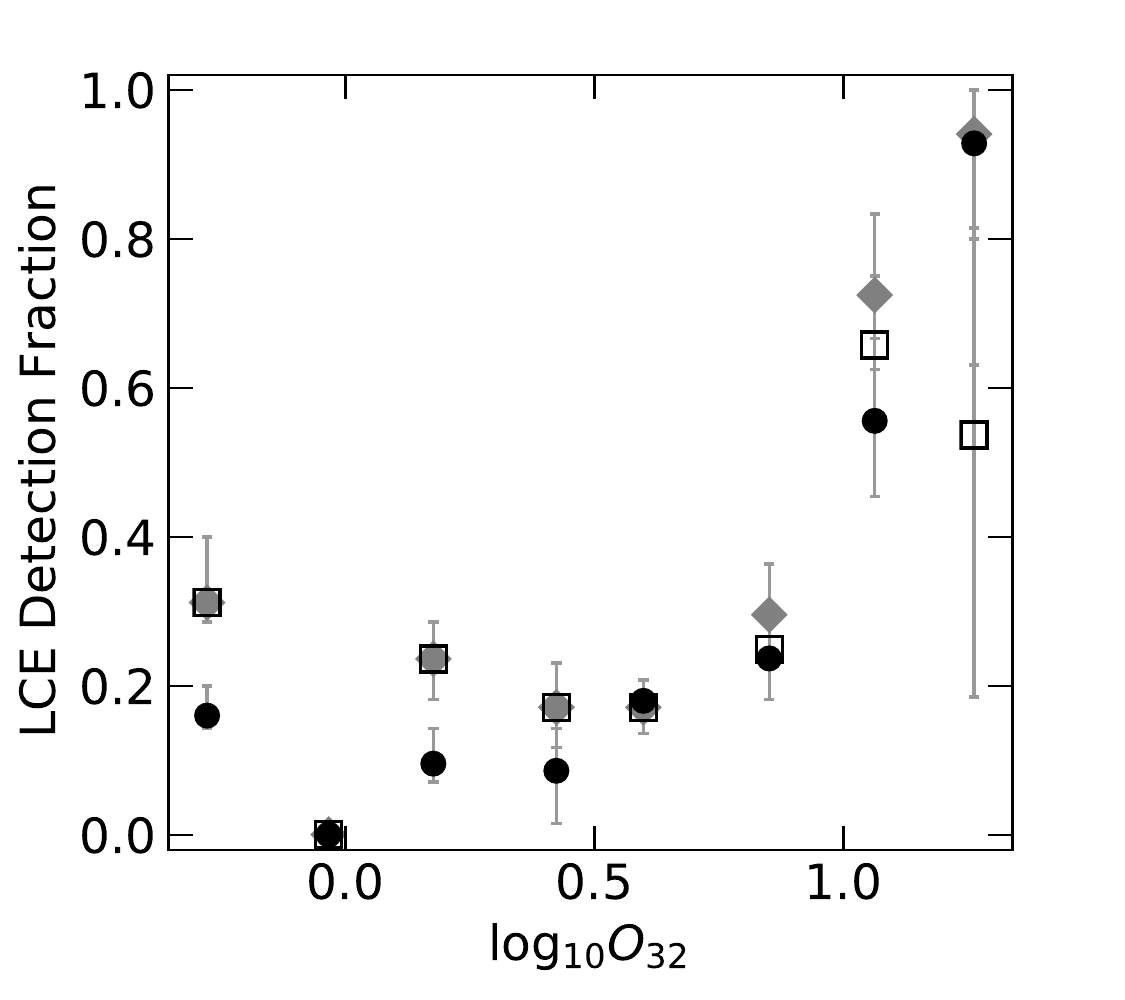}
    \caption{Same as Figure \ref{fig:lya_lce_frac} but for \orat.
    \label{fig:o32_lce_frac}}
\end{figure}

As shown in Figure \ref{fig:o32}, we see increasing average and maximum \fesclyc\ with increasing \orat. { The total combined sample exhibits a significant ($|\tau|>0.2$ at $>3\sigma$ confidence) correlation with \orat\ in all three \fesclyc\ metrics (see Table \ref{tab:corrcoef}). The \fesclyc\ relation proposed in \citet[][dashed line in Figure \ref{fig:o32}]{2018MNRAS.474.4514I} serves as a description of the envelope in \fesclyc\ values, particularly in the case of the UV-derived escape fraction (right panel of Figure \ref{fig:o32}). For the other two \fesclyc\ metrics, several strong LCEs have \fesclyc\ in excess of the predicted relation, notably the two LCEs from  \citet{2019ApJ...885...57W}. These outlying galaxies have higher stellar masses $>10^9$ M$_\odot$, which makes values life \fesclycrel\ and \fesclyc(H$\beta$) more sensitive to the star formation history. Because the majority of the outlying objects exhibit agreement with this envelope when using the UV \fesclyc\ metric (which assumes a non-parametric star formation history), we interpret these outliers as having \fesclyc\ systematically biased by this effect rather than evidence that the \citet{2018MNRAS.478.4851I} prescription is not appropriate at low \orat.}

{Scatter of \fesclyc\ below the envelope could be caused by a number of effects related to line-of-sight optical depth and other galaxy properties. Galaxy-to-galaxy variations in opening angle, covering fraction, burst age, ionization parameter, star-formation rate, metallicity, extinction law and dust, and/or optical depth may indicate that additional physical properties are relevant to determining \fesclyc.} Unfortunately the degenerate dependence of \orat\ on metallicity and ionization parameter substantially complicates any physical interpretation of this result \citep{Sawant2021}. { While the mass-metallicity relation appears to evolve with redshift \citep{2021ApJ...914...19S}, the coupling of $\rm 12 + \log(O/H)$ to ionization parameter observed at low redshift persists at $z\sim2-4$ \citep{2020MNRAS.491.1427S}. Therefore, the underlying properties which affect \orat\ remain related in the same way regardless of epoch, supporting the extension of the \orat\ diagnostic to higher redshifts. Regardless of the physical connection between \orat\ and \fesclyc, the scatter in this diagnostic indicates that an isotropic density-bounded escape scenario alone is likely incompatible with the observations. We discuss interpretations and caveats of \orat\ as an \fesclyc\ diagnostic in greater detail in \S\ref{sec:2d-diag} and \S\ref{sec:discussion}.}

{ With these nuances to \orat\ in mind, we do find that \fesclyc\ correlates strongly with \orat, particularly for \fesclyc\ derived from the UV spectrum. Further evidence for this relationship is the LCE fraction shown in Figure \ref{fig:o32_lce_frac}. The highest \orat\ galaxies in the combined sample (\orat$\ga$10) have LCE fractions $\ga0.5$. However, LCE fractions $>0.1$ persist across the full range of \orat\ values regardless of \fesclyc\ indicator. As we demonstrate in Figure \ref{fig:o32}, while most galaxies with \orat$>5$ in the combined sample are strong LCEs along the line of sight, some galaxies with lower \orat\ can still be strong LCEs.} Two scenarios might contribute to this trend with \orat: separate populations of LCEs with LyC escaping under different conditions or an evolutionary sequence in which LyC escape recurs at later times when fewer early type stars persist.

\subsection{H$\beta$ Equivalent Width}\label{sec:hbew}

\begin{figure*}
    \centering
    \includegraphics[width=0.32\textwidth]{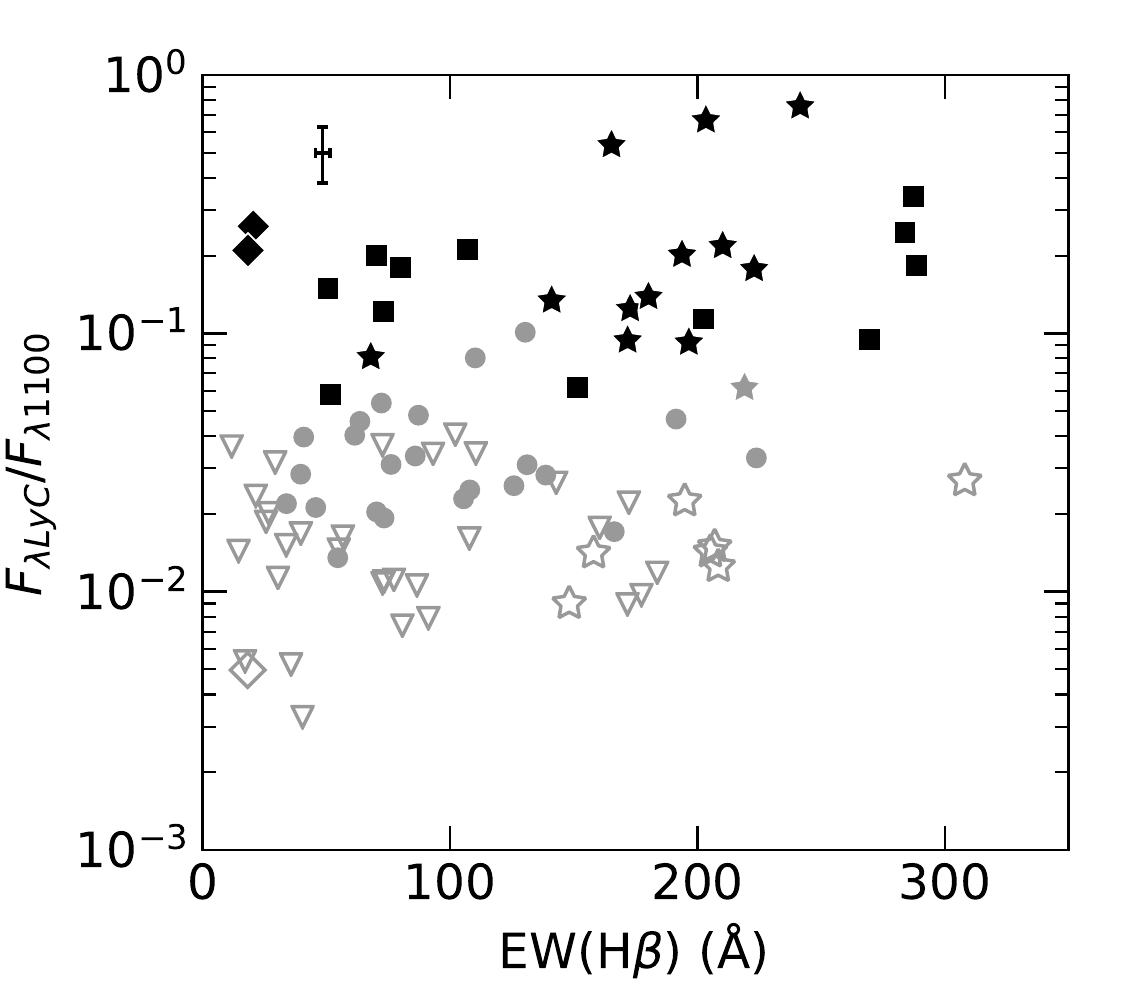}
    \includegraphics[width=0.32\textwidth]{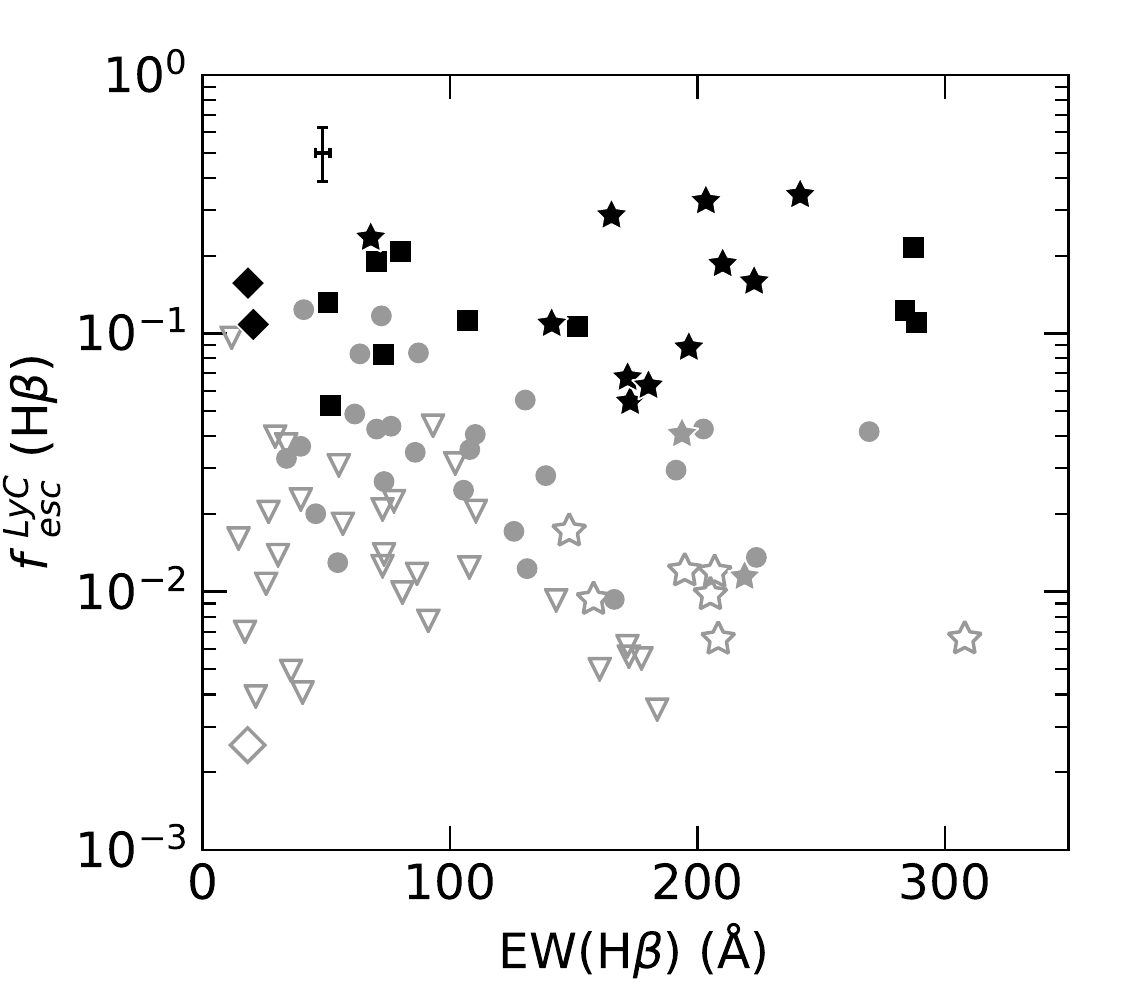}
    \includegraphics[width=0.32\textwidth]{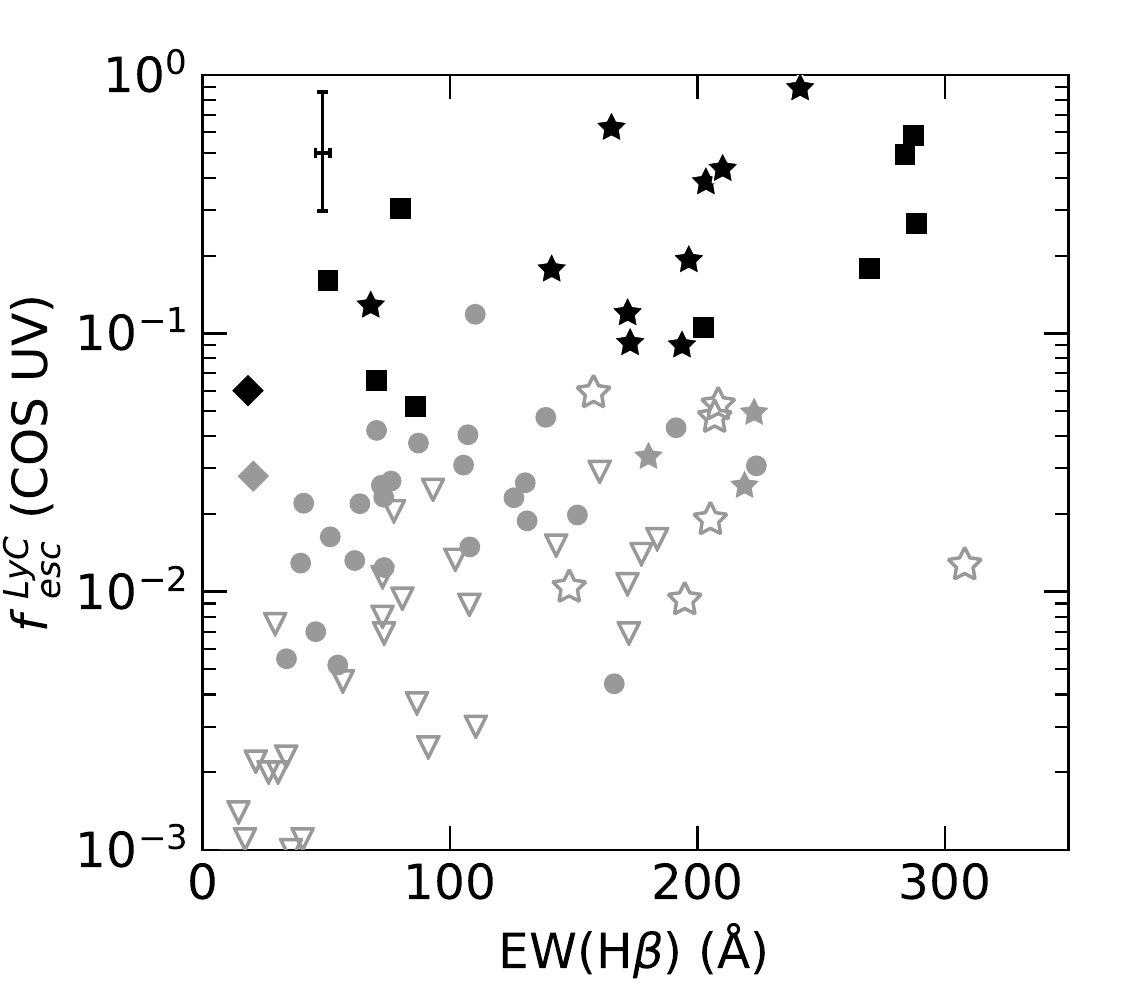}
    \caption{Same as Figure \ref{fig:fesc_lya} but for the rest-frame H$\beta$ EW.}
    \label{fig:ew_hb}
\end{figure*}

\begin{figure}
    \centering
    \includegraphics[width=\columnwidth]{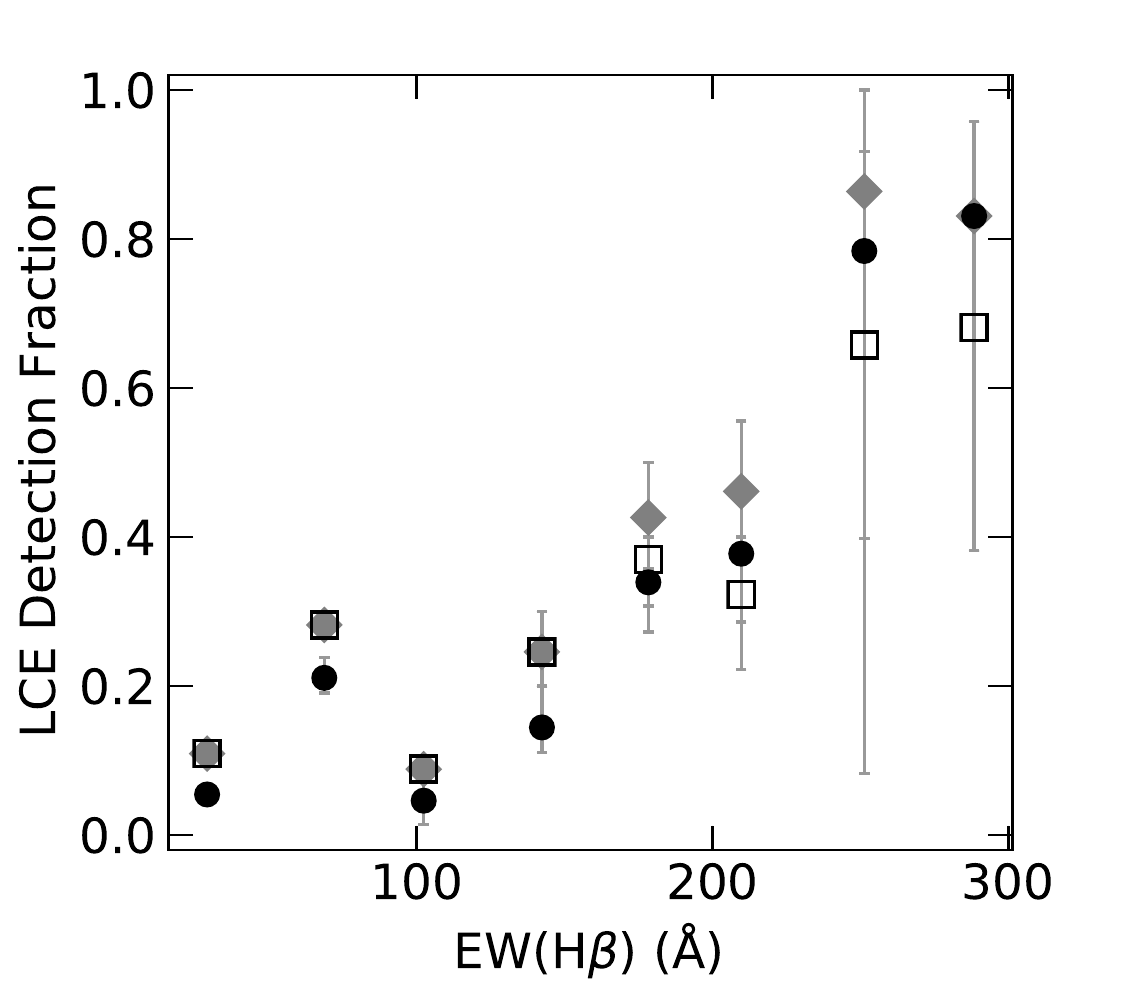}
    \caption{Same as Figure \ref{fig:lya_lce_frac} but for rest-frame EW(H$\beta$).}
    \label{fig:ew_hb_lce_frac}
\end{figure}

As evident in Figure \ref{fig:ew_hb}, the average and maximum values of \fesclyc\ increase with H$\beta$ EW. {\fesclyc(UV) has a significant and strong correlation with H$\beta$ EW, indicating at least a moderate relationship between the two.}
The distribution of \fesclyc\ appears to occupy two distinct ranges in EW (see also Figure \ref{fig:ew_hb_lce_frac}), one with high ($\ga150$ \AA) EW containing the most prodigious strong LCEs (\fesclyc$\ga0.1$) and one with low ($\la150$ \AA) EW containing additional strong LCEs and the majority of weaker LCEs. Two thirds of the galaxies in our combined sample with H$\beta$ EW $>150$ \AA\ have detected LyC. Similarly, 28 of the 39 non-detections appear concentrated at EWs below 150 \AA.

Quantifying these results, the fraction of galaxies that are strong LCEs appears to increase with EW, which we demonstrate in Figure \ref{fig:ew_hb_lce_frac}. This may indicate a preference for young, strong bursts of star formation. { As noted above, though, strong LCEs have a wide range in EWs. Thus, the increase in strong LCE fraction with EW demonstrates a dearth of weak LCEs and non-emitters rather than an increase in the prevalence in LCEs at high EW. In other words, high H$\beta$ EW indicates a strong LCE, but high EW galaxies do not account for the full population of LCEs.} Some scatter persists in LCE fraction, particularly at lower EW, depending on the \fesclyc\ indicator. Uncertainties in the LCE fraction are large enough to suggest any change in LCE fraction between the various indicators may not be significant. However, our results suggest a significant occurrence of LCEs for EWs below 150 \AA, consistent with the \fesclyc\ trends in Figure \ref{fig:ew_hb}.

The prevalence of LCEs at high EW demonstrates that H$\beta$ EW is more sensitive to star-formation history { or sSFR than} it is to optical depth. Studies at higher redshifts find similar results in combined H$\beta$+[\ion{O}{3}] EWs \citep[e.g.,][]{2017ApJ...839...73C,2021MNRAS.tmp..470E,2021MNRAS.tmp.3408S}, suggesting that this trend persists in earlier cosmological epochs. As stellar mass and starburst age dominate the H$\beta$ EW, LCEs with high EWs are likely young starbursts and/or low mass galaxies. { The lack of weak LCEs and non-emitters at high EWs may indicate that young ages and/or low masses better facilitate high LyC escape fractions. However, a subset of between 6 and 10 strong LCEs have EWs below 150 \AA, indicating that strong LCEs occupy a wide range of ages, star-formation histories, and/or sSFRs.} These low-EW LCEs are likely more evolved galaxies with higher stellar mass and metallicity. We discuss age and mass further in \S\ref{sec:2d-diag}.

\subsection{FUV Magnitude}
\label{sec:m_fuv}

\begin{figure*}
    \centering
    \includegraphics[width=0.32\textwidth]{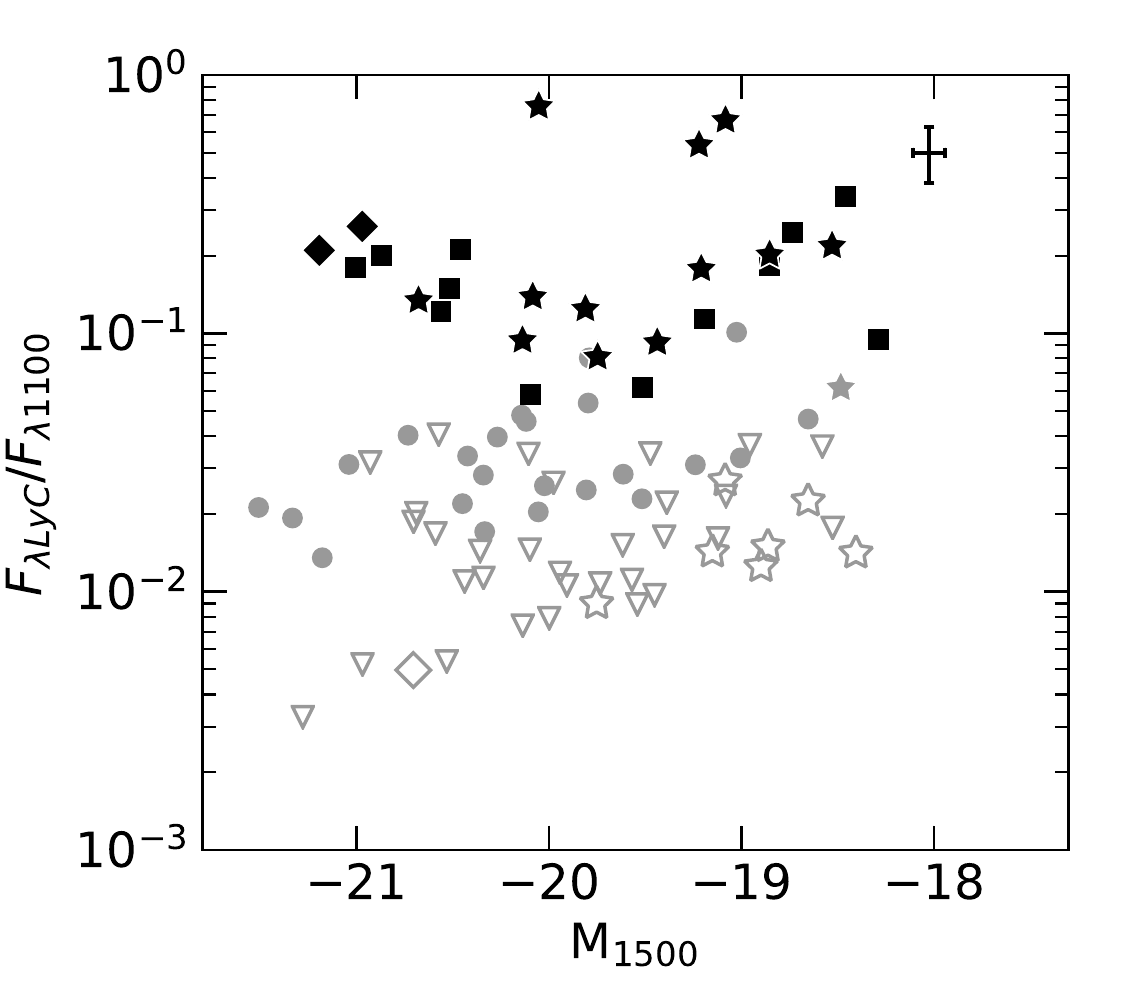}
    \includegraphics[width=0.32\textwidth]{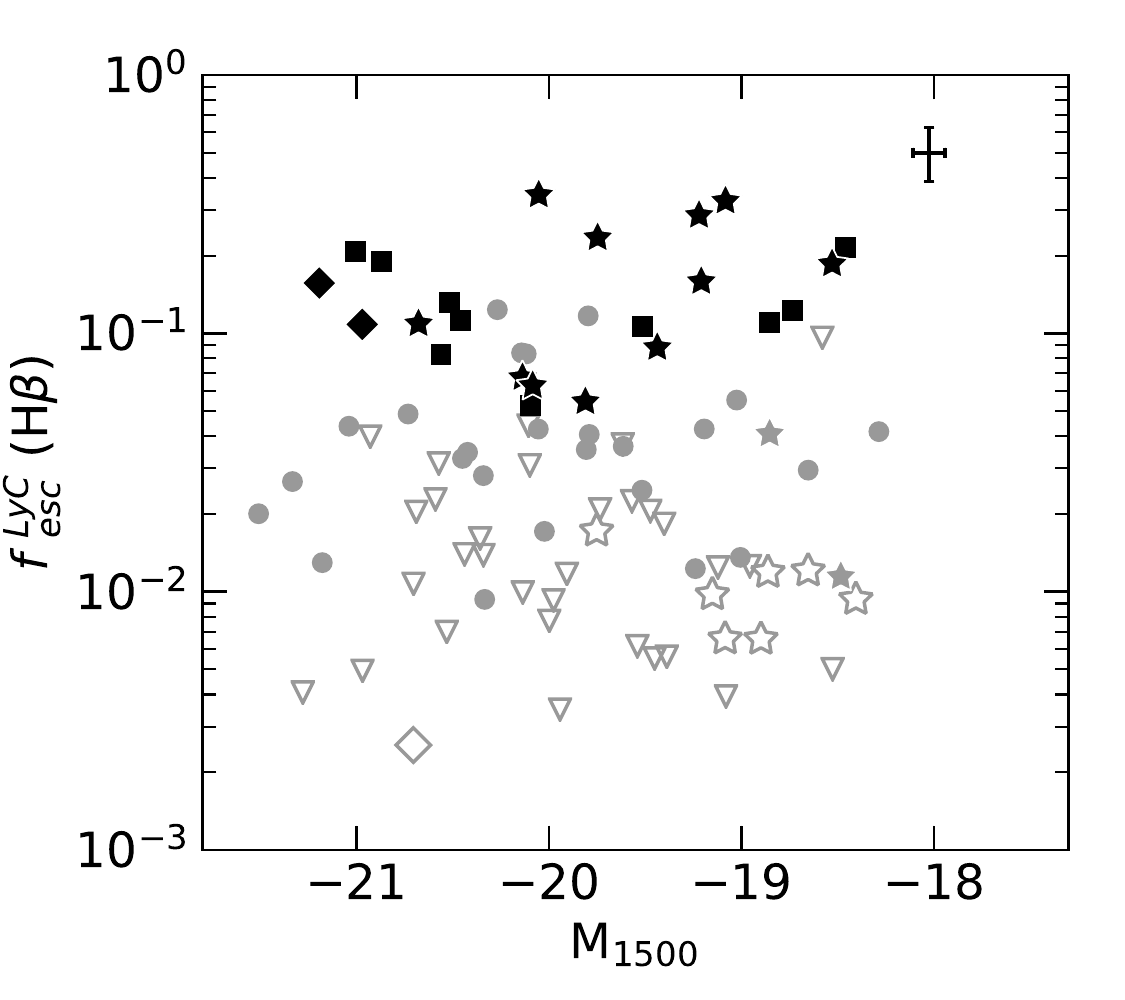}
    \includegraphics[width=0.32\textwidth]{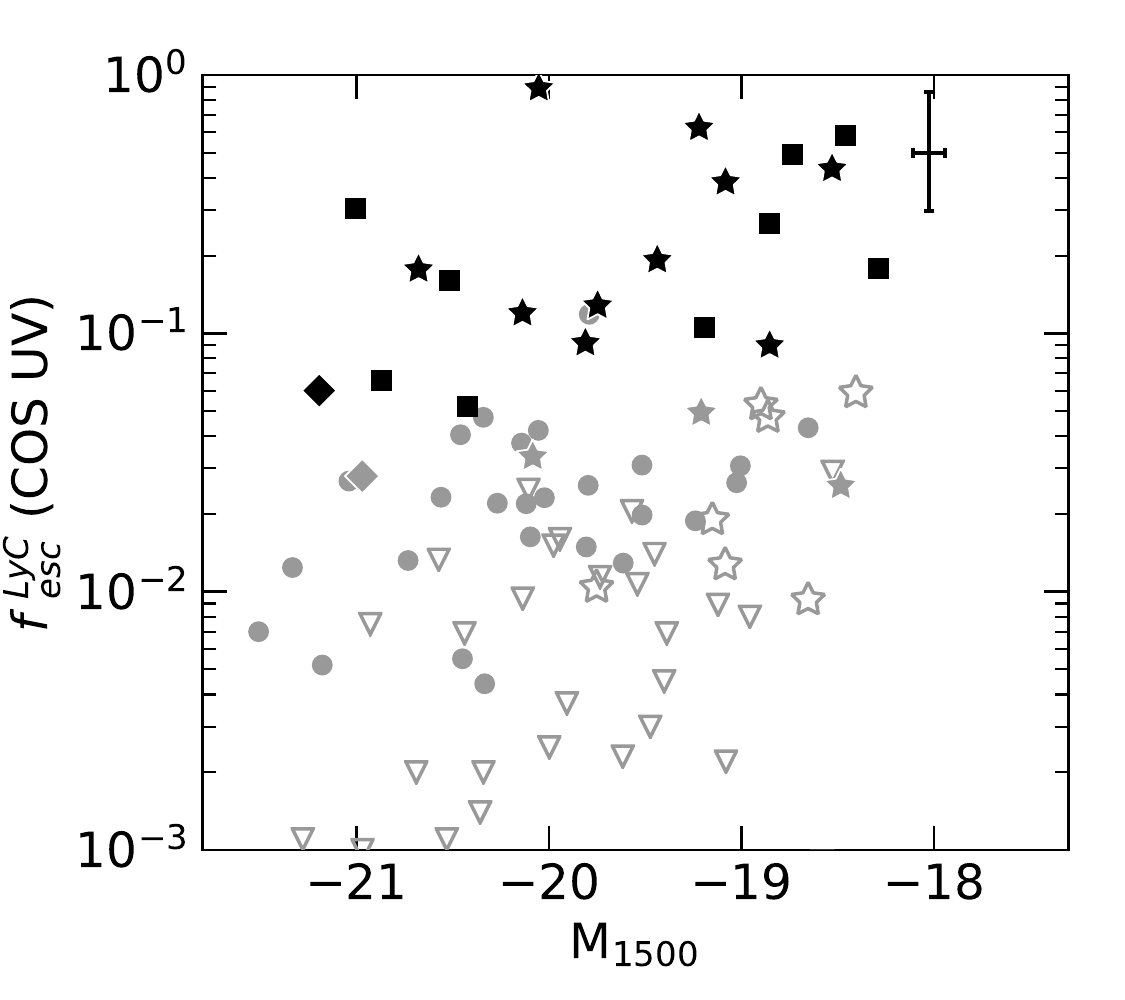}
    \caption{Same as Figure \ref{fig:fesc_lya} but for M$_{1500}$.}
    \label{fig:muv}
\end{figure*}

\begin{figure}
        \centering
        \includegraphics[width=\columnwidth]{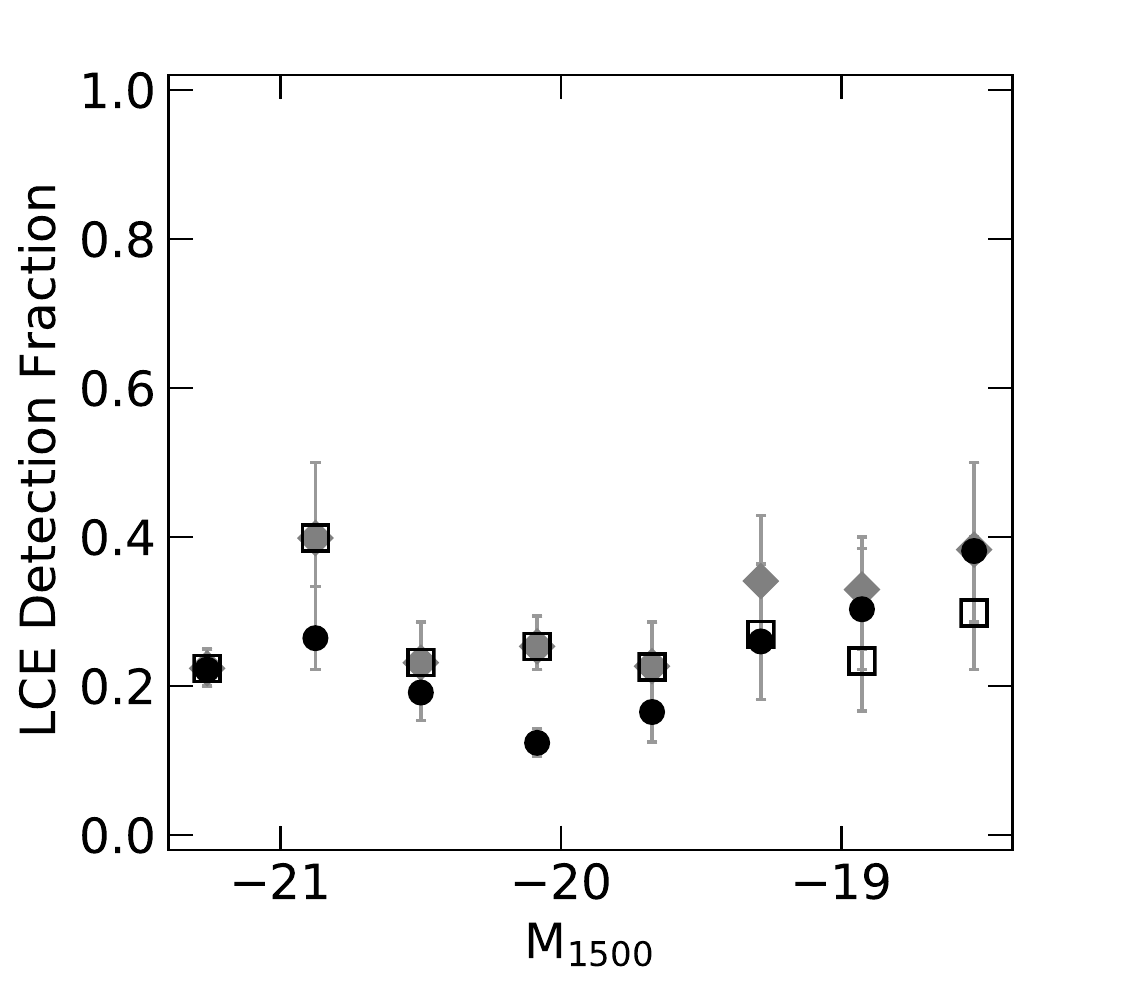}
        \caption{Same as Figure \ref{fig:lya_lce_frac} but for M$_{1500}$.}
        \label{fig:muv_lce_frac}
\end{figure}

{ The immense scatter in Figure \ref{fig:muv} anticipates the weak, insignificant correlation coefficients for observed $M_{1500}$ in all three \fesclyc\ metrics. Strong leakers persist below $M_{1500}=-20.25$, implying a population of strong LCEs at higher stellar mass or higher UV luminosity. Figure \ref{fig:muv_lce_frac} illustrates this persistence of LCEs across a range of $M_{1500}$. The LCE fraction is relatively constant with observed FUV magnitude with a slight increase at the faint end of the distribution ($M_{1500}>-19.25$), especially in the UV \fesclyc\ where LCEs have \fesclyc$\ga0.02$ and 9/15 LCEs are strong LCEs. This mild increase may indicate that strong LCEs are more likely fainter and/or less massive galaxies, particularly given the slight trend in \fesclyc\ upper bounds with $M_{1500}$. }

{ The persistence of LCEs at higher UV luminosities could imply that LCEs consist of multiple epochs of recent star formation and/or higher stellar masses. Indeed, EW H$\beta$ declines with increasing UV luminosity such that, for the UV \fesclyc, 5 of the 6 strong LCEs and 9 of the 12 weak LCEs with $M_{1500}<-20.25$ have H$\beta$ EWs $<100$ \AA.
A major caveat to this interpretation is the effect of dust on the observed UV magnitude. With the LzLCS, we find \fesclyc\ depends both on dust \citep[][]{2022arXiv220111800S} and age (\S\ref{sec:hbew}) but less-so on mass (\S\ref{sec:mass}). Such effects complicate direct interpretation of $M_{FUV}$ as an \fesclyc\ diagnostic without ancillary information. We discuss $M_{1500}$ in a high redshift context further in \S\ref{sec:disc_highz}.}

\subsection{UV $\beta_{1200}$}
\label{sec:uv_beta}

\begin{figure*}
    \centering
    \includegraphics[width=0.32\textwidth]{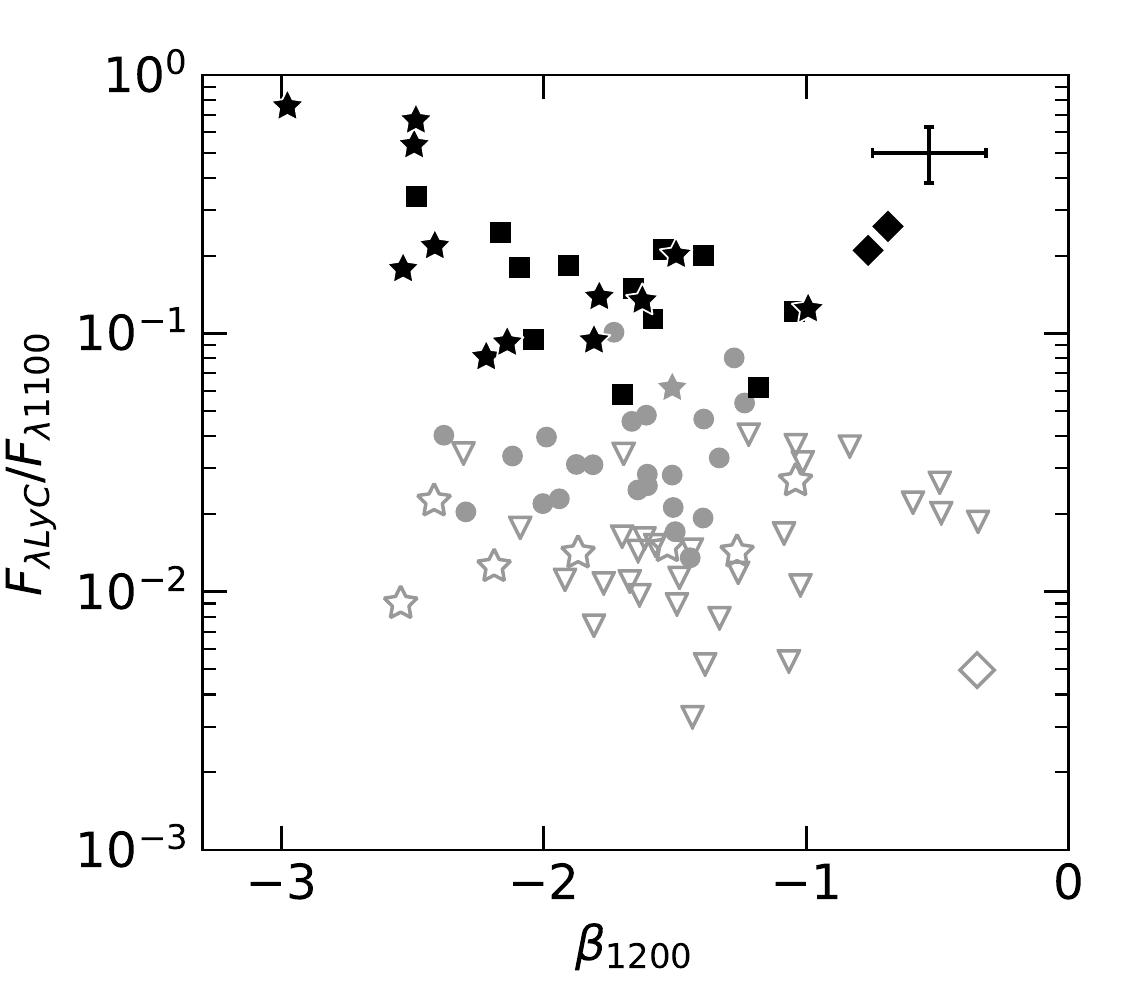}
    \includegraphics[width=0.32\textwidth]{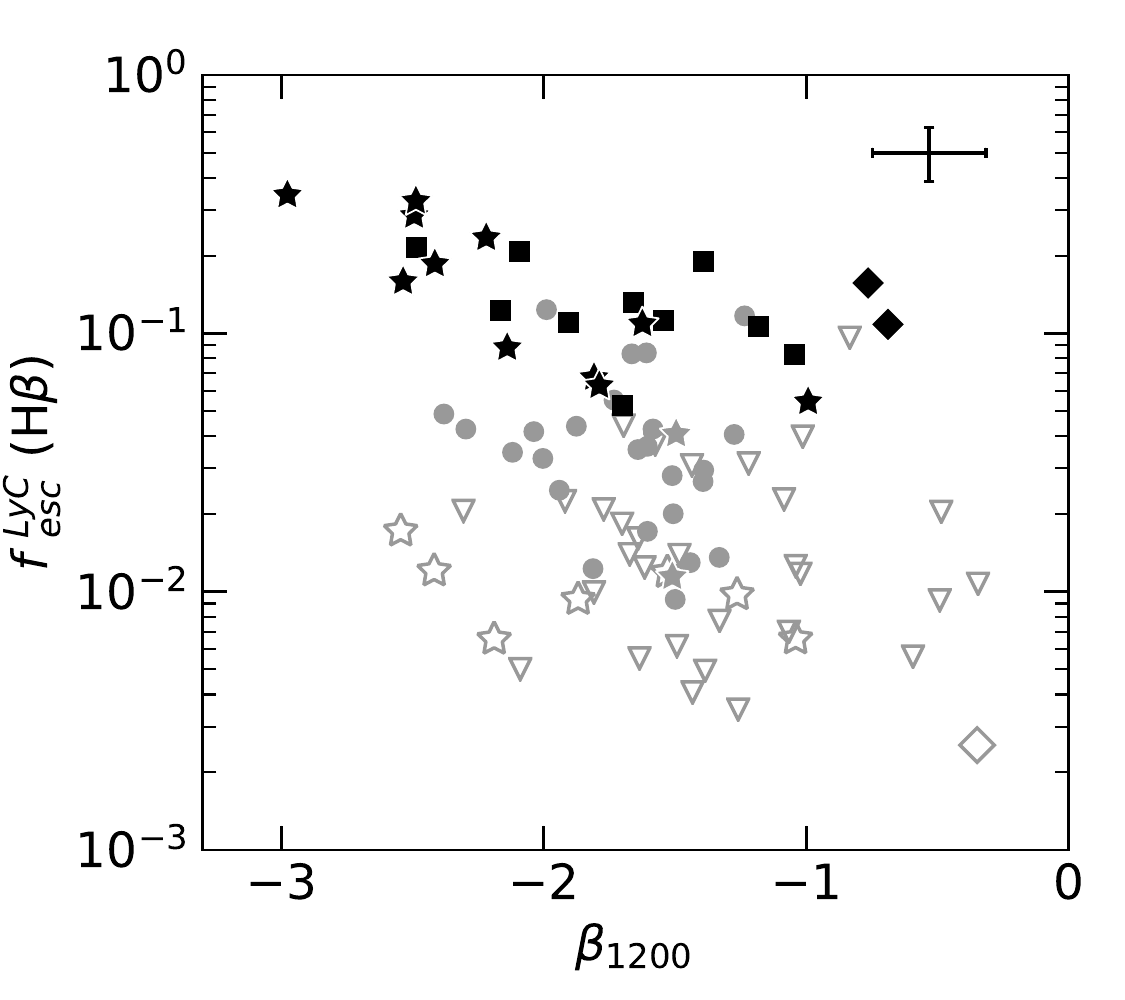}
    \includegraphics[width=0.32\textwidth]{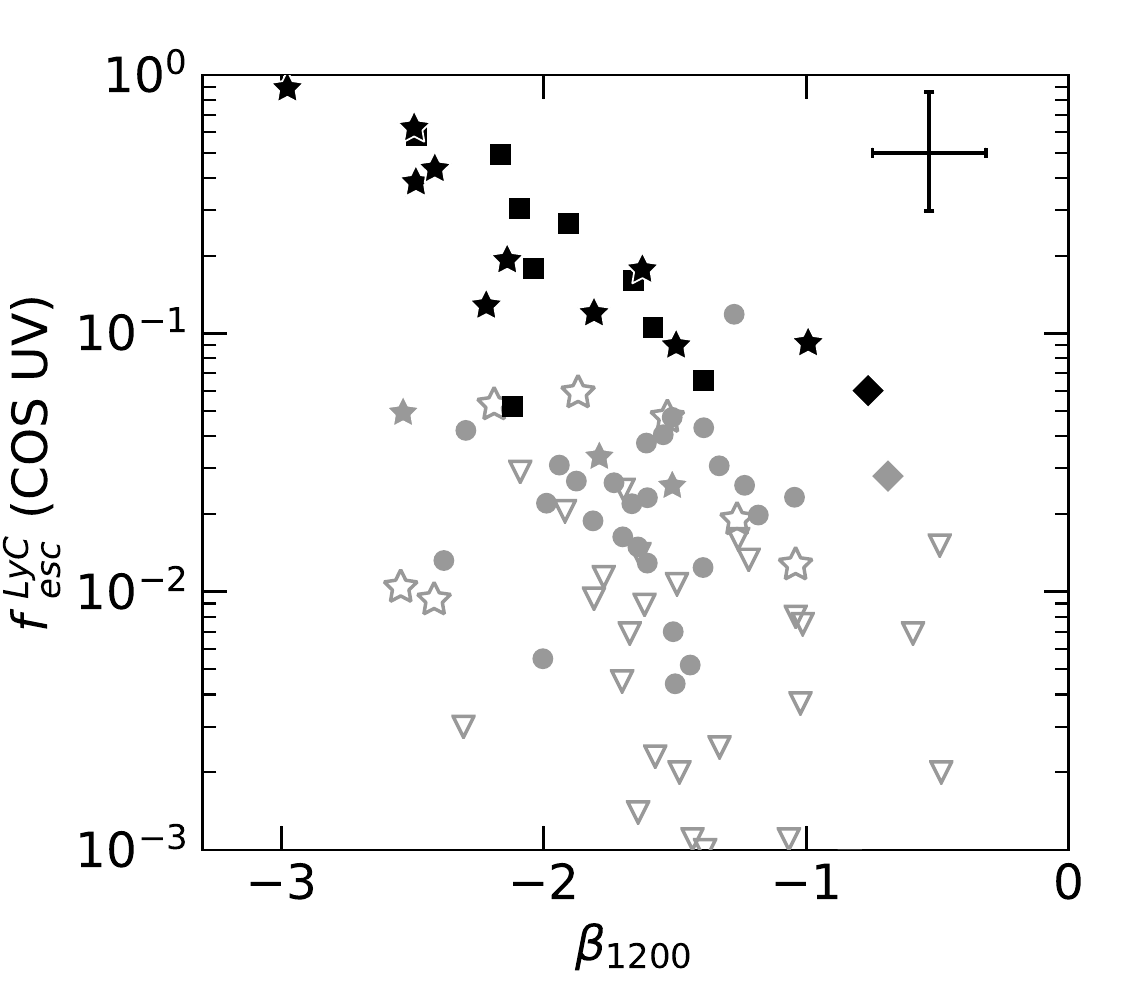}
\caption{Same as Figure \ref{fig:fesc_lya} but for the UV spectral index $\beta_{1200}$.}
    \label{fig:uv_beta}
\end{figure*}

{ \citet{2022arXiv220111716F} measured the UV slope $\beta_{1200}$ from the COS spectra over a range of 1050-1350 \AA. While we present results for this measurement below, we note that $\beta_{1200}$ can be quite different from the commonly used $\beta_{1500}$. Therefore, $\beta_{1200}$ provides only a relative sense of the relationship between $\beta_{1500}$ and \fesclyc.}

\begin{figure}
        \centering
        \includegraphics[width=\columnwidth]{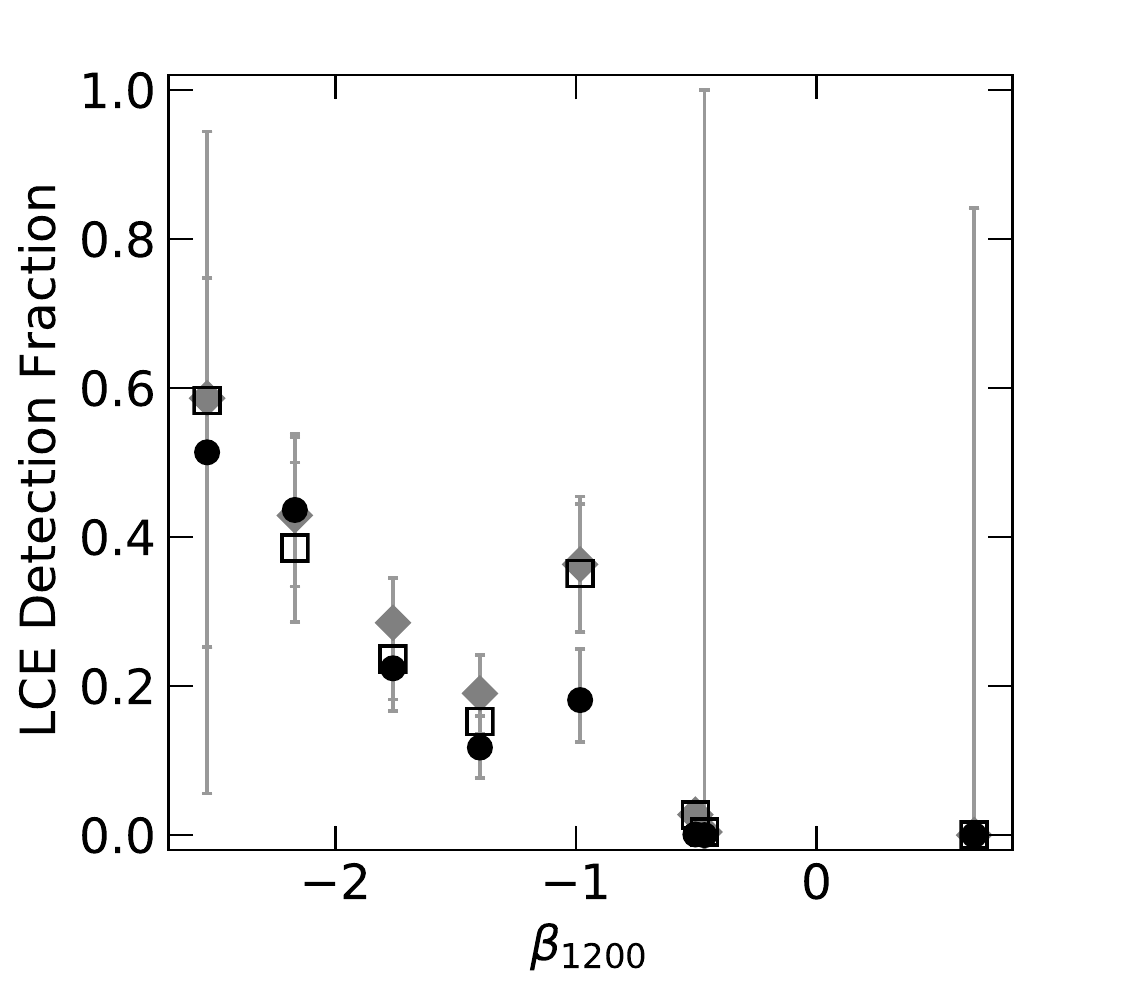}
        \caption{Same as Figure \ref{fig:lya_lce_frac} but for UV $\beta_{1200}$.}
        \label{fig:uv_beta_lce_frac}
\end{figure}

Figure \ref{fig:uv_beta} demonstrates that the highest \fesclyc\ values occur for the steepest values of $\beta_{1200}$. Indeed, the highest values of \fesclyc\ all occur where $\beta\leq-2$ while a distinct envelope in \fesclyc\ describes \fesclyc\ values for $\beta\ga-2$. { Though strong LCEs can persist across a wide range of $\beta\in[-3,-1]$, the UV slope has a significant correlation coefficient for two of the three \fesclyc\ indicators ($\tau\sim-0.25$, see Table \ref{tab:corrcoef}). As implied by the correlation coefficients and Figure \ref{fig:uv_beta}, bluer galaxies tend to have higher \fesclyc. The distribution of LCE fraction over UV $\beta_{1200}$ in Figure \ref{fig:uv_beta_lce_frac} is equally informative. LCE fraction increases as the slope steepens from $\sim0$ at $\beta_{1200}>-1$ to nearly 0.6 at $\beta_{1200}\sim-2.5$, suggesting that the strongest LCEs contain young stellar populations without substantial dust attenuation.}

{ These trends in $\beta_{1200}$ make sense because early type stars produce the majority of the LyC and low dust extinction better facilitates LyC escape. However, some galaxies with $-2<\beta<-1$ are strong LCEs, with the \citet{2019ApJ...885...57W} strong LCEs being even redder than -1. One or a combination of scenarios could describe LCEs with $\beta_{1200}>-2$: (i) a patchy dust screen that reddens the UV spectrum but still allows FUV and LyC photons to escape \citep[the so-called ``picket fence'' model, e.g.,][]{2001ApJ...558...56H,2022arXiv220111800S} and/or (ii) multiple generations of stars that, through feedback processes, facilitate LyC escape by clearing out gas while bolstering the redder part of the UV spectrum with emission from older stellar populations (see discussion in \citealt{2017ApJ...845..165M} of this effect in Mrk 71). We discuss the relationship between $\beta_{1200}$, dust, and stellar population age further in \S\ref{sec:zackrisson}.}

\subsection{Stellar Mass\label{sec:mass}}

\begin{figure*}
    \centering
    \includegraphics[width=0.32\textwidth]{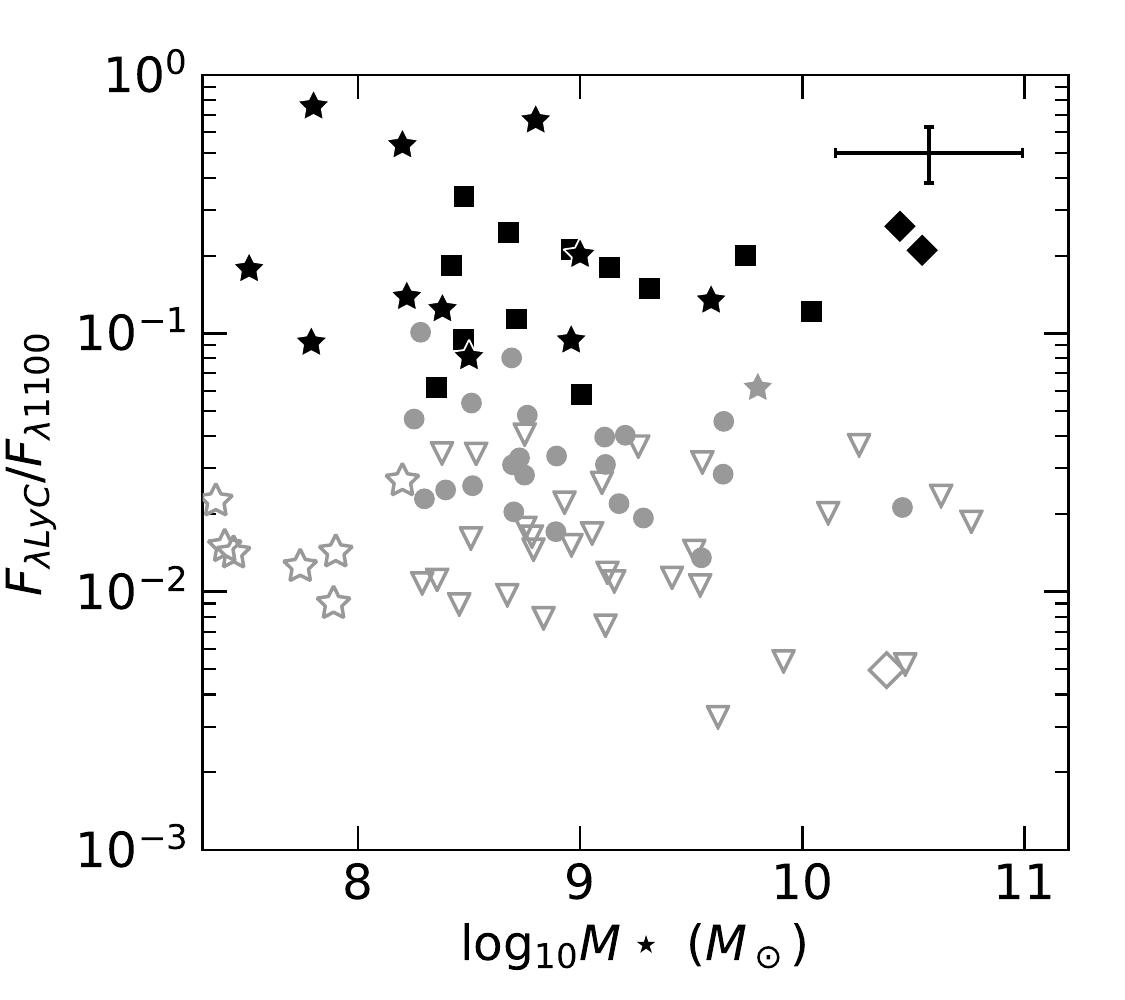}
    \includegraphics[width=0.32\textwidth]{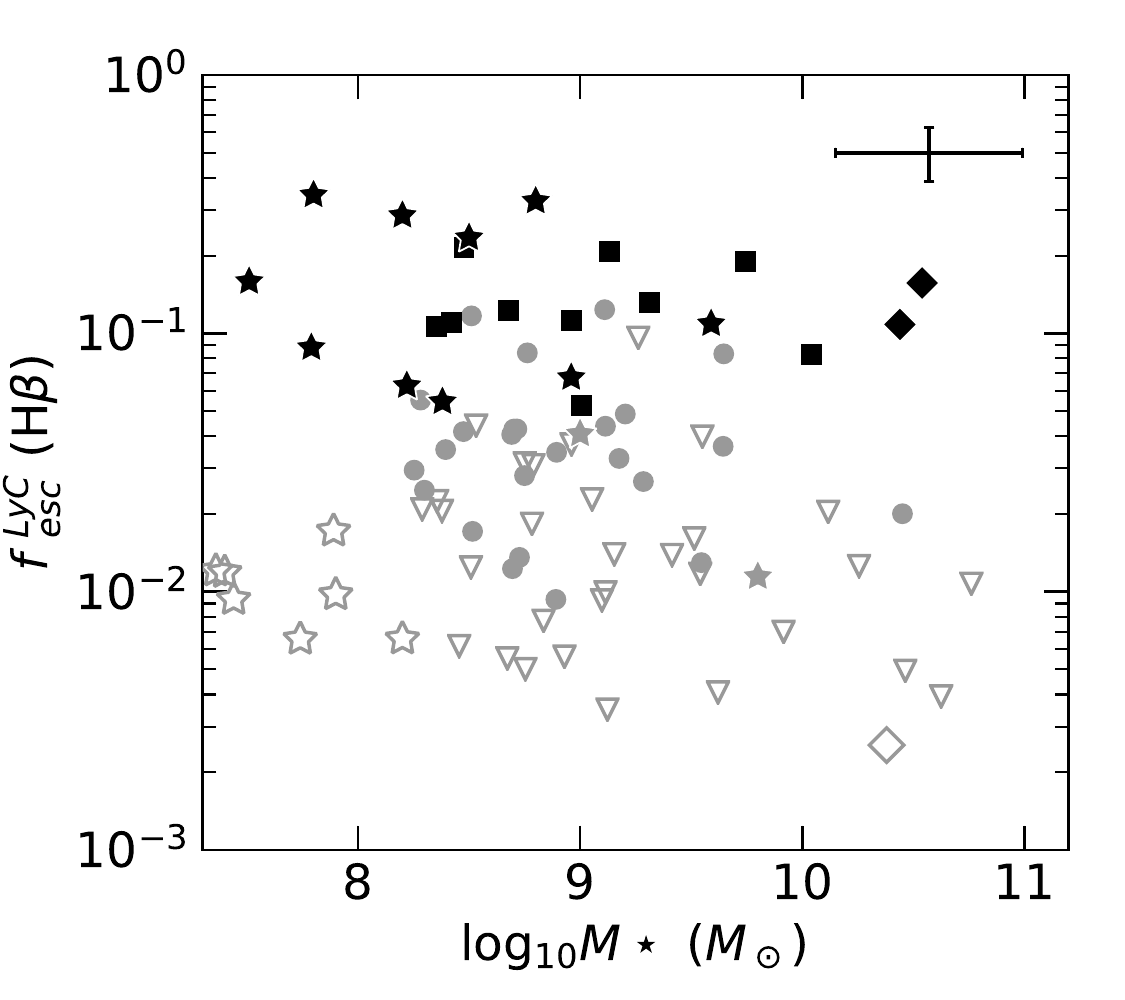}
    \includegraphics[width=0.32\textwidth]{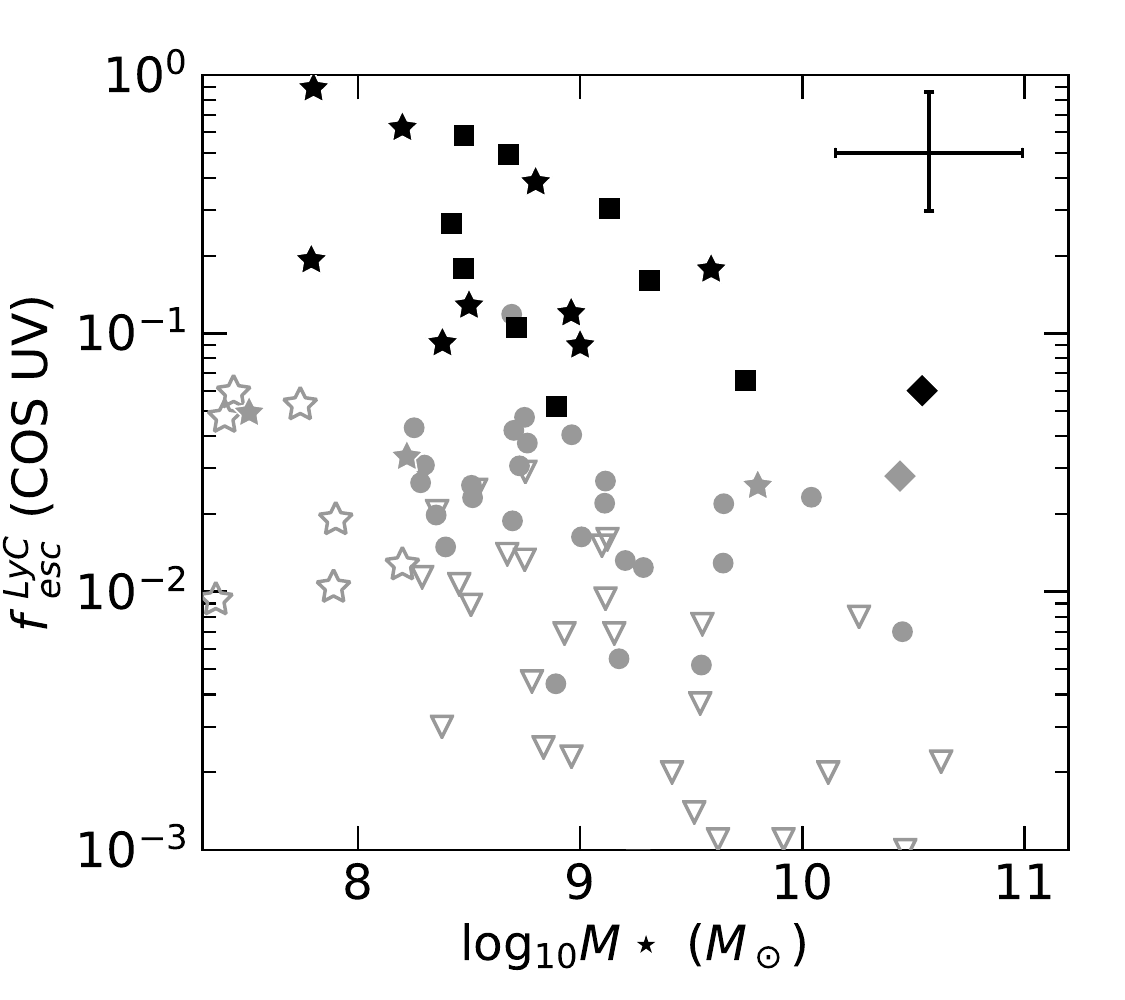}
\caption{Same as Figure \ref{fig:fesc_lya} but for stellar mass.}
    \label{fig:mstar}
\end{figure*}

Simulations disagree as to whether dwarf or more massive star-forming galaxies are the major source of LyC photons for reionization. Figure \ref{fig:mstar} suggests that dwarf galaxies ($M_\star < 10^9$ $M_\odot$) are more likely to be strong LCEs. More than half (31/52) of the dwarf galaxies in the combined sample are LCEs, 13 of which are strong LCEs.

\begin{figure}
    \centering
    \includegraphics[width=\columnwidth]{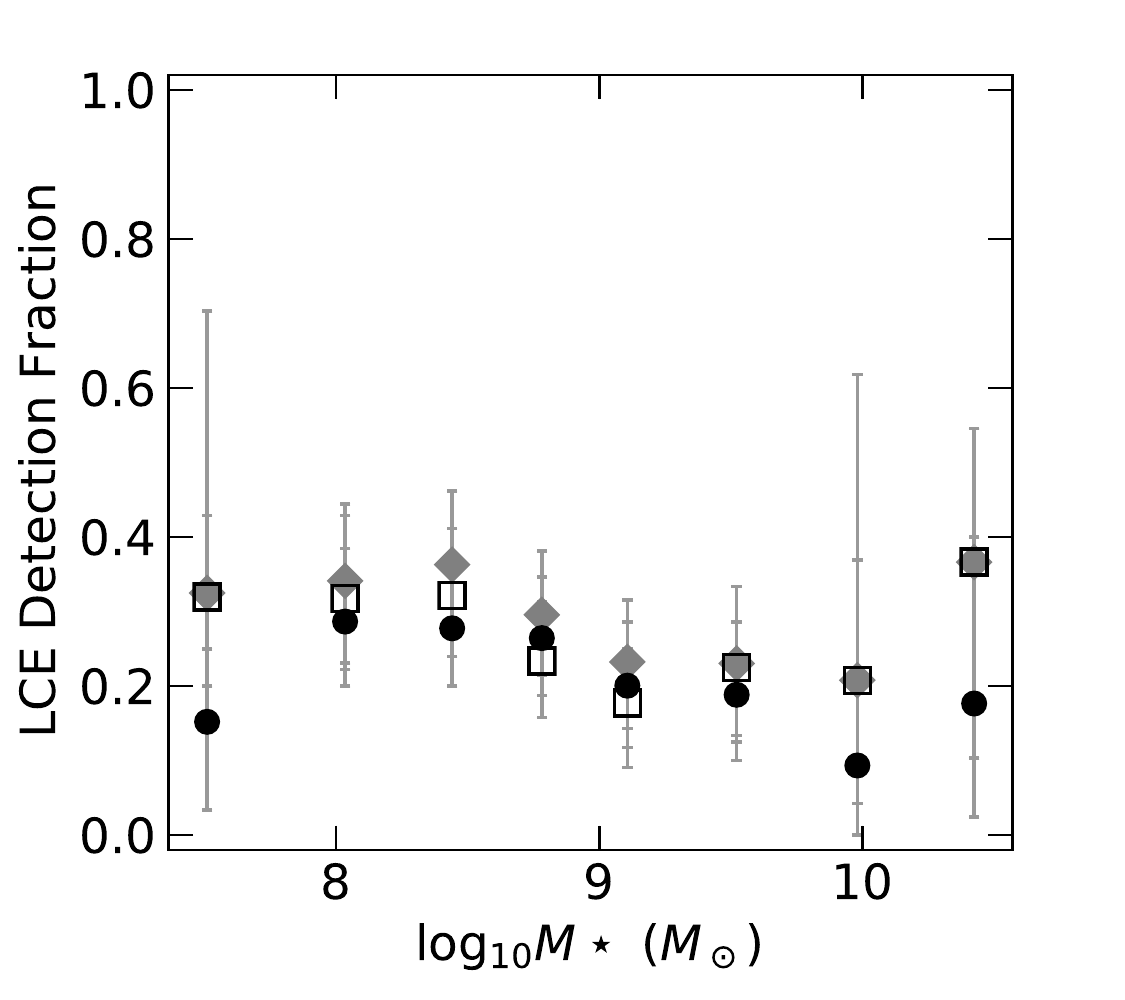}
    \caption{Same as Figure \ref{fig:lya_lce_frac} but for $M_\star$.}
    \label{fig:mstar_lce_frac}
\end{figure}

{ With increasing mass, the number of strong LCEs and corresponding \fesclyc(UV)\ values appear to decrease. Even so, galaxies with $M_\star>10^9$ M$_\sun$ can still be prodigious LCEs, some even having \fesclyc$>0.1$. The $\tau$ coefficient and corresponding $p$ value suggest the correlation is neither strong nor significant for any \fesclyc\ metric. Furthermore, the other two \fesclyc\ metrics do not indicate much of an envelope in \fesclyc\ with $M_\star$. As evident in Figure \ref{fig:mstar_lce_frac}, the LCE fraction distribution is also relatively flat over the range of stellar masses. Like $M_{1500}$, the distribution of \fesclyc\ with respect to $M_\star$ suffers from substantial scatter. Unlike $M_{1500}$, the stellar masses have relatively large uncertainties. These large uncertainties may contribute to the apparent scatter and even to the flat LCE fraction distribution.}

\subsection{\sigsfr, \ssfr, and Half-light Radius}\label{sec:sfr}

\begin{figure*}
    \centering
    \includegraphics[width=0.32\textwidth]{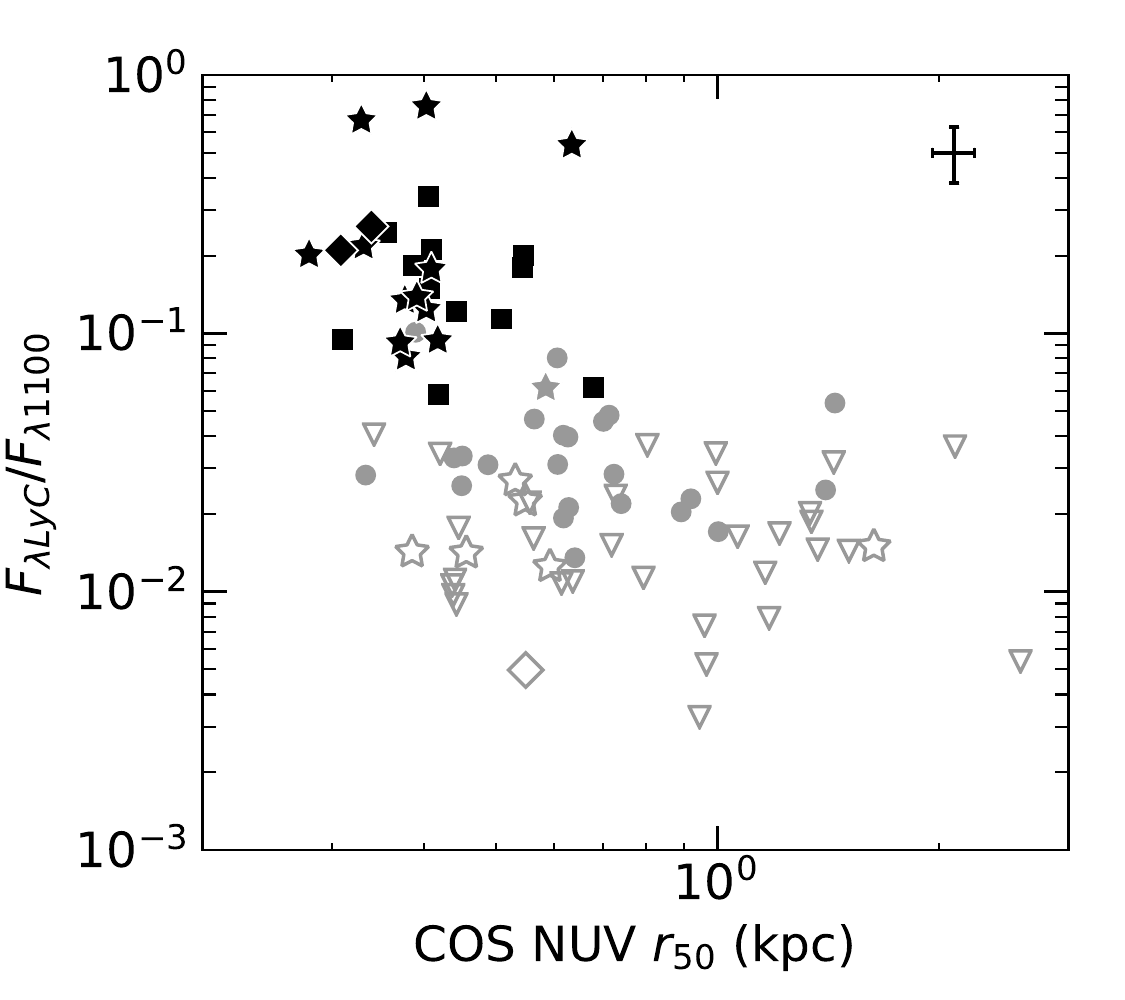}
    \includegraphics[width=0.32\textwidth]{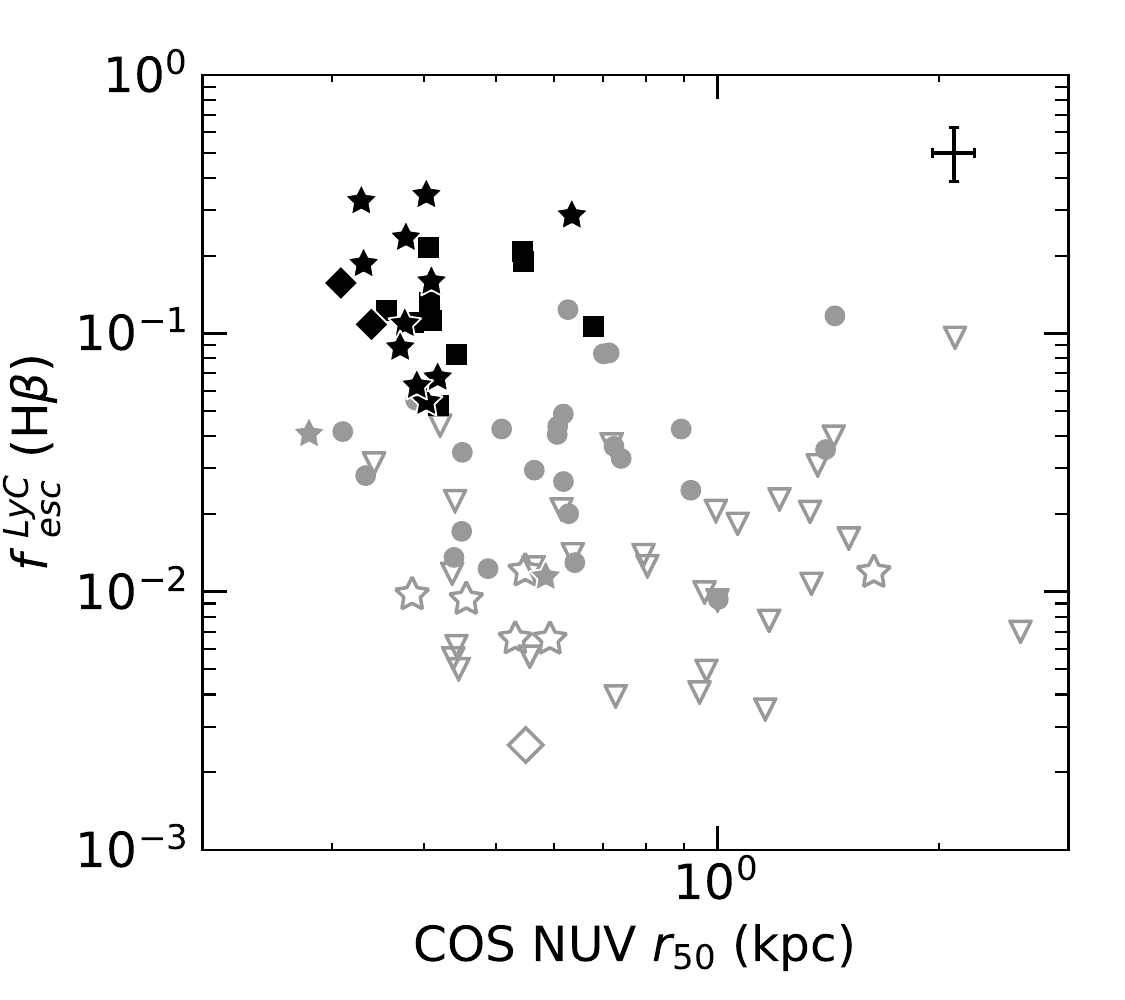}
    \includegraphics[width=0.32\textwidth]{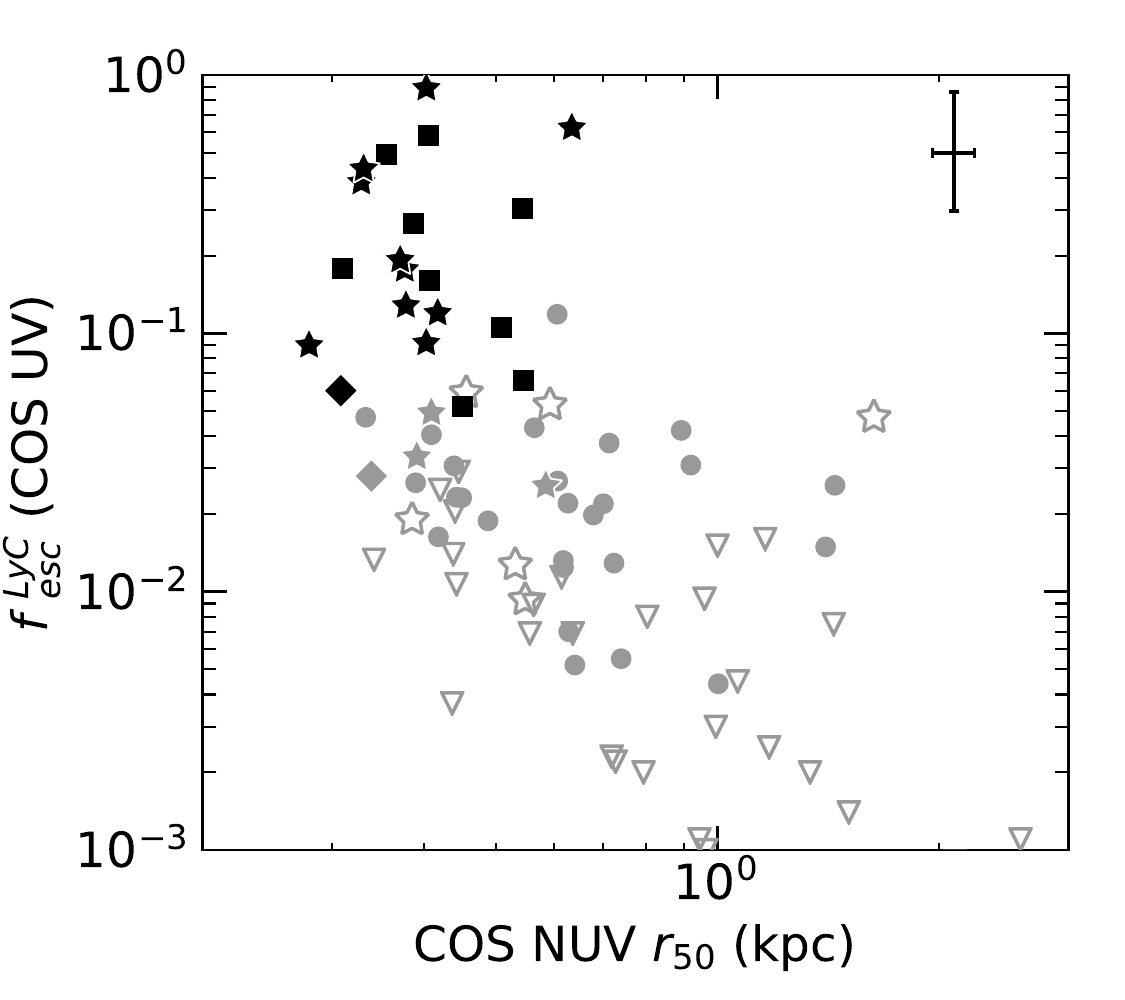}
    \includegraphics[width=0.32\textwidth]{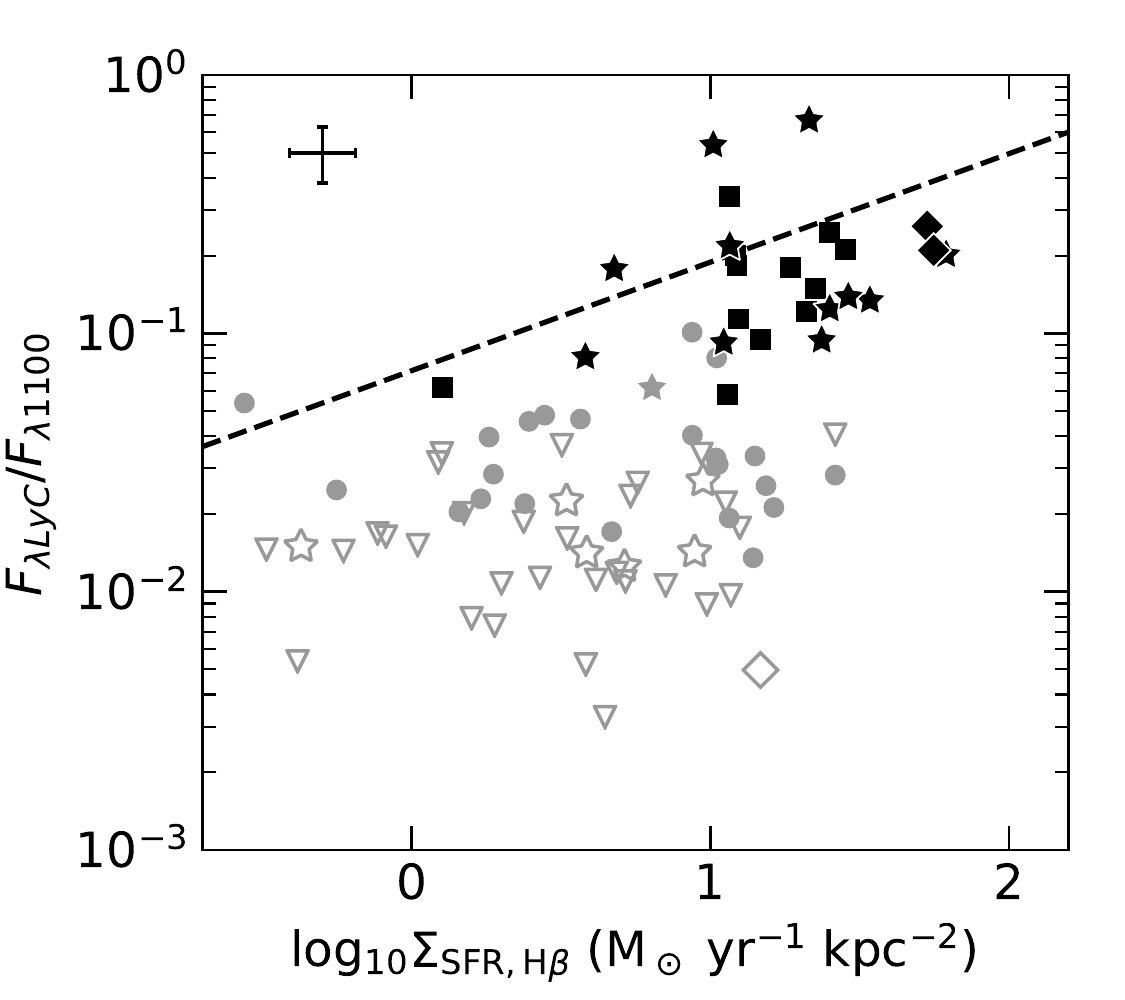}
    \includegraphics[width=0.32\textwidth]{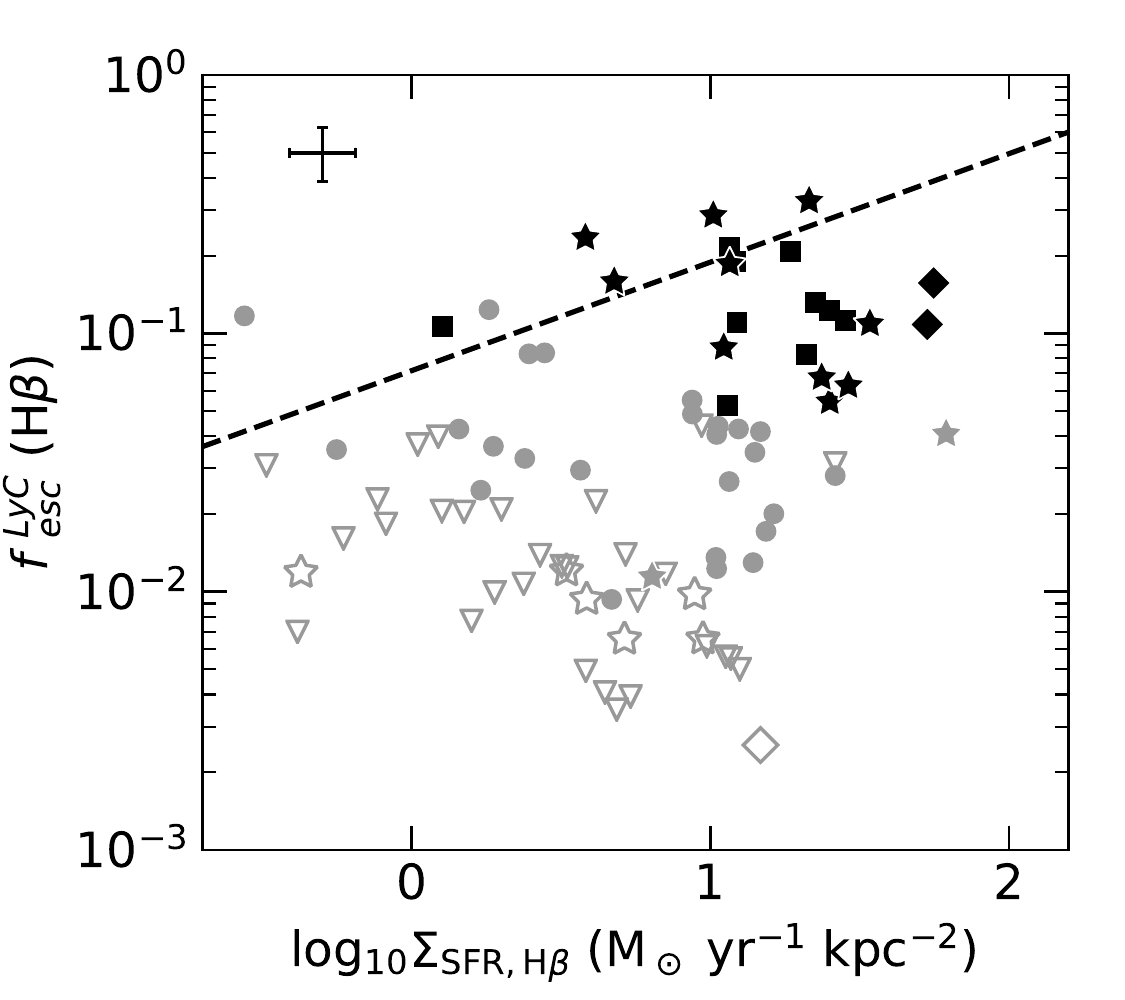}
    \includegraphics[width=0.32\textwidth]{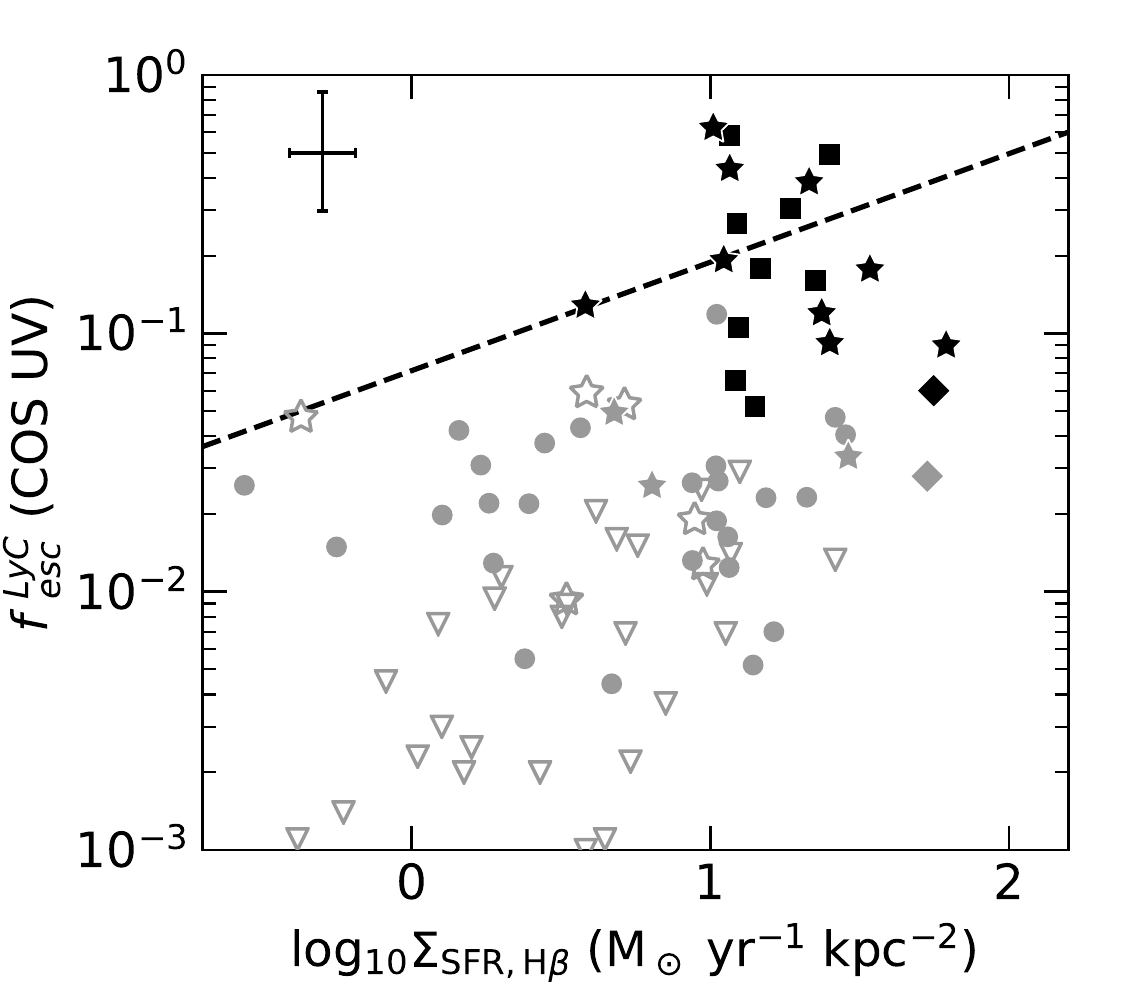}
    \includegraphics[width=0.32\textwidth]{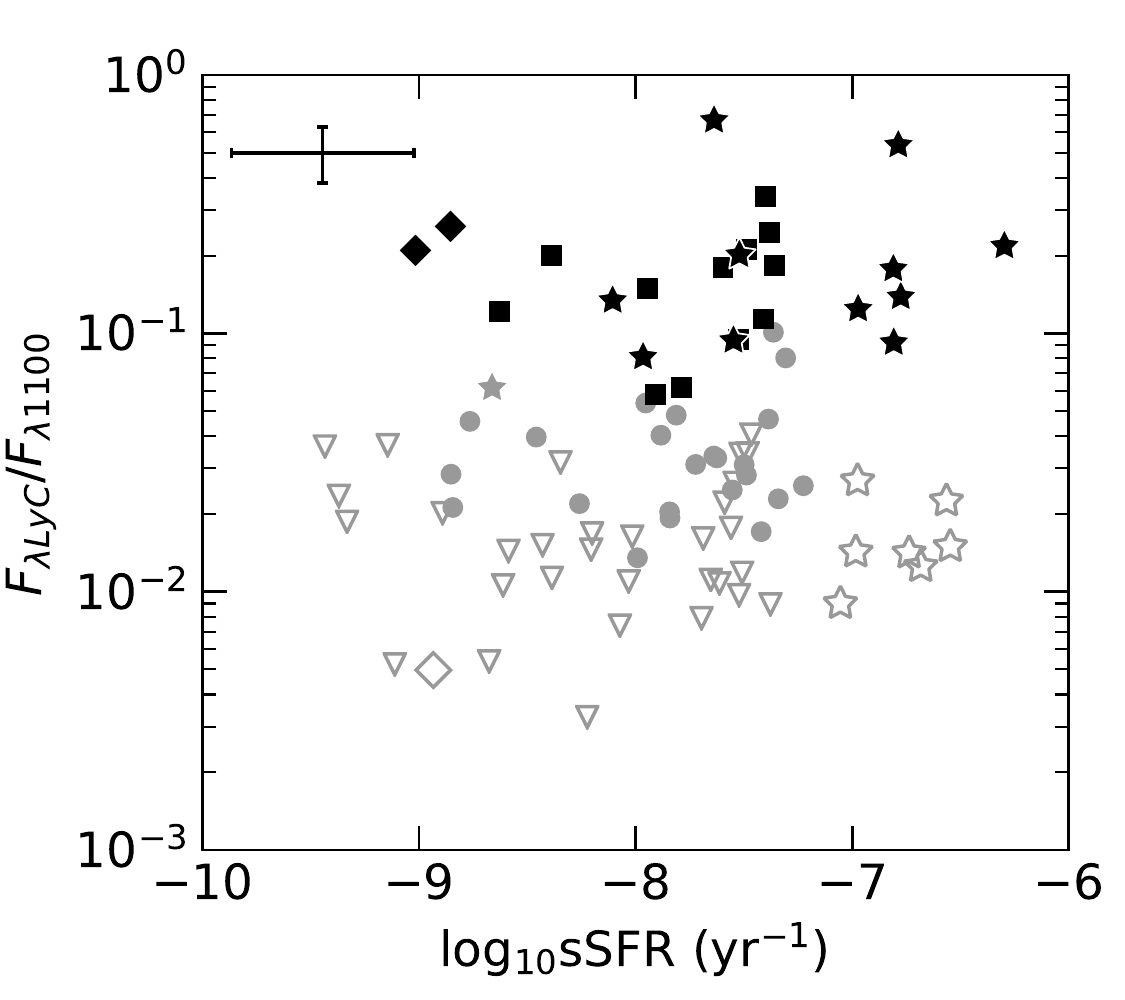}
    \includegraphics[width=0.32\textwidth]{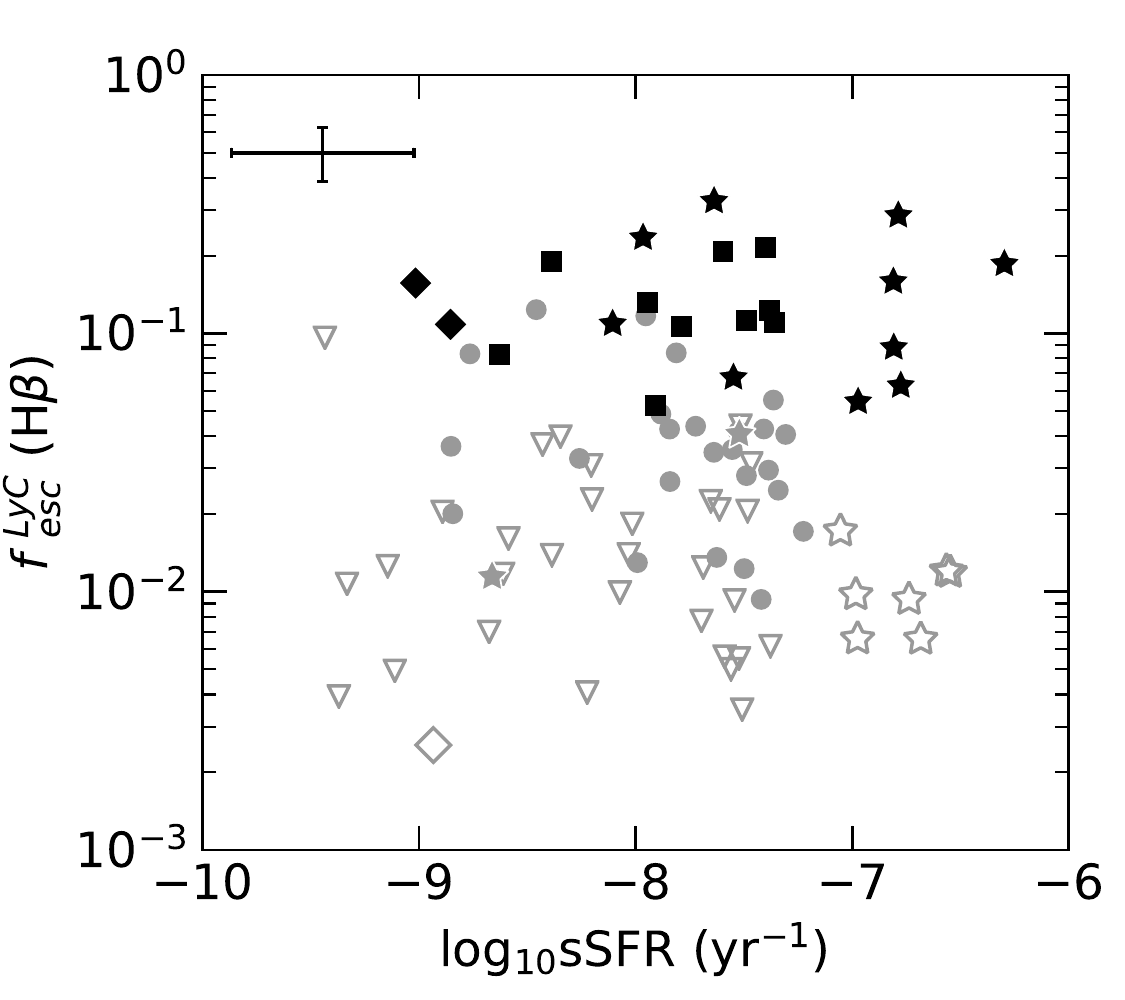}
    \includegraphics[width=0.32\textwidth]{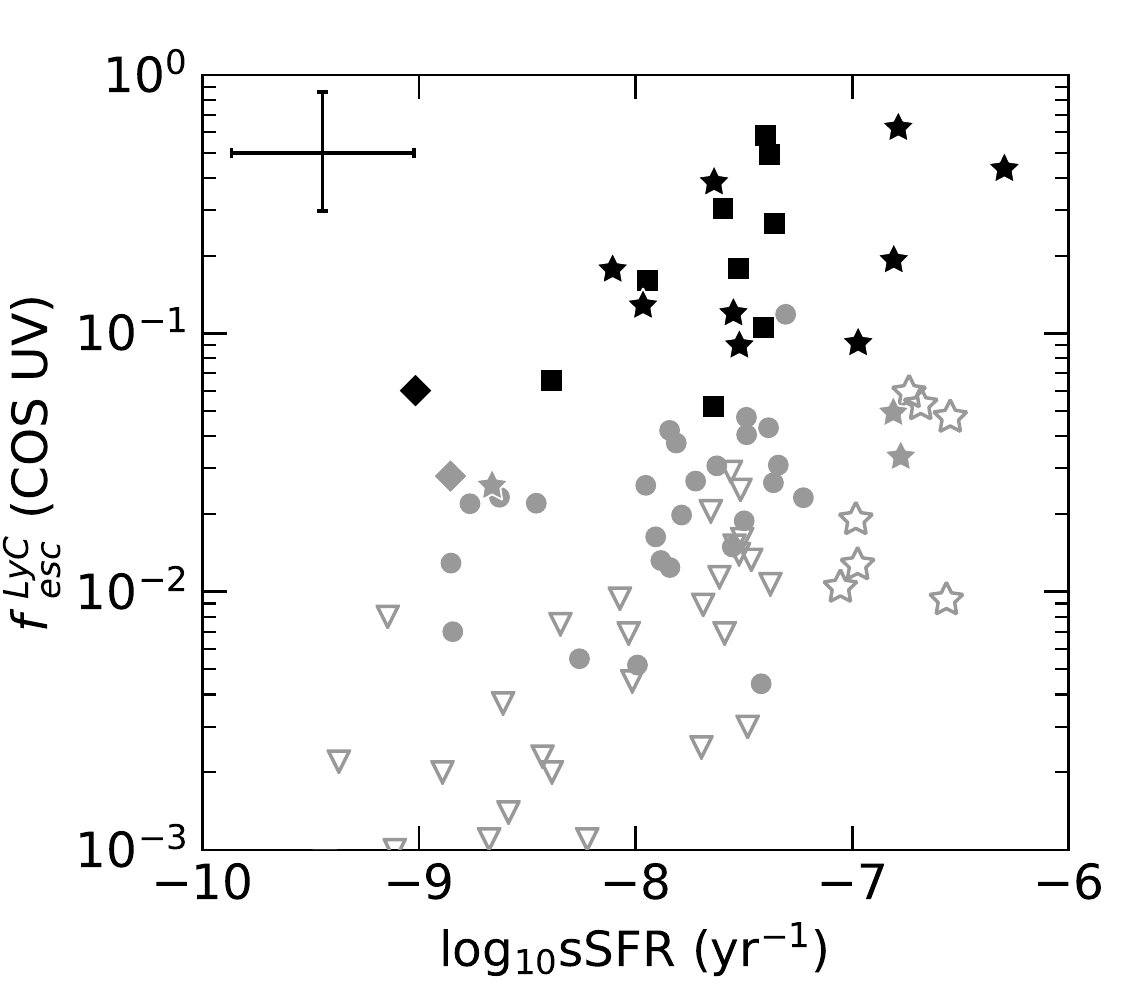}
    \caption{Same as Figure \ref{fig:fesc_lya} but for UV half-light radius $r_{50}$, \sigsfr, and \ssfr. Dashed line in \sigsfr\ is the \fesclyc$\propto\Sigma_{\rm SFR}^{0.4}$ relation from \citet{2020ApJ...892..109N}.}
    \label{fig:sigma_sfr}
\end{figure*}

\sigsfr\ exhibits a threshold of \sigsfr$\sim10$ M$_\odot$ yr$^{-1}$ kpc$^{-2}$ above which nearly all strong LCEs appear, as evidenced by Figure \ref{fig:sigma_sfr}. We see scatter in \fesclyc\ similar to other diagnostics examined here, again implying the effects of orientation or variations in host galaxy properties. For comparison, we also show \fesclyc\ as it varies with half-light radius in Figure \ref{fig:sigma_sfr}. The prominent transition in \fesclyc\ at $r_{50}\approx0.7$ kpc suggests the concentration of star formation dominates \sigsfr\ in strong LCEs.

This affinity of strong LCEs for high \sigsfr\ and low $r_{50}$ suggests higher concentrations of star formation provide the feedback necessary to clear LyC escape paths in the ISM. Consistent with this interpretation, we find that the maximum observed value of \fesclyc\ increases with \ssfr. Although less pronounced, this envelope in \fesclyc\ is similar to the \fesclyc$\propto$\sigsfr$^{0.4}$ relation described by \citep{2020ApJ...892..109N}.  These trends suggest concentrated star formation may be a good indicator of \fesclyc\ and could be pivotal to understanding the origin of reionization.

\begin{figure*}
    \centering
    \includegraphics[width=0.32\linewidth]{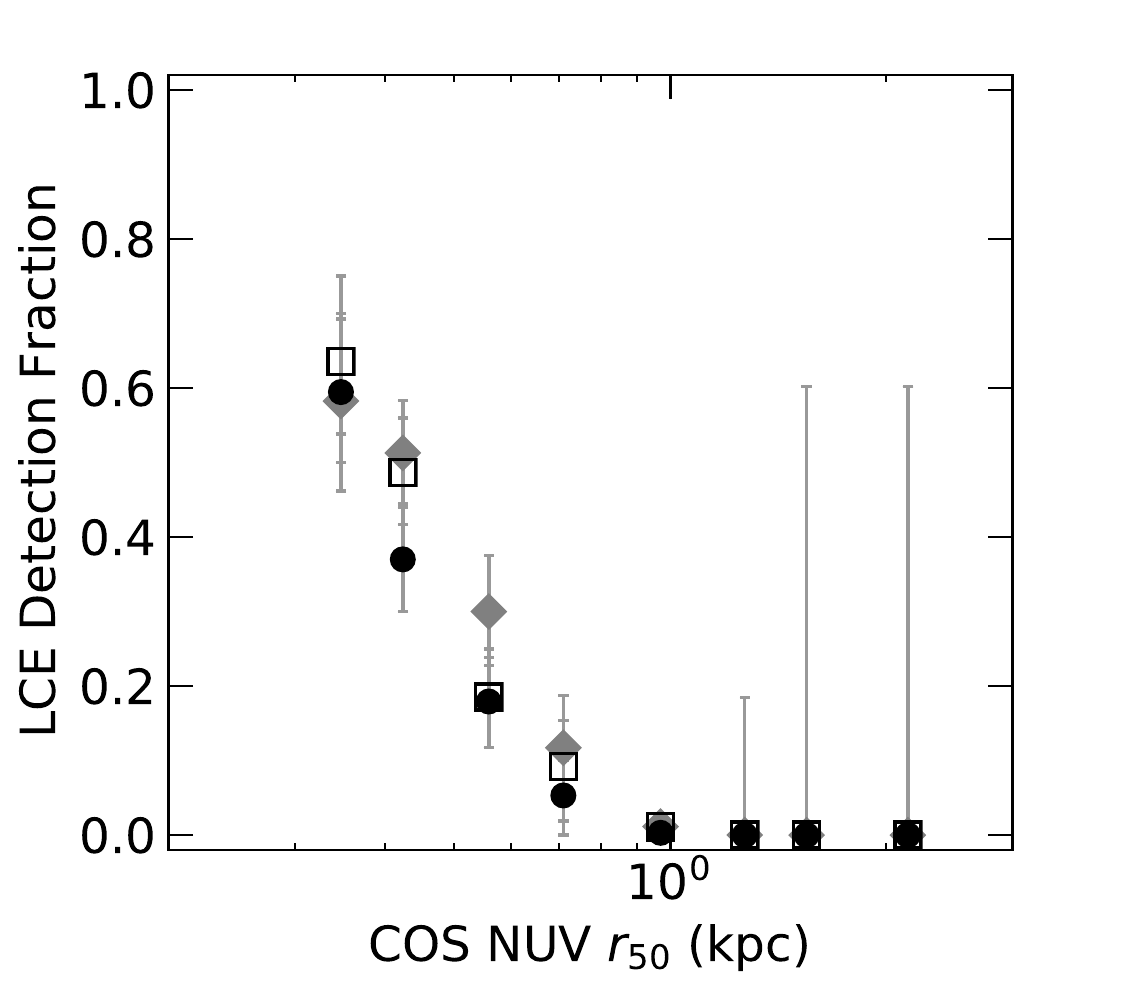}
    \includegraphics[width=0.32\linewidth]{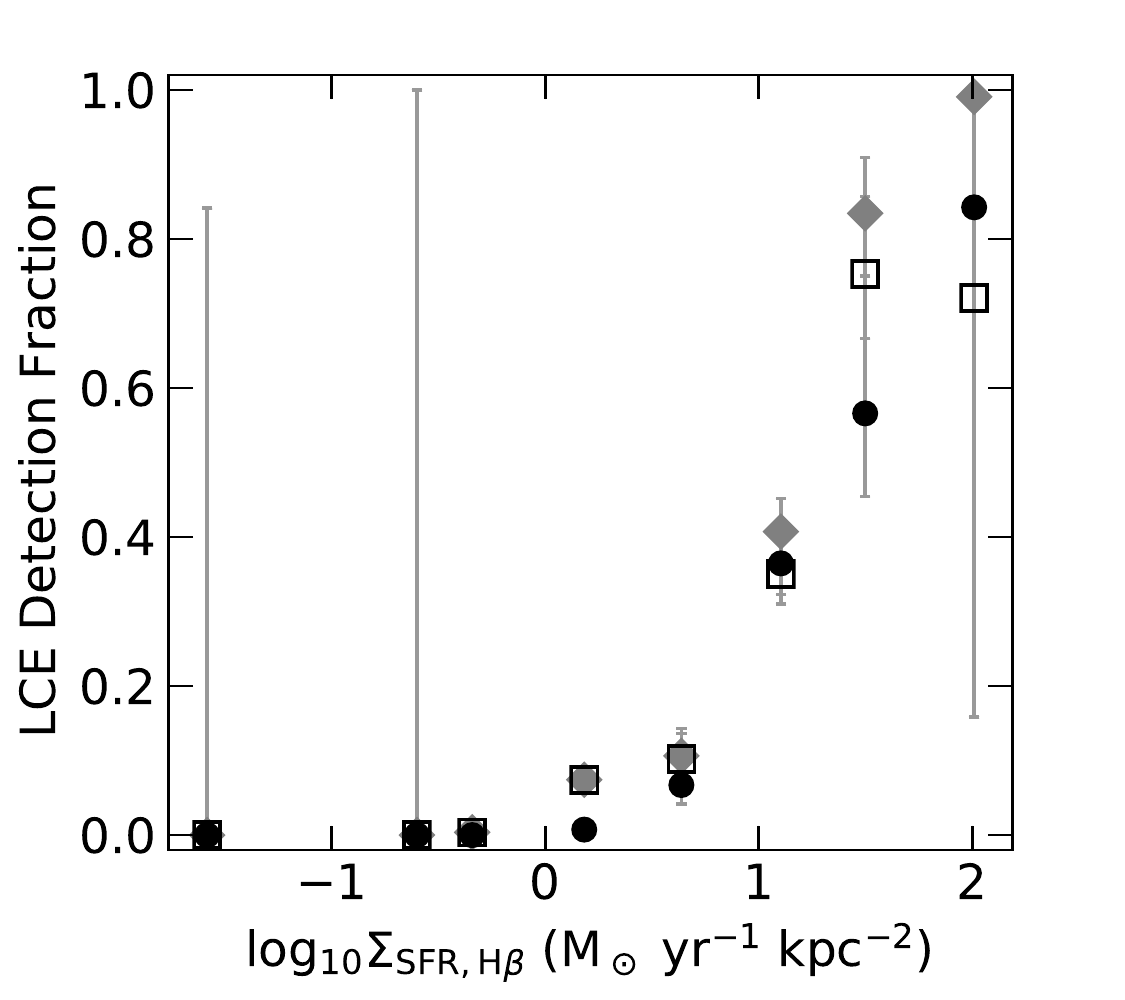}
    \includegraphics[width=0.32\linewidth]{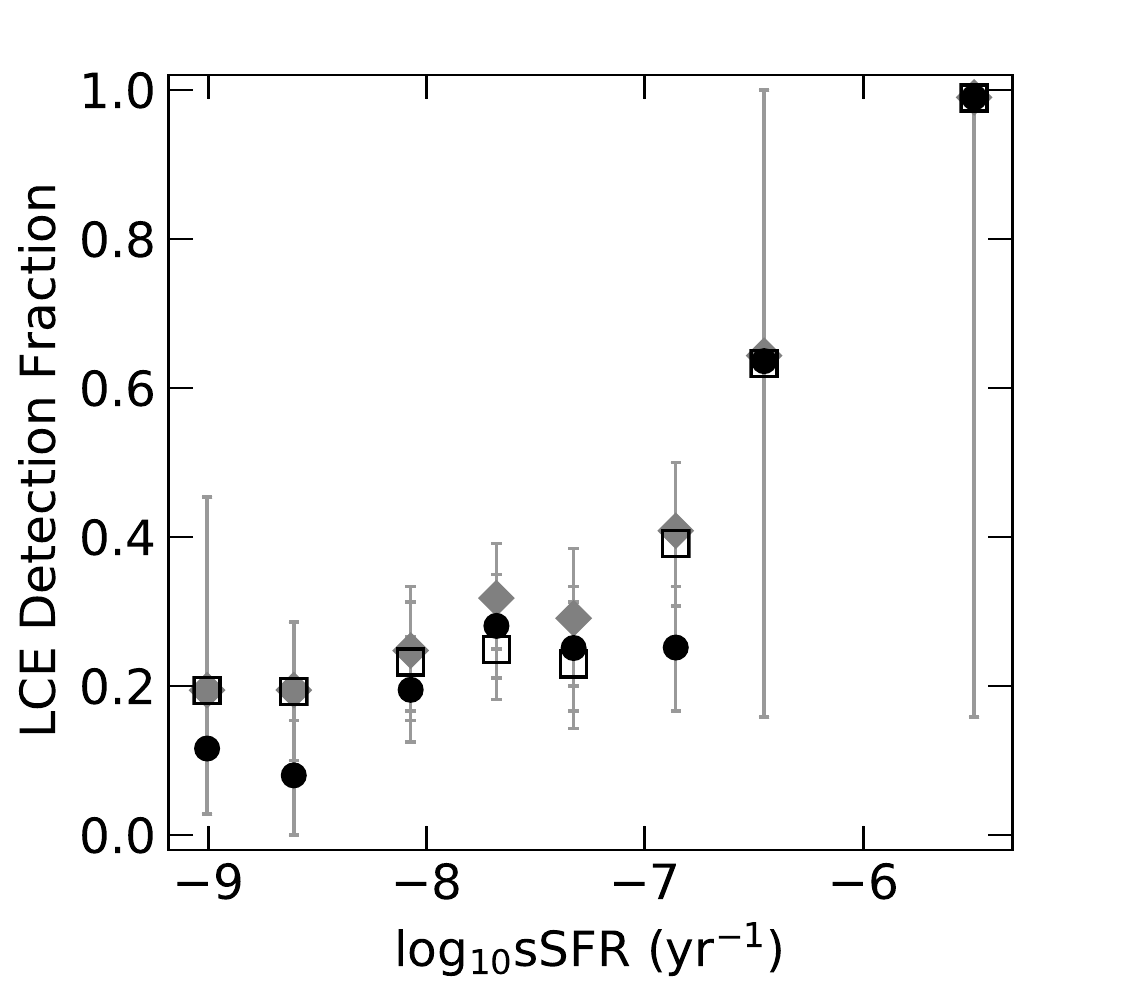}
    \caption{Same as Figure \ref{fig:lya_lce_frac} but for $r_{50}$, \sigsfr, and \ssfr.}
    \label{fig:sigma_sfr_lce_frac}
\end{figure*}

We show the LCE fraction with respect to both \sigsfr\ and \ssfr\ in Figure \ref{fig:sigma_sfr_lce_frac}. In both cases, LCE fraction clearly increases with the concentration of star formation. The LCE fraction changes relatively little with \sigsfr\ until reaching $\approx10$ M$_\odot$ yr$^{-1}$ kpc$^{-2}$, at which point LCE fraction increases from $\sim10$\% to 60\% over half a decade in \sigsfr. { LCE fraction increases more gradually with \ssfr, suggesting that the concentration of star formation is more important than the effects of stellar mass.
}

{ Comparing \sigsfr, sSFR, and $\rm\Sigma_{sSFR}=$\sigsfr$/M_\star$ highlights this point as \sigsfr\ and $\rm\Sigma_{sSFR}$ yield much higher $\tau$ values than sSFR. The $\tau$ values of \sigsfr\ and $\rm\Sigma_{sSFR}$ are also quite comparable, which demonstrates that factoring in stellar mass does not produce a better diagnostic. This result comes with the caveat that the LzLCS stellar masses have high uncertainties, which could mitigate any effects stellar mass may have on the observed trends.}

High \sigsfr, like those seen for GP galaxies, need not imply unusually strong SNe feedback \citep[e.g.,][]{2017A&A...605A..67C,2017ApJ...851L...9J}. Because \sigsfr\ is derived from H$\beta$, the apparent envelope in \fesclyc\ could instead arise from an increased compactness of \ion{H}{2} regions. Indeed, $\tau$ is higher and more significant for $r_{50}$ than for \sigsfr, sSFR, or $\rm\Sigma_{sSFR}$. Such concentrated star forming regions could deplete the surrounding gas so that LyC photons can leave the local \ion{H}{2} region into the more diffuse host galaxy and potentially escape into the IGM.

\subsection{Metallicity}

\begin{figure*}
    \centering
    \includegraphics[width=0.32\textwidth]{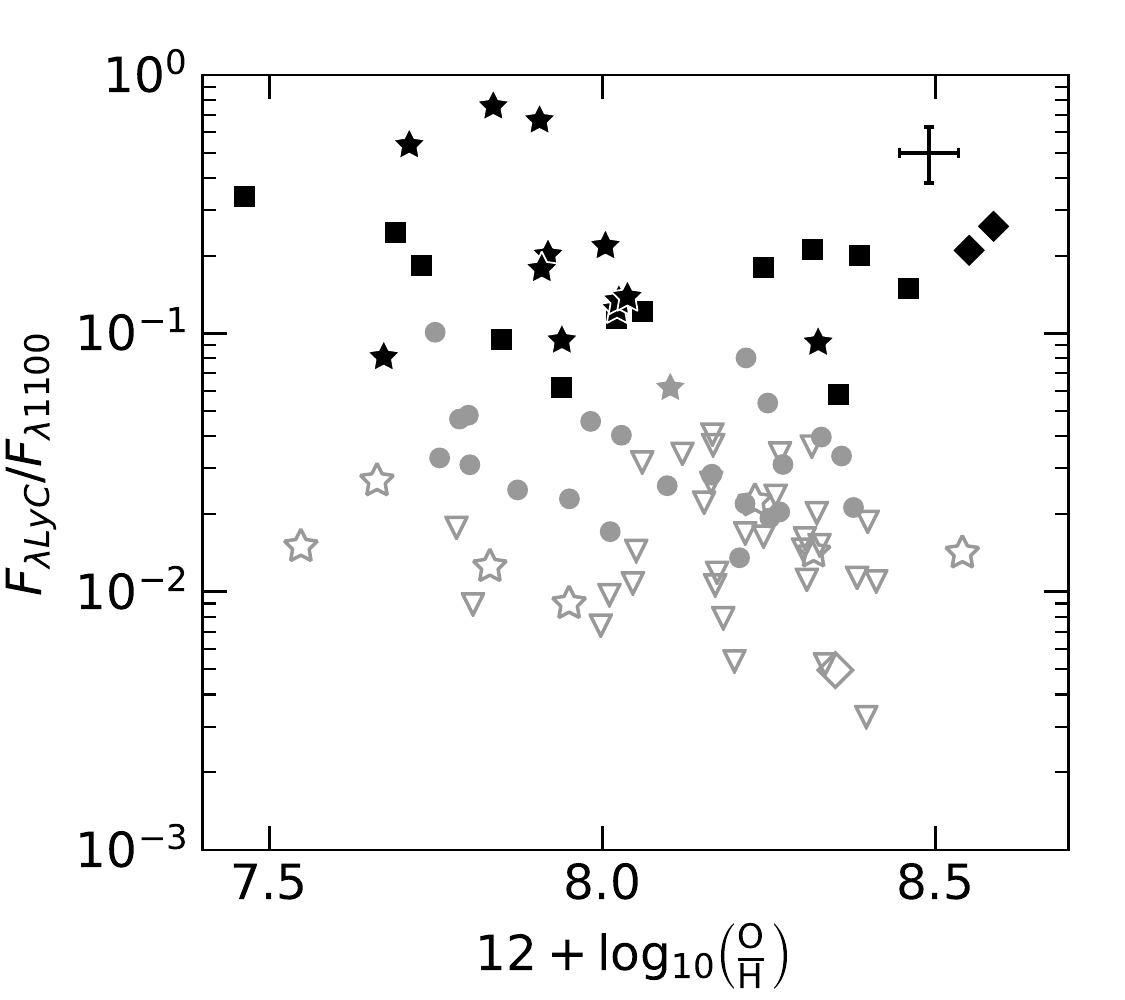}
    \includegraphics[width=0.32\textwidth]{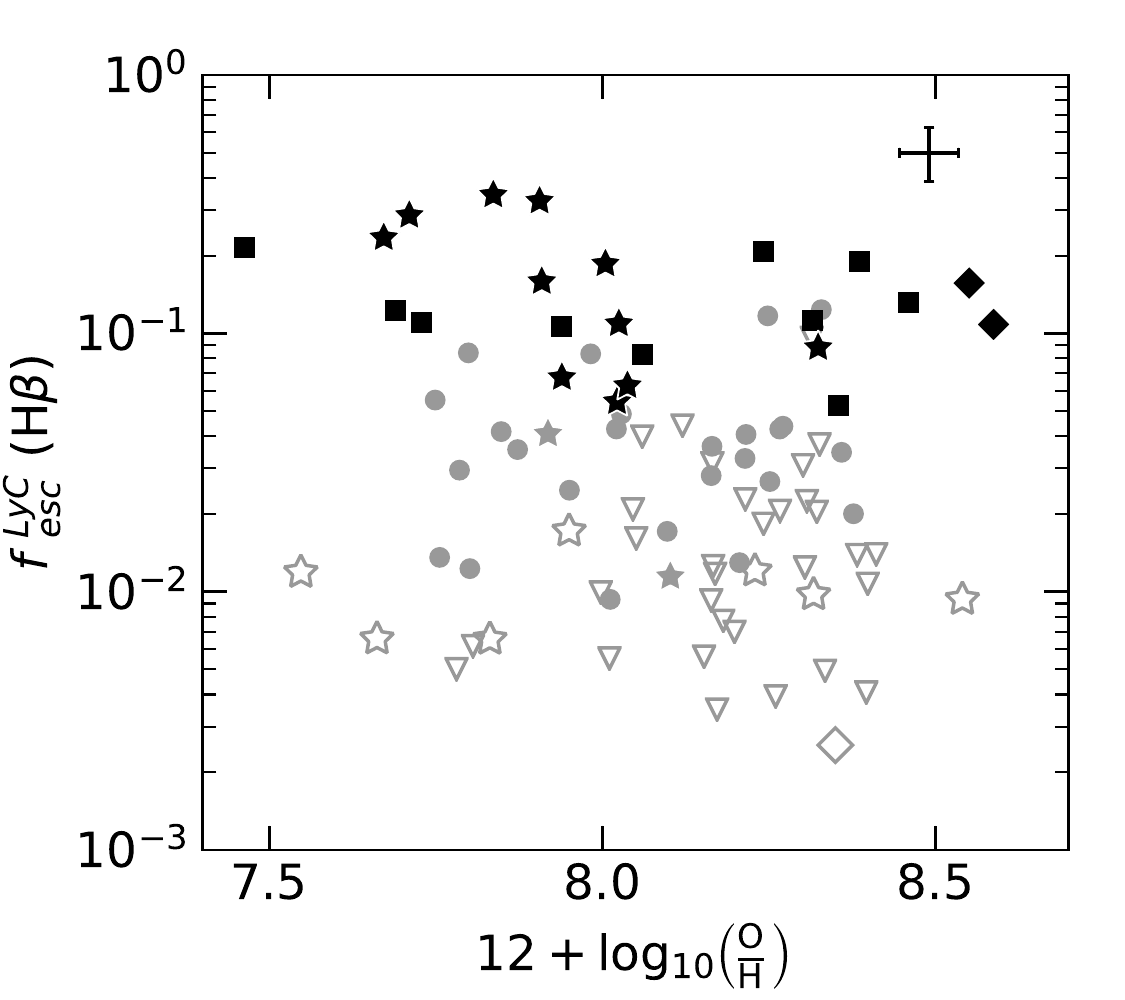}
    \includegraphics[width=0.32\textwidth]{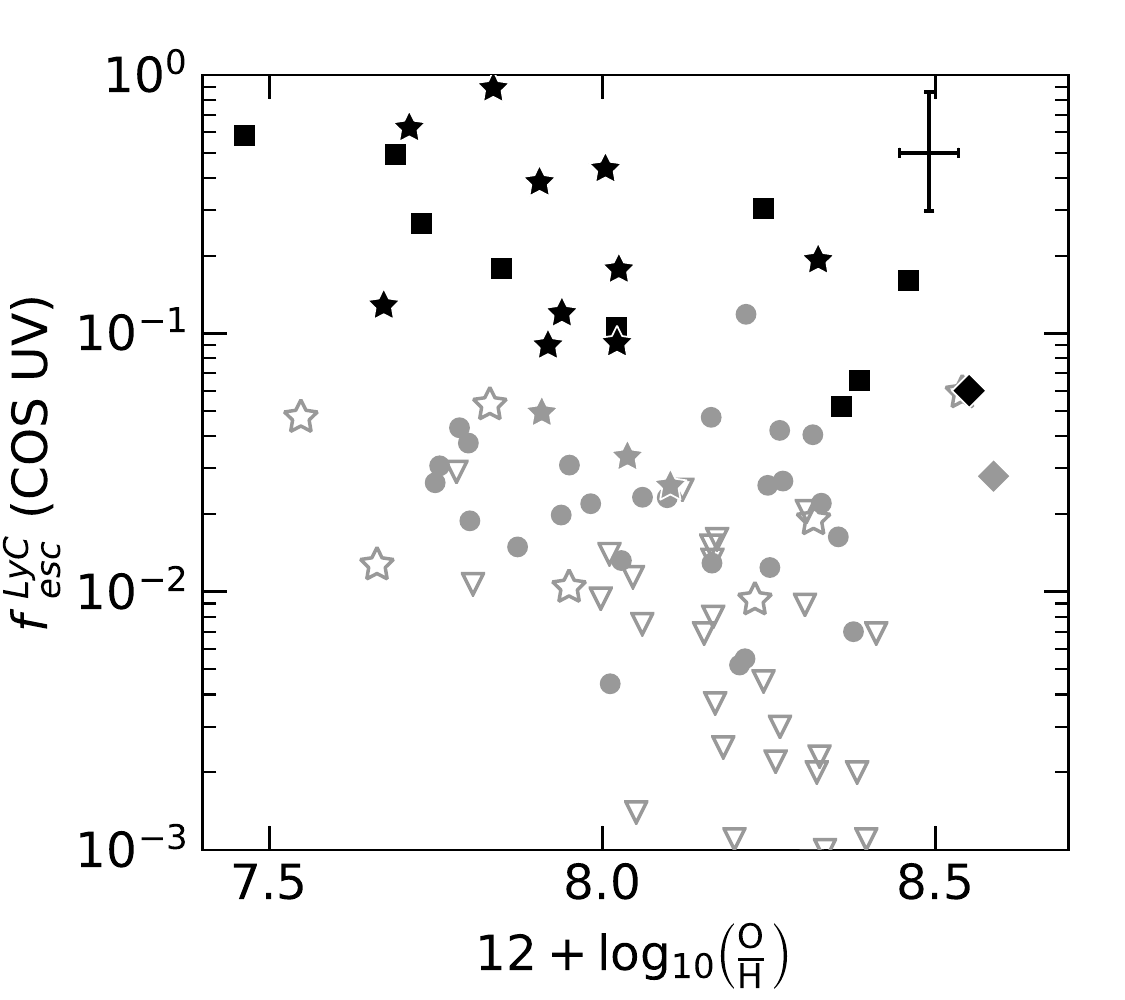}
\caption{Same as Figure \ref{fig:fesc_lya} but for gas phase metallicity \oh.}
    \label{fig:oh12}
\end{figure*}

{ In Figure \ref{fig:oh12}, we compare \fesclyc\ to the gas phase metallicity \oh. The LCEs with \fesclyc$\ga10$\% all reside below \oh$\sim 8.1$. Much like the case of $M_\star$, the strong LCEs with \fesclyc$\la10$\% and the weak LCEs are more evenly distributed from \oh$=7.7-8.5$. While substantial scatter appears in all three diagnostics, roughly two thirds (28/39) of the non-LCEs fall \emph{above} \oh$\sim 8.1$. With a correlation coefficient of $\tau\sim-0.2$ at about 2$\sigma$ significance, \fesclyc\ does not have a significantly strong correlation with \oh; however, the preferences of the most prodigious LCEs for lower metallicities and non-LCEs for higher metallicities is promising as a diagnostic. }

\begin{figure}
    \centering
    \includegraphics[width=\columnwidth]{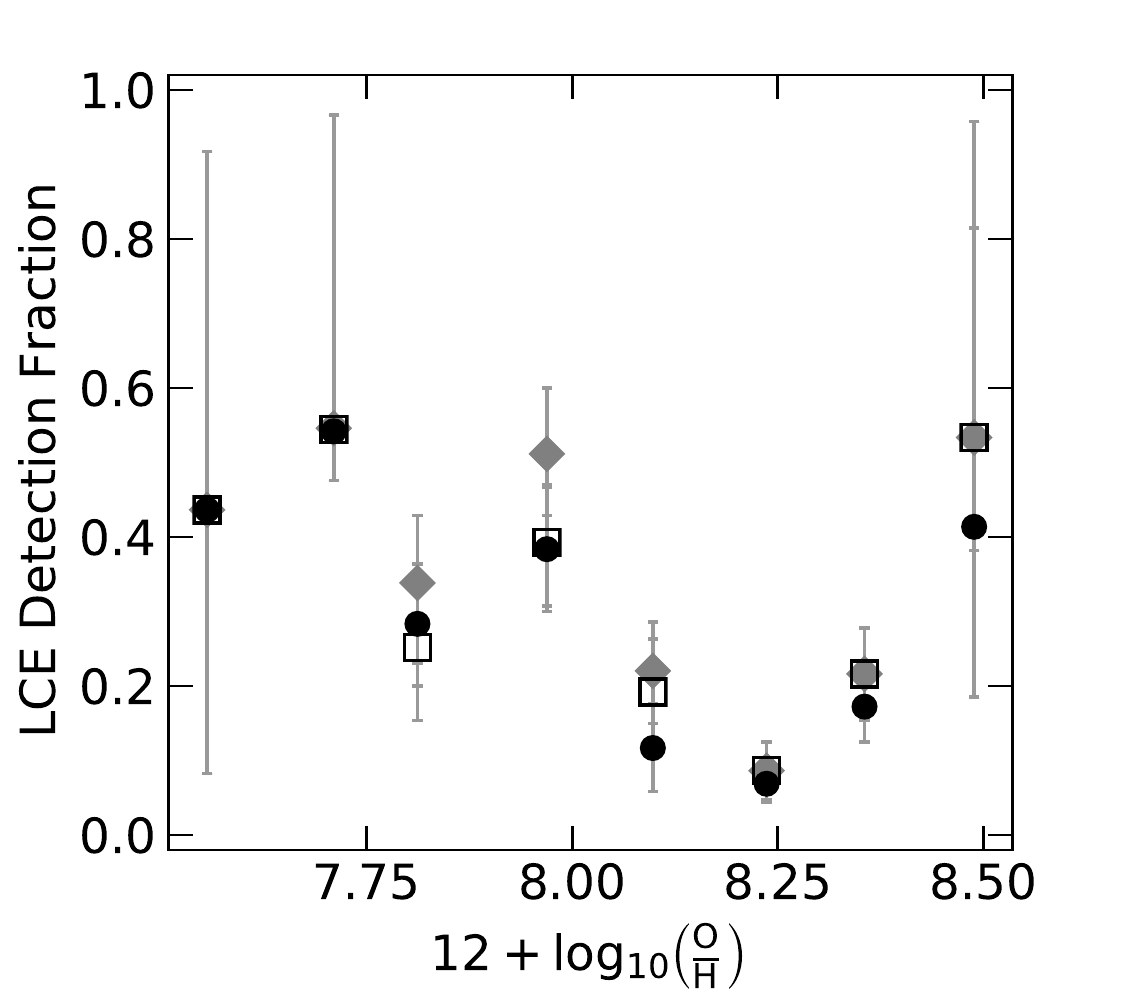}
    \caption{Same as Figure \ref{fig:lya_lce_frac} but for \oh.}
    \label{fig:oh12_lce_frac}
\end{figure}

The LCE fraction shown in Figure \ref{fig:oh12_lce_frac} also suggests \oh\ is a promising, albeit complicated, diagnostic. As with the \fesclyc\ trends shown in Figure \ref{fig:oh12}, we see strong LCEs both above and below \oh$\sim 8.1$. Given the uncertainty in the LCE fraction, it is unclear whether the lack of LCEs at intermediate values of \oh$\sim 8.1$ implies the existence of distinct LCE populations separated by metallicity.

\subsection{Summary of Correlations}

{ Ten of the seventeen indirect \fesclyc\ diagnostics considered above are significantly strong ($|\tau|>0.2$ with $p<0.00135$, $3\sigma$ significance): \fesclya, EW \lya, $v_{sep}$, \orat, EW H$\beta$, $\beta_{1200}$, $M_{1500,int}$, NUV $r_{50}$, \sigsfr, and $\rm\Sigma_{sSFR}$.
The \fesclya, EW \lya, $\beta_{1200}$, NUV $r_{50}$, and \sigsfr\ diagnostics are significantly strong across both \fesclyc\ metrics. These five diagnostics are thus the most compelling for identifying LCEs and inferring \fesclyc. H$\beta$ EW and $M_{1500,int}$ are significantly strong correlations only with the COS UV \fesclyc. In general, \fesclyc\ derived from fitting the UV spectrum gives the strongest, most significant correlations for nearly all the indirect diagnostics. The success of this particular LyC escape indicator is particularly compelling because, as discussed in \S\ref{sec:lzlcs}, it is the most reliable and direct measure of \fesclyc.}

{The indirect \fesclyc\ diagnostics motivated by optical depth contain the majority of the strong correlations. Likely owing to their unique sensitivity to \ion{H}{1}, the diagnostics based on \lya--\fesclya, \lya\ EW, and profile peak separation--have the strongest correlations as measured by Kendall's $\tau$ for censored data. However, the $v_{sep}$ correlations lack the significance of the other \lya\ correlations, in part because of insufficient data. While not as strongly correlated with \fesclyc\ as the \lya\ properties, the \orat\ flux ratio still exhibits a strong, significant correlation. The [\ion{O}{1}] flux relative to other emission lines' is less successful as an \fesclyc\ diagnostic, likely owing to the faintness of [\ion{O}{1}].}

{Of the remaining diagnostics, which are motivated by the physical mechanism behind LyC escape, the half-light radius and \sigsfr\ have the strongest, most significant correlations with \fesclyc. Indeed, strong LCEs exhibit clear associations with highly concentrated star formation, distinguishing which sorts of galaxies or regions within galaxies are emitting LyC. Both H$\beta$ EW and UV $\beta_{1200}$ also correlate strongly and significantly with \fesclyc. While stellar mass and attenuated UV magnitude lack strong, significant correlations, these two diagnostics still exhibit envelopes in \fesclyc\ and suggest the strongest LCEs have $M_\star<10^9$ M$_\odot$. Such envelopes in \fesclyc\ indicate that multiple properties contribute to an ideal escape scenario, including starburst age, ionization parameter, star formation rate, and orientation of optically-thin channels.}

\section{Relationships Between Diagnostics}\label{sec:2d-diag}

The results of these indirect diagnostics, particularly the envelopes in \fesclyc, prompt further investigation of two-dimensional diagnostics to determine trends in \fesclyc\ with multiple parameters and explore sources of the scatter apparent in many diagnostics. { Below, we consider relationships between diagnostics regarding stellar population age ($\beta_{1200}$ vs H$\beta$ EW, \orat\ vs H$\beta$ EW), star formation history (\orat\ vs $M_\star$, \orat\ vs \oh), ionization structure (\orat\ vs $O_{31}$), and ionization vs mechanical feedback (\sigsfr\ vs \orat). Given fewer systematic uncertainties associated with the UV SED \fesclyc, we proceed with this particular \fesclyc\ metric in our subsequent analysis and discussion.}

\subsection{UV $\beta_{1200}$ vs H$\beta$ EW}\label{sec:zackrisson}

\begin{figure}
    \centering
    \includegraphics[width=\columnwidth]{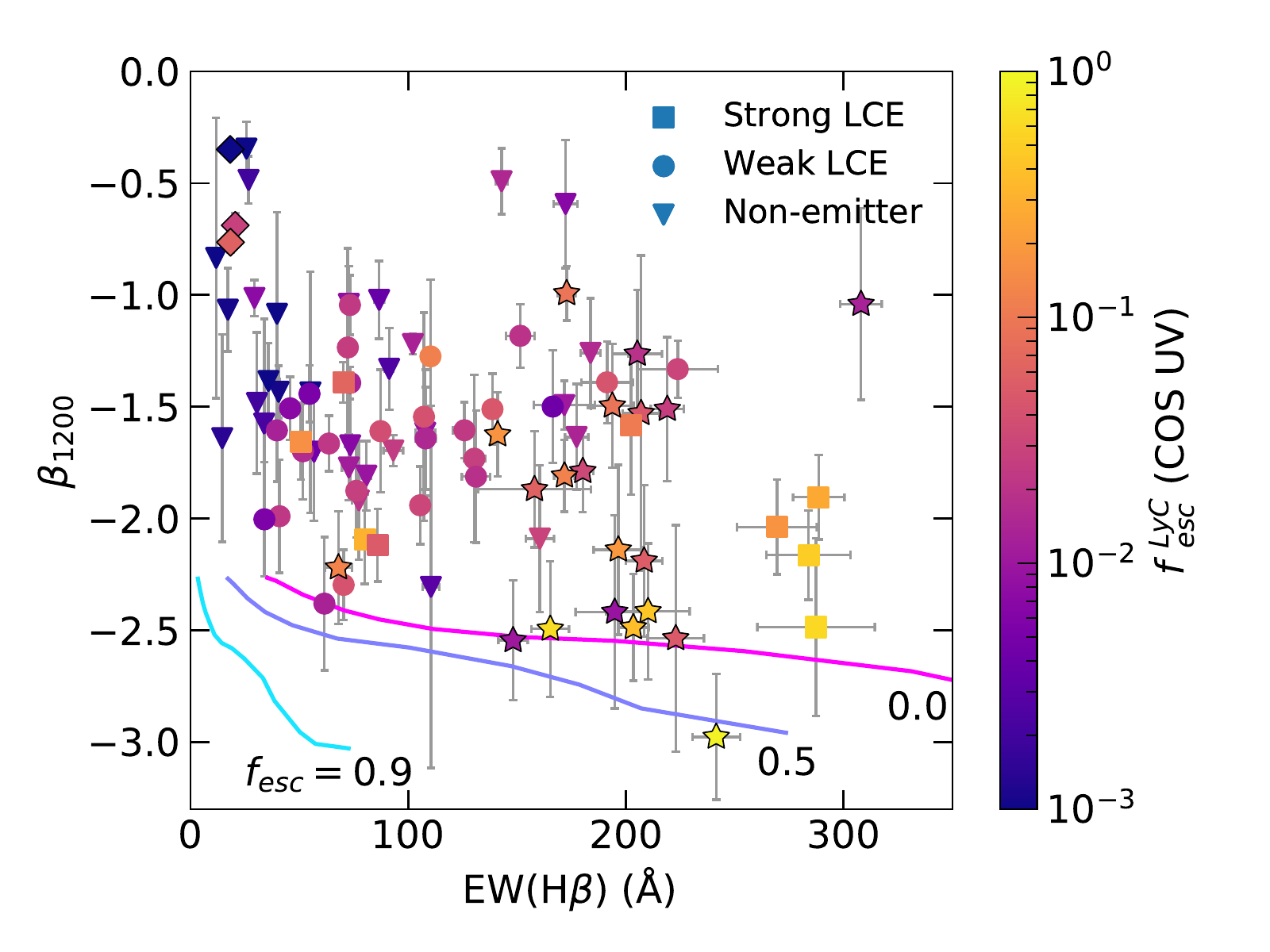}
\caption{UV $\beta_{1200}$ vs H$\beta$ EW for the LzLCS strong (squares) and weak (dots) LCEs and non-emitters (triangles) color-coded by \fesclyc\ derived from the UV spectrum.
Solid lines are the \citet{2013ApJ...777...39Z} $\sim30$\% solar metallicity picket fence model predictions for \fesclyc$=0.0$ (magenta), $0.5$ (purple), and $0.9$ (cyan). There appears to be no association of \fesclyc\ with a particular sequence in the UV $\beta_{1200}$ - EW(H$\beta$) plane. {While symbols are color-coded by the UV-fit \fesclyc, the observed trends here and in subsequent figures are similar if one of the other \fesclyc\ metrics are used.}}
    \label{fig:zackrisson}
\end{figure}

Although most galaxies with H$\beta$ EWs $<150$ \AA\ do not appear associated with LyC escape (\S\ref{sec:hbew}), \citet{2013ApJ...777...39Z} propose the UV $\beta_{1200}$ -- H$\beta$ EW plane could constrain \fesclyc. We compare their model predictions for \fesclyc$=0.0$, $0.5$, and $0.9$ to the LzLCS results in Figure \ref{fig:zackrisson} for $\sim30$\% solar-metallicity, picket fence nebulae over a range of $10^6$ to $7\times10^8$ year burst ages. Our combined results do not agree with their predictions. We do see a mild gradient in \fesclyc\ over UV $\beta_{1200}$ that appears to shift to steeper continua with increasing EW, as their predictions suggest. However, the observed locus appears shifted to higher spectral slopes: only one previously published LCE falls into the \fesclyc$>0$ space predicted by \citet{2013ApJ...777...39Z}.

A major caveat to this apparent disagreement is the effect of dust, which may substantially affect the predicted sequences in the UV $\beta_{1200}$ - H$\beta$ EW plane \citep{2017ApJ...836...78Z}. Moreover, the star formation history also determines where a galaxy resides in Figure \ref{fig:zackrisson} as underlying older stellar populations can strongly affect the optical continuum and thus the H$\beta$ EW. LzLCS galaxies appear to be older and/or dustier than the \citet{2013ApJ...777...39Z} models, which places our sample at lower H$\beta$ EW and higher UV $\beta_{1200}$ than their predictions. { Interpreting the discrepancy comes with an additional complication that the UV $\beta$ predicted in the \citet{2013ApJ...777...39Z} is measured at redder wavelengths than those accessible in the COS spectrum. Because an attenuated young stellar SED peaks at $\sim1100-1200$ \AA, $\beta_{1200}$ is necessarily different from the \citet{2013ApJ...777...39Z} UV $\beta$ predictions. That being said, we find no distinguishable trends in \fesclyc\ in the $\beta_{1200}$-EW(H$\beta$) plane.}

Whereas the \citet{2013ApJ...777...39Z} models suggest that strong LCEs should have low H$\beta$ EWs and low UV $\beta_{1200}$, the strongest LCEs have high ($>100$ \AA) H$\beta$ EWs while those LCEs with lower EWs have higher UV $\beta_{1200}$. The latter group of LCEs may contain porous \ion{H}{2} regions with optically thin channels cleared by SN feedback or turbulence (see \S\ref{sec:discussion}). In particular, a two-stage starburst or dusty young starburst could be responsible for the high UV $\beta_{1200}$, low H$\beta$ EW exhibited by some of the strong LCEs.

\subsection{\orat: Age, Mass, Metallicity}
\label{sec:o32_agemass}

\begin{figure}
    \centering
    \includegraphics[width=\columnwidth]{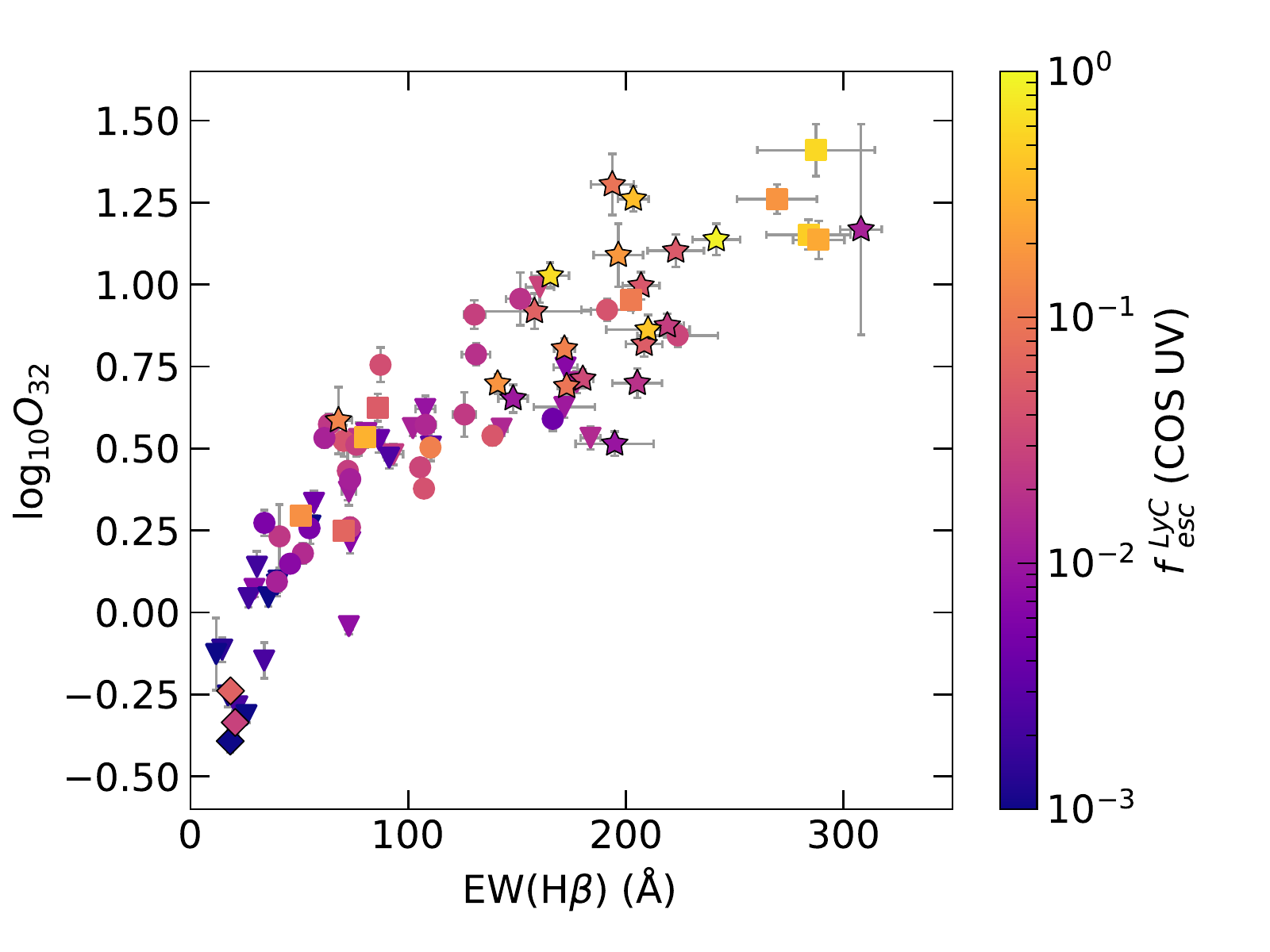}
    \caption{\orat\ vs H$\beta$ EW. Symbols as in Figure \ref{fig:zackrisson}. A strong correlation is evident, owing to the coupling of ionization parameter with starburst age.}
    \label{fig:o32_ewhb}
\end{figure}

{Starburst age and ionization parameter are related since the earliest O stars die off within the first 2-4 million years. Because these stars dominate the LyC flux, properties sensitive to ionization parameter like \orat\ will consequently depend, at least in part, on the age of the starburst \citep[e.g.,][]{2013ApJ...766...91J,2017MNRAS.471..548I} in addition to other properties. Figure \ref{fig:o32_ewhb} demonstrates a tight relationship between \orat\ and H$\beta$ EW.}

The trend in Figure \ref{fig:o32_ewhb} could result from a variety of processes. Decreasing H$\beta$ flux due to fewer ionizing photons would decrease the EW in tandem with \orat\ as the O$^{+2}$ zone declines relative to O$^{+}$. Due to the mass-metallicity correlation for galaxies \citep[e.g.,][]{2004ApJ...613..898T}, increased continuum flux due to higher galaxy mass would decrease the EW while increasing the metallicity, the latter serving to decrease \orat. We find that \orat\ and EW decrease with increasing mass, but because mass correlates with \oh, it is not clear whether mass dominates this trend.

{We find that LCEs occupy a wide range in \orat\ and H$\beta$ EW.} Roughly half of the LCEs occur at $\log_{10}$\orat$>1$ or H$\beta$ EW$>100$ \AA. Non-emitters do not appear in this part of the diagram while all the \fesclyc$>0.2$ leakers reside here. The remaining LCEs have lower H$\beta$ EWs ($\approx 50-150$ \AA) where many of the non-emitters also reside. These separate sets of LCEs may imply different starburst ages or stellar mass populations at which LyC escape occurs.

Escaping LyC photons should affect both H$\beta$ EW and \orat. The former should decrease due to a drop in H$\beta$ flux while the latter should increase due to a drop in [\ion{O}{2}] flux. A lack of preference of \fesclyc\ for any part of the distribution demonstrates that age and ionization parameter mitigate any effects of escaping LyC photons on the measured properties.

\begin{figure}
    \centering
    \includegraphics[width=\columnwidth]{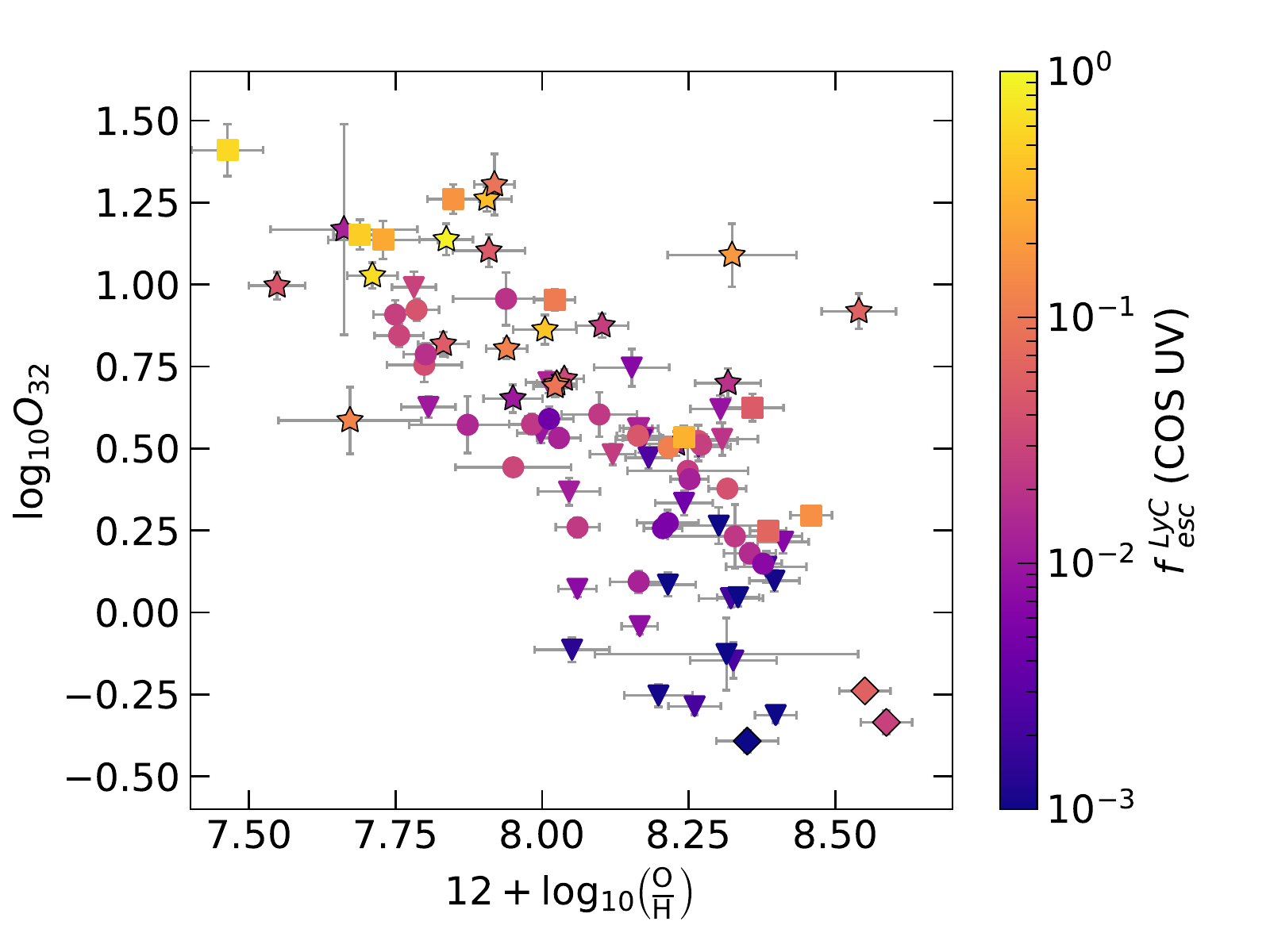}
    \includegraphics[width=\columnwidth]{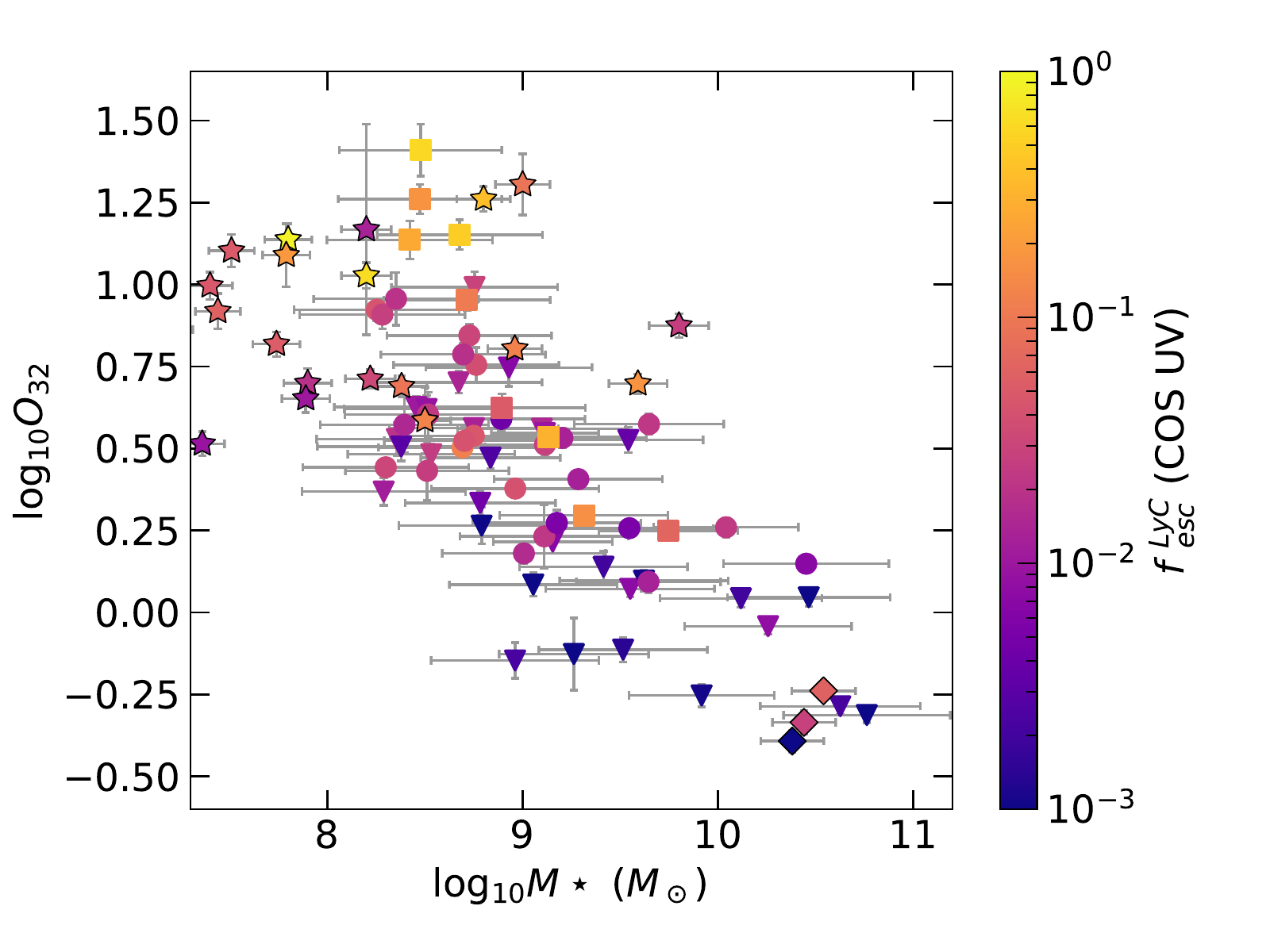}
    \caption{\orat\ compared to metallicity indicator \oh\ and SFH indicator $M_\star$. Symbols as in Figure \ref{fig:zackrisson}.}
    \label{fig:o32_oh12_mstar}
\end{figure}

{Gas-phase and stellar metallicities also affect \orat. As collisionally excited lines, [\ion{O}{3}] and [\ion{O}{2}] depend on electron temperature, which is higher in low metallicity gas due to decreased cooling by metal emission lines. Higher stellar metallicity changes the opacity of stellar interiors and thus affects their photosphere temperatures. Secondarily, high stellar metallicity increases line blanketing in the LyC of O and B stars. These metallicity effects on stellar atmospheres serve to decrease the ionizing photon budget. These dependencies on gas-phase and stellar metallicity indicate that \orat\ correlates with an abundance indicator like \oh, which we show in Figure \ref{fig:o32_oh12_mstar}.}

While there is some scatter, a trend persists between \orat\ and \oh\ over nearly an entire decade. Although high \fesclyc\ LCEs tend to have higher \orat\ and lower \oh, LCEs populate the majority of the distribution in Figure \ref{fig:o32_oh12_mstar}. {Harder ionizing spectra at lower metallicity could facilitate a density-bounded escape scenario. Weaker feedback from low-metallicity stars may also generate a dense, clumpy gas geometry that enables LyC escape \citep{2019ApJ...885...96J}, perhaps owing to low covering fractions \citep[e.g.,][]{2020A&A...639A..85G,2022arXiv220111800S}. Mechanical feedback from stars is stronger at higher metallicity, which might explain the subpopulation of high LCEs at higher metallicity. However, higher metallicity is also linked to higher UV attenuation \citep[e.g.,][]{2019ApJ...885...57W}, thereby limiting LyC escape. The reduced \fesclyc\ of LCEs at higher metallicity may also be due to the fact that, at higher galaxy masses, increased gravity may reduce the effectiveness of feedback and thus limit LyC escape.}

Following the known mass-metallicity relation, \orat\ ought to anti-correlate with $M_\star$ just as it does with \oh. Indeed, in Figure \ref{fig:o32_oh12_mstar}, we see such a trend until $M_\star\la10^9$ M$_\odot$ (corresponding to \orat$\approx2-3$), at which point the two properties appear to decouple. { With weaker gravitational potentials, low $M_\star$ galaxies also undergo more bursty star formation \citep[e.g.,][]{2012ApJ...744...44W,2016ApJ...833...37G}. Such episodic star formation results in a range of stellar feedback and ionization parameters, which would lead to a wide range in \orat.}

For a given metallicity or stellar mass, strong LCEs tend to have the highest \orat\ values. The leakers with \fesclyc$>0.2$\ concentrate at low \oh, low $M_\star$, and high \orat. Thus, the strength of a starburst relative to the mass of a galaxy may determine its likelihood of leaking LyC.

\subsection{Oxygen Ionization Structure}

\begin{figure}
    \centering
    \includegraphics[width=\columnwidth]{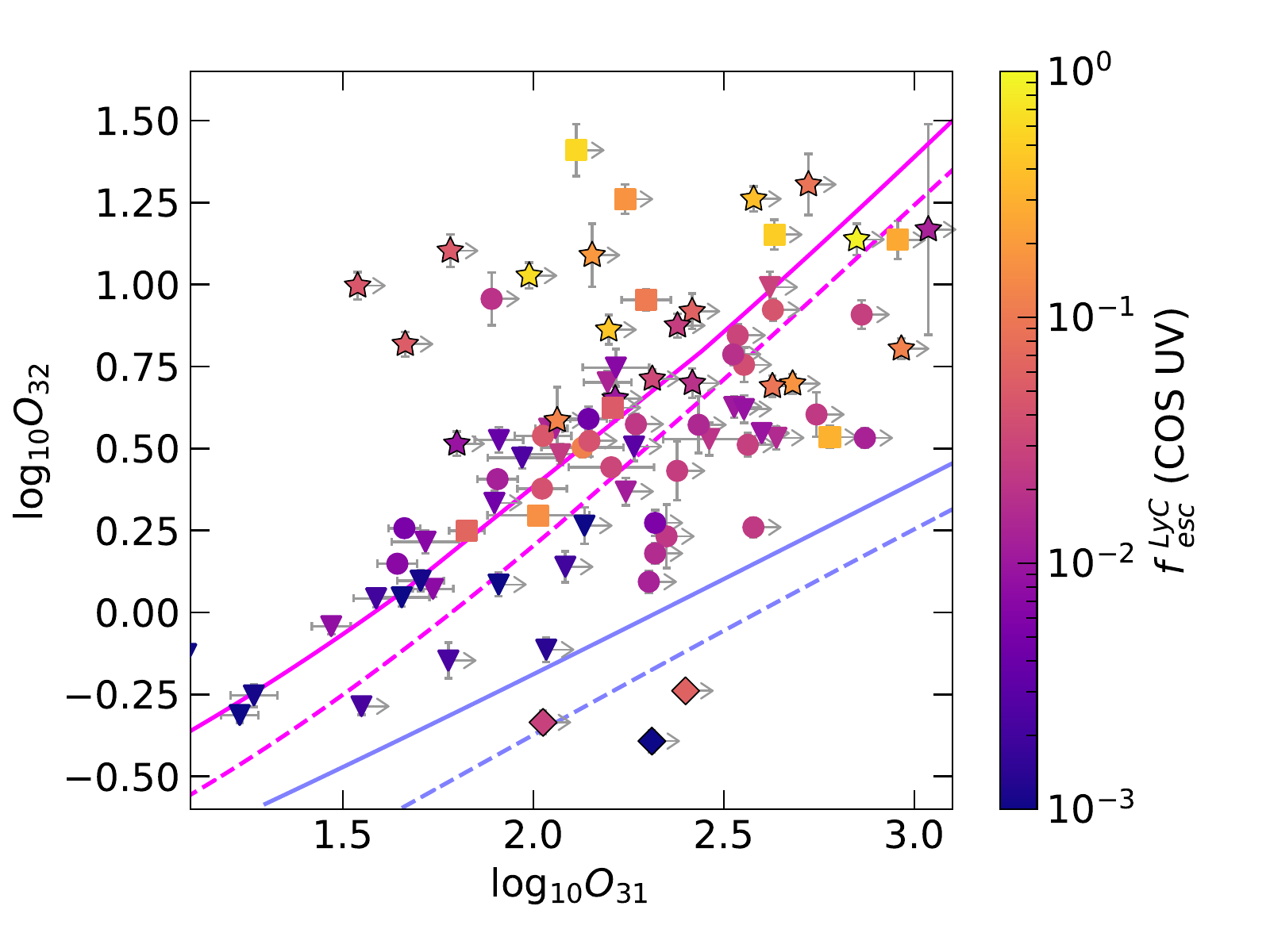}
    \caption{\orat\ vs $O_{31}$. Symbols as in Figure \ref{fig:zackrisson}. \citet{2015A&A...576A..83S} predictions for a 2 Myr starburst in a filled sphere (solid) and a spherical shell (dashed) nebula bounded by radiation (magenta) and density (purple). As a burst ages, \orat\ decreases at a fixed $O_{31}$.}
    \label{fig:stasinskao3o1}
\end{figure}

Photoionization models predict that $O_{31}$ should increase by 0.5 dex or more as the LyC optical depth decreases from ionization to density boundary conditions at a fixed \orat\ \citep[e.g.,][]{2015A&A...576A..83S,2019MNRAS.490..978P,2020A&A...644A..21R}. We compare our results to the predicted sequences for radiation- and density-bounded nebulae models from \citet{2015A&A...576A..83S} in Figure \ref{fig:stasinskao3o1}. The combined samples reside on or above the radiation-bounded sequence. We find no relationship between \fesclyc\ and any offset from the locus. Because the strongest LCEs have only upper limits in [\ion{O}{1}], these galaxies may in fact be consistent with density-bounded or picket fence scenarios. LCEs with detected [\ion{O}{1}] reside along the radiation-bounded model predictions, which may suggest that an isotropic density-bounded escape scenario is not realized. Indeed, \citet[][]{2020A&A...644A..21R} predict that density-bounded channels within an otherwise optically-thick medium can describe the distribution of \orat\ and $O_{31}$ flux ratios we observe for LCEs with \fesclyc$\la0.1$. The [\ion{O}{1}] flux from LCEs with high \orat\ and disproportionately low $O_{31}$ may also be contaminated by shocks, causing $O_{31}$ to shift to lower values than other LCEs \citep[][]{2015A&A...576A..83S,2019MNRAS.490..978P}.

\subsection{\sigsfr: Ionization \& Age}

\begin{figure}
    \centering
    \includegraphics[width=\columnwidth]{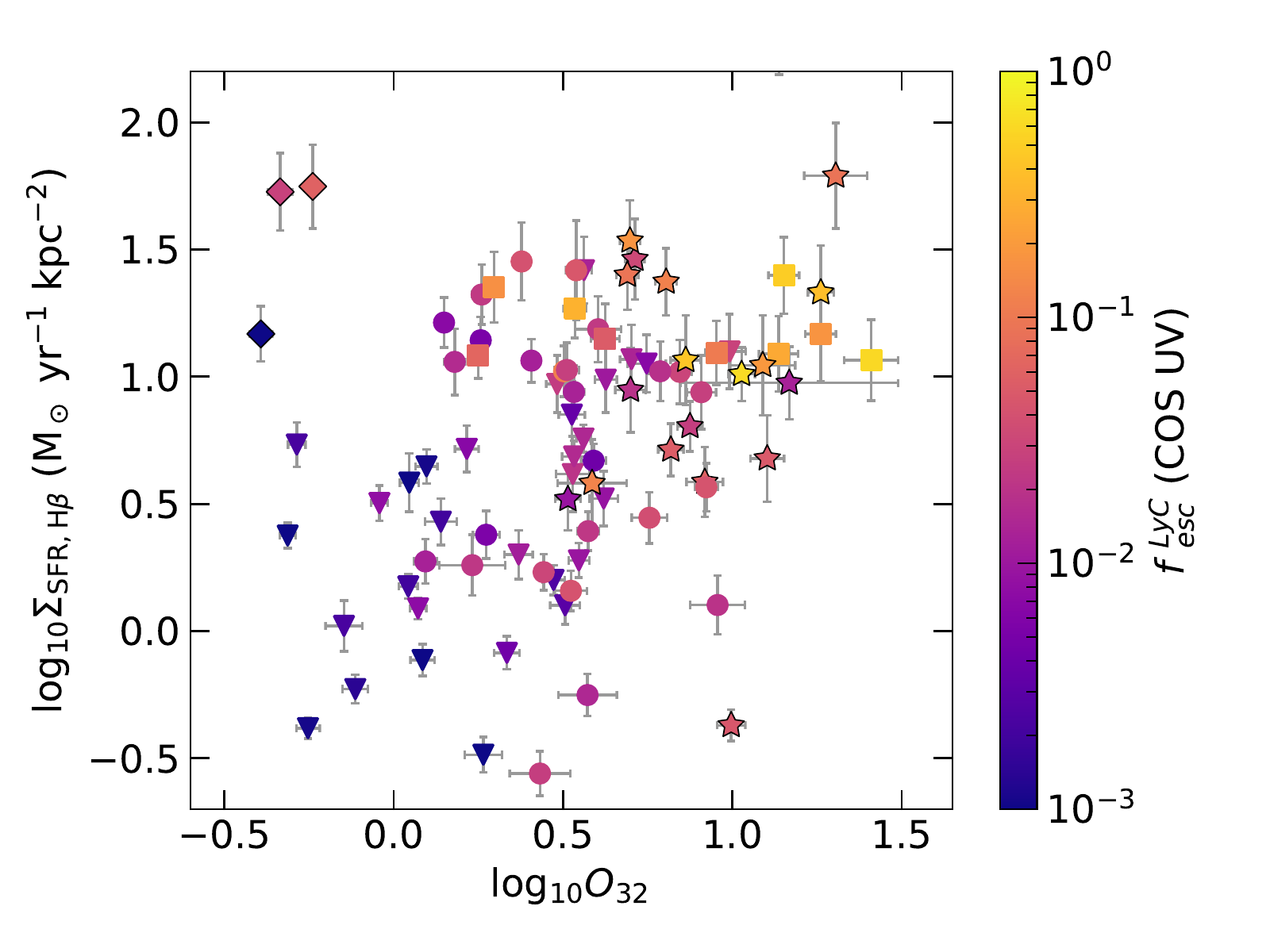}
    \caption{\sigsfr\ vs \orat. Symbols as in Figure \ref{fig:zackrisson}.}
    \label{fig:sfr_o32}
\end{figure}

\sigsfr\ describes the concentration of O and B stars. As a result, \sigsfr\ influences the ionization parameter, which depends strongly on the number density of early type stars. At least in the case of density-bounded LyC escape, we anticipate higher \orat\ for LCEs than for non-emitters of the same \sigsfr\ due to a deficit in [\ion{O}{2}] flux.

We find LCEs prefer high \sigsfr\ and high \orat, although only one of the two is necessary to demonstrate the likelihood of LyC escape. Figure \ref{fig:sfr_o32} demonstrates that, with few exceptions, LCEs have high \orat, high \sigsfr, or a combination thereof. The LCE population with $\log_{10}\Sigma_{SFR}<1$ $M_\odot$ yr$^{-1}$ kpc$^{-2}$ is sparse. Thus, high \orat\ and high \sigsfr\ are each indicative of LCEs, in particular {\it strong} LCEs, with the most extreme LCEs having both high \orat\ and high \sigsfr.

Combined with the results of UV $\beta_{1200}$ and H$\beta$ EW, the combination of \sigsfr\ and \orat\ could suggest two different types of strongly star-forming LCEs. One population is younger with high \orat, high H$\beta$ EW, and low metallicity. The other population has lower \orat, lower H$\beta$ EW, and higher metallicity, and resides in the upper left of Figure \ref{fig:sfr_o32}. The latter suggests mechanical feedback, while the former population may point to the additional role of ionization.

\section{Discussion}\label{sec:discussion}

\subsection{Insight Into Lyman Continuum Escape}

The substantial scatter evident in all considered diagnostics could be attributed to one or both of two physical explanations. First, orientation could cause the observed scatter since \fesclyc\ likely depends on line of sight in an anisotropic, radiation bounded escape scenario \citep[e.g.,][]{2013ApJ...779...76Z,2015ApJ...801L..25C}. { Alternatively, the observed scatter might be due to a time delay between the onset of star formation and the escape of the LyC since feedback requires sufficient time to clear out escape channels, thus decoupling properties from \fesclyc\ to some extent \citep[e.g.,][]{2017MNRAS.470..224T,2020arXiv201000592B,2022MNRAS.510.4582N}.} Whatever might cause this scatter, it is likely physical because an upper envelope in \fesclyc\ persists across several diagnostics. These envelopes suggest genuine trends do exist between \fesclyc\ and different parameters, but orientation, time delays, and/or starburst properties like age and ionization parameter might influence the values of \fesclyc, causing the observed scatter.

{As implied by the envelopes and correlations in the \fesclyc\ diagnostics, the strongest correlations are \lya\ EW, \fesclya, and $v_{sep}$ for the diagnostics motivated primarily by optical depth. Other diagnostics with strong correlations include \orat, H$\beta$ EW, \sigsfr, and $r_{50}$, which may give insights into the physical mechanisms responsible for LyC escape. Some properties, notably \orat\ and H$\beta$ EW, exhibit a broad range in galaxies with \fesclyc$>0.05$. Different LyC escape scenarios \citep[e.g.,][]{2019arXiv190502480K,2020A&A...639A..85G} could populate different parts of these diagnostics because the corresponding conditions (e.g., optical depth, porosity of the ISM, or feedback mechanism driving LyC escape) correspond to different ranges of the observable parameter space of \orat\ and H$\beta$ EW.}

Investigation of two-dimensional diagnostics sheds light on these possible explanations. At higher mass and higher metallicity, the lower \orat, lower H$\beta$ EW LCEs likely clear out their birth cloud by mechanical feedback from early type stars. This serves to more uniformly distribute the dense gas and dust leftover from star formation into an optically thick bubble around the O and B stars. SNe and turbulent feedback subsequently clear optically thin channels in this bubble.

The younger, lower mass, lower metallicity nature of the high \orat, high H$\beta$ EW LCEs indicates this second class of LyC leakers are like the GP galaxies. Such galaxies, while strongly star forming, likely lack the high velocity outflows necessary to expel the remnant birth cloud material and form a uniform obscuring medium \citep{2019ApJ...885...96J}. However, they may have sufficient radiative feedback to evacuate cavities in the ISM \citep{2019arXiv190502480K}. As a result, these GP LCEs might contain an ISM that is optically thin except where dense clouds cause optically thick patches of relatively small covering fractions. We speculate that the range of parameter spaces occupied by LCEs indicates a shift in conditions and escape mechanism from young GP starbursts with large optically thin cavities to older, optically thick star forming galaxies with a porous ISM.

While these putative differences in LCEs spurred our comparison of \fesclyc\ proxies, the two-dimensional diagnostics also demonstrate how different parameters depend on one another. These comparisons show that metallicity, mass, \sigsfr, age, and ionization parameter are inextricably linked. Furthermore, we find that age and ionization parameter dominate any optical depth effects on properties like \orat\ and H$\beta$ EW. In other words, young ages and strong radiation boost \orat\ and H$\beta$ EW in the strongest LCEs and may be critical for high levels of LyC escape. It is also plausible that a feedback-induced time delay introduces LCE ``stragglers" into the distributions of \fesclyc\ \citep[e.g.,][]{2017MNRAS.470..224T,2020arXiv200715012S}.

\subsection{Comparison with Simulations}

Cosmological simulations provide a critical link between the low redshift LCEs and high redshift surveys. Predictions from simulations connect parameters such as stellar mass and \sigsfr\ to \fesclyc\ at the epoch of reionization. Because LzLCs galaxies and other GP-like galaxies are reasonable analogs of high-redshift galaxies, comparing our results to these predictions from simulations can be useful both for translating indirect diagnostics to high redshift and for interpreting the distributions we observe.

One of the most contentious issues regarding reionization is the mass regime of galaxies that provide the ionizing photons. Some simulations predict that \fesclyc\ decreases with stellar mass \citep[e.g.,][]{2014ApJ...788..121K} and UV luminosity \citep[e.g.,][]{2017MNRAS.470..224T,2020arXiv200715012S} while others predict that \fesclyc\ increases with stellar mass and UV luminosity \citep[e.g.,][]{2020arXiv200305945M}. The strongest LCEs in the combined sample of local LCEs are smaller, fainter galaxies with $M_\star\la10^9$ M$_\odot$ and $M_{FUV}\ga-19$. Reionization is likely driven by galaxies with similar properties.

The LzLCS samples the upper end of the simulated $M_\star$-\sigsfr\ distribution of LCEs in \citet{2020ApJ...892..109N}. We find that the combined sample of LCEs is largely consistent with their predictions, including the range in \fesclyc. However,
our average \fesclyc\ toward higher stellar mass bins is about 0.1 dex lower than theirs even if considering only the strongest LCEs. { Moreover, although we find qualitative agreement between the envelope in \fesclyc\ with respect to \sigsfr\ and the \fesclyc$\propto\Sigma_{SFR}^{0.42}$ relation prescribed by \citet{2020ApJ...892..109N}, the $M_{1500}<-20$ LCEs in the LzLCS are not strong leakers at the same rate as their $M_{1500}>-19$ counterparts.} While seemingly contrary to predictions by \citet{2020ApJ...892..109N}, the fact that brighter, higher mass galaxies are less likely to be prodigious LCEs in our sample may be consistent with their simulations as intermediate-mass galaxies at $z>4$ tend to have emission-line properties similar to those of the $z\sim0.3$ low-mass galaxies of this study. The \fesclyc-\sigsfr\ envelope also agrees with the sharp increase in \fesclyc\ predicted by \citet{2017MNRAS.468.2176S}, although we do not see the 20\% \fesclyc\ values they suggest for $M_\star>10^9$ $M_\odot$. This might suggest that properties in addition to \sigsfr\ (e.g., metallicity, dust) affect LyC escape.

\begin{figure}
    \centering
    \includegraphics[width=\columnwidth]{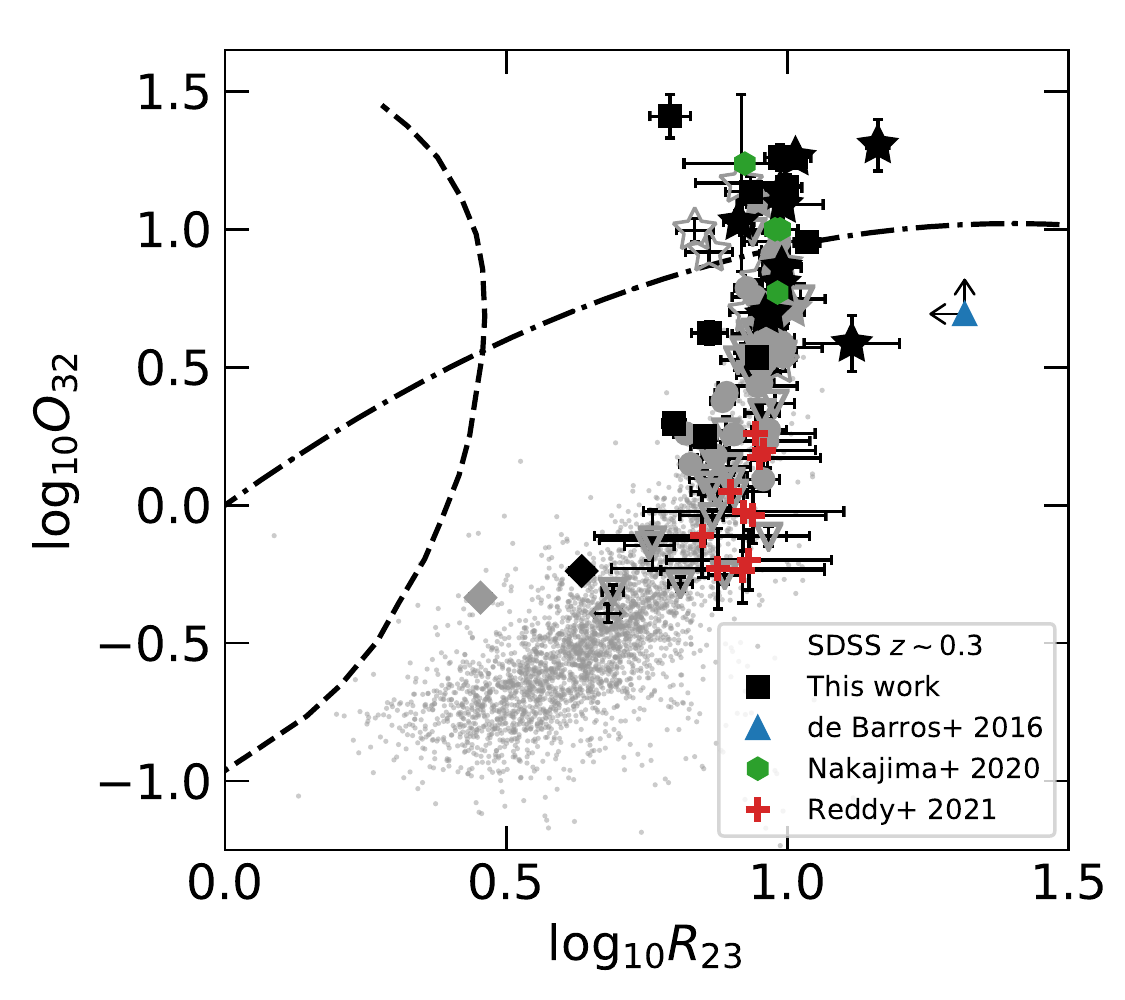}
    \caption{The \orat\ vs $R_{23}$ diagnostic for the combined sample. Symbols as in Figures \ref{fig:fesc_lya}-\ref{fig:oh12} with the addition of LCEs from \citet{2020ApJ...889..161N} with measured \orat\ and $R_{23}$ (green hexagons), the $z=3.2$ LCE from \citet[blue triangle]{2016A&A...585A..51D}, { and the $z=1.85-3.49$ LBGs from \citet[red crosses]{2021arXiv210805363R}.} For reference, we show SDSS star-forming galaxies (grey dots) selected from the value-added catalog of \citet{2013MNRAS.431.1383T} with requirements that $z\in[0.2,0.4]$ and that strong emission lines have $F/\sigma>3$. We show the locus of high-redshift LCEs predicted from cosmological simulations by \citet[dash-dotted]{2020MNRAS.498..164K} and \citet[dashed]{2020arXiv201000592B}.}
    \label{fig:o32_r23}
\end{figure}

\orat\ is a more empirical measurement than $M_\star$ or \sigsfr\ and directly traces the ionization parameter. Together, these attributes have favored \orat\ as an \fesclyc\ diagnostic in the past. Unfortunately, properties like \sigsfr\ and metallicity can also contribute to \orat, thus making interpretation of \orat\ unclear \citep[see][]{2014MNRAS.442..900N}. As with the LzLCS, recent studies have found substantial scatter in the \fesclyc-\orat\ diagnostic \citep[e.g.,][]{2018MNRAS.478.4851I,2019MNRAS.483.5223B}, which \citet{2019MNRAS.483.5223B} attribute to a combination of orientation and opening angle effects. With these physical and empirical caveats in mind, recent cosmological hydrodynamic simulations have predicted where high redshift LCEs should appear in the \orat-$R_{32}$ diagnostic \citep[e.g.,][]{2020MNRAS.498..164K}. We show the distribution of LzLCS, \pubsamp, and published LCEs \citep{2016A&A...585A..51D,2020ApJ...889..161N} and LAEs \citep{2021arXiv210805363R} at $z\sim2-3$ in Figure \ref{fig:o32_r23} along with $z\in[0.2,0.4]$ star-forming galaxies taken from the \citet{2013MNRAS.431.1383T} SDSS catalog for comparison. Simulations by \citet{2020arXiv201000592B} suggest the entire distribution of LCEs shifts 0.5 dex lower in $R_{23}$ at high redshift due to decreases in metallicity while maintaining high \orat\ through high ionization parameter.
On the other hand, simulations by \citet{2020MNRAS.498..164K} suggest LCEs at high redshift shift to a locus $\sim1$ dex higher in \orat\ than that of the SDSS galaxies, intersecting the low-metallicity ``tail'' where the combined low-redshift LCEs reside.

{ In contrast to these theoretical predictions, LCEs at moderate redshift ($z\sim3$, \citealt{2020ApJ...889..161N}) still reside in this tail, despite having fesc values in excess of 10\%. Thus, even in earlier epochs, properties like age, ionization parameter, and metallicity may continue to outweigh the effects of LyC escape in setting \orat. Star-forming galaxies from the MOSDEF survey at $z\sim2-3$ have \orat\ values more consistent with non-LCEs and typical SDSS star-forming galaxies. These galaxies are likely analogous to Lyman break galaxies (LBGs), suggesting that LBGs are only weakly leaking, if at all. \citet{2021arXiv210805363R} find that these galaxies have older burst ages (continuous star formation ages of $\sim100$ Myr), moderate ionization parameters ($\log U\sim-3$), and relatively higher metallicities ($\sim$30-40\% solar), which could readily account for the observed \orat\ values. These same properties may be associated with low levels of LyC escape.
The LzLCS results suggest that \orat\ and \fesclyc\ may be indirectly related, as both \orat\ and \fesclyc\ may depend on the same physical properties such as ionization parameter and metallicity. As such, these underlying properties will predominantly determine the observed \orat\ flux ratio at $z\sim3$ and at $z\sim6$.}

\subsection{Implications for High Redshift\label{sec:disc_highz}}

{ One of the principal objectives of the LzLCS is to ascertain which diagnostics are best suited to identifying LCEs and inferring their \fesclyc. While \fesclya, EW \lya, $v_{sep}$, \orat, $\beta_{1200}$, NUV $r_{50}$, and \sigsfr\ all exhibit strong significant correlations with \fesclyc\ at $z\sim0.3$, comparisons with observations of these properties at higher redshifts are necessary to extend our results to the epoch of reionization.}

\lya\ offers one of the most promising indicators of LyC escape: both \fesclya\ and \lya\ EW correlate well with \fesclyc, at least in part because \lya\ is also sensitive to the line-of-sight neutral hydrogen column density. However, this sensitivity complicates the detection of \lya\ because the neutral hydrogen fraction in the IGM increases dramatically with redshift \citep[][]{2019MNRAS.489.2572T,2021A&A...650A..63T}, accounting for the steep drop in LAEs observed at $z\ga 6$ \citep[e.g.,][]{2011ApJ...743..132P,2013ApJ...775L..29T,2014ApJ...795...20S}. Although \lya\ falls in the optical observing window at redshifts of $z\sim6$, few \lya\ emitters may be observable (an effect exacerbated by the shift of \lya\ into the near infrared at $z>7$). But it does suggest that LCEs are frequently LAEs. The few LAEs detected at $z>6$ may be leaking LyC. Indeed, the size of ionized bubbles necessary to produce observed double-peaked \lya\ at $z\sim6$ imply LyC escape \citep[][]{2021MNRAS.500..558M}.

{ Star-forming galaxies observed at higher redshifts allow a more direct comparison with the LCEs from our sample. The $z\sim2-3$ LBG-like star-forming galaxies from \citet{2021arXiv210805363R} have lower \fesclya\ and EW \lya\ than the $z\sim0.3$ LCEs, suggesting that LBGs are either extremely weak LCEs or non-emitters.
The $z\sim3$ individual LCEs from \citet{2020ApJ...889..161N} and $z\sim2-3$ stacked spectra LCEs with \fesclyc$\geq0.2$ from \citet{2018ApJ...869..123S}, \citet{2021MNRAS.505.2447P}, and \citet{2022MNRAS.510.4582N} exhibit \lya\ EWs of $30-110$ \AA\ and LyC escape fractions slightly larger than those of $z\sim0.3$ LCEs at similar EWs. As shown in Figure \ref{fig:ew_lya}, the \citet{2021MNRAS.505.2447P} relation between \fesclyc\ and EW \lya\ \citep[based on the $z\sim3$ stacks from][]{2018ApJ...869..123S} is qualitatively consistent with the \fesclyc-\lya\ envelope in the LzLCS galaxies.}

{ At higher redshifts, the \citet{2018A&A...614A..11M} analysis of stacked spectra of $z=3.5-4.3$ LAEs indicates that galaxies with \lya\ EWs $\ge50$ \AA\ have relative \fesclyc\ significantly higher than 0.05 while those with \lya\ EWs $<50$ \AA\ do not. The individual LyC detection in \emph{Ion3} at $z=4$ exhibits quadruply-peaked \lya\ with an EW of $\sim 40$ \AA\ and a relative escape fraction of 0.6 \citep{2018MNRAS.476L..15V}. These $z\sim2-4$ results are consistent with the fact that 22/26 of the LzLCS strong LCEs have EW \lya\ $>40\ \AA$. Thus, EW \lya\ as a LCE identifier and \fesclyc\ diagnostic should be readily extensible to the epoch of reionization.}

{ Studies by \citet{2018ApJ...869..123S} and \citet{2021MNRAS.505.2447P} at $z\sim3$ suggest that fainter galaxies have higher \fesclyc. As shown in \S\ref{sec:m_fuv}, we only find weak correlations with $M_{1500}$, although we do see a possible dependence of maximal \fesclyc\ on $M_{1500}$. That being said, our sample probes fainter magnitudes than studies at higher redshifts. \citet{2022arXiv220111800S} found that individual LCE detections at higher redshifts are consistent with LzLCS results for $F_{\lambda\rm LyC}/F_{\lambda1500}$ (see their Figure 16). LyC measurements using stacked spectra from \citet{2018ApJ...869..123S} suggest that a possible envelope in $F_{\lambda\rm LyC}/F_{\lambda1500}$ persists out to $M_{1500}\sim-22$. Despite scatter, the increase in max $F_{\lambda\rm LyC}/F_{\lambda1500}$ with magnitude suggests that $M_{1500}$ may still be a useful tool to estimate galaxy contributions to reionization. However, additional observations of UV-faint LCEs at $z\sim3$ are necessary to confirm that the dependence of maximal \fesclycrel\ on $M_{1500}$ at $z\sim3$ is similar to that observed in the combined LzLCS at $z\sim0.3$.}

Several indirect diagnostics driven by physical mechanism also show promise at high redshift. Star-forming galaxies at $z\sim6-8$ have UV slopes and star formation rates comparable to those of LCEs in this study \citep[e.g.,][]{2021ApJ...909..144B}, which may suggest that the $\beta_{1200}$ and \sigsfr\ diagnostics at $z\sim0.3$ can be used to identify LCEs and even infer \fesclyc\ at the epoch of reionization. \citet{2015ApJS..219...15S,2019ApJ...871..164S} find that star-forming galaxies become increasingly compact, approaching UV half-light radii of $0.5$ to $0.7$ kpc at $z\sim6$. Such compactness is typical for the $z\sim0.3$ LCEs in this study { and LAEs in other studies \citep[e.g.,][]{2021ApJ...914....2K}.} Moreover, this UV concentration corresponds to an average increase in \sigsfr\ up to values $\ga10$ M$_\odot$ yr$^{-1}$ kpc$^{-2}$ at $z\sim6$ \citep{2015ApJS..219...15S}, which is comparable to the \sigsfr\ of $z\sim0.3$ LCEs. Such concentrated star formation at the epoch of reionization may be indicative of LyC escape. Star-forming galaxies at $z\sim5-6$ could have higher-velocity outflows than galaxies of the same star-formation rates and stellar masses at $z\sim0$ \citep[][]{2019ApJ...886...29S}. Such extreme outflows might indicate augmented stellar feedback, in turn improving LyC escape indicators like \sigsfr, $M_\star$, and UV $\beta$ at $z\sim6$ relative to $z\sim0.3$.

{ The effect of gravity on \fesclyc\ diagnostics is not straightforward. In \S\ref{sec:mass}, we found no correlation of \fesclyc\ with $M_\star$. We did find a significant trend between \fesclyc\ and \ssfr\ in \S\ref{sec:sfr}, which may indicate that strong feedback better facilitates LyC escape in weaker gravitational potentials. At $z\sim2.3$ \citep{2016ApJ...816...23S}, $z\sim3$ \citep{2021arXiv210805363R}, and $z=7$ \citep{2021MNRAS.tmp..470E}, galaxies tend to have \ssfr\ in the 1-10 Gyr$^{-1}$ range, which corresponds to weaker LCEs and non-emitters at $z\sim0.3$. Taking the \S\ref{sec:sfr} results at face value, these higher redshift galaxies might not be prodigious LCEs. However, the analysis by \citet{2021MNRAS.tmp.3408S} of individual LCEs at $z\sim3$ suggests galaxies with lower \ssfr\ values of 1-10 Gyr$^{-1}$ have \fesclyc\ in excess of 10\% at higher redshift. No $z\sim0.3$ LCEs with \fesclyc$>10$\% have sSFR $<10$ Gyr$^{-1}$. This striking difference at $z\sim3$ may point to an increase in burst-enabled LyC escape with redshift.}

{ Other properties may readily persist at higher redshifts (e.g., H$\beta$ EW, [\ion{O}{3}]$\lambda5007$ EW, \orat, $R_{23}$, cf. \citealt{2020ApJ...889..161N,2021arXiv211015858B,2021MNRAS.tmp..470E,2021arXiv210805363R,2021MNRAS.tmp.3408S}) but are difficult to observe at $z\ga 6$ without space-based facilities like {\it JWST}. However, as we enter the {\it JWST} era, indirect indicators of \fesclyc\ will prove immensely useful via the trends presented in \S\ref{sec:diagnostics}.} Indeed, unlike observing the LyC, the reionization of the IGM is relatively insensitive to the anisotropic escape of LyC photons. The broad range of LCE properties may still complicate interpreting some diagnostics, as might the interdependent nature of parameters, since it is not clear how either will differ at $z\sim6$.

LCEs at the epoch of reionization could be intrinsically unlike those in the LzLCS. Dust and the corresponding attenuation laws might alter the optical depth of the ISM at the epoch of reionization. Low metallicities in the early universe could suppress fragmentation during star formation, resulting in a top-heavy IMF with higher concentrations of O and B stars than at low redshift for the same \sigsfr. Thus, a more physically complete understanding of the ISM and star formation at high redshift are paramount to successfully interpreting the signatures of LyC escape. Nevertheless, we find that young ages, concentrated star formation, and high ionization parameter are important for LyC escape.

\section{Conclusion}

{Using the results for the 66 galaxies at $z\sim0.3$ in the LzLCS, we have conducted the first thorough statistical test of indirect diagnostics of \fesclyc\ and LyC escape. { We find that \fesclya, EW \lya, $v_{sep}$, \orat, EW H$\beta$, $\beta_{1200}$, $M_{1500,int}$, NUV $r_{50}$, \sigsfr, and $\rm\Sigma_{sSFR}$ provide the best correlations with \fesclyc. However, many of these diagnostics exhibit an upper limit in the values of \fesclyc\ that varies with the property in question. We interpret these results as underlying trends that are obscured by a combination of line-of-sight effects and variations in other physical properties. Indirect diagnostics arising from stellar feedback mechanisms (e.g., stellar winds, SNe) tend to suffer from this effect more so than those based on optical depth.} Such mechanism-based trends are still useful to infer \fesclyc\ while also providing insight into the nature of LyC escape. We note that the apparent success of each diagnostic may depend on the \fesclyc\ metric used, with templates fit to the COS spectra giving the strongest, most significant correlations of properties with \fesclyc. Since many of the trends show significant scatter for any given diagnostic parameter, we also perform a quantitative assessment of the fraction of galaxies which are LCEs.}

{Our statistical tests include the following diagnostics: fraction of \lya\ photons (\fesclya) which escape the host galaxy, EW(\lya), the velocity separation ($v_{sep}$) between the \lya\ profile peaks, the [\ion{O}{1}]$\lambda6300$/H$\beta$ flux ratio, the [\ion{O}{3}]$\lambda5007$/[\ion{O}{1}]$\lambda6300=O_{31}$ flux ratio, the [\ion{O}{3}]$\lambda5007$/[\ion{O}{2}]$\lambda3726,3729=O_{32}$ flux ratio, EW(H$\beta$), $M_{1500}$, UV power law slope $\beta_{1200}$, host galaxy stellar mass $M_\star$, half-light radius $r_{50}$, \sigsfr, and sSFR=SFR$/M_\star$.}

{ Of these diagnostics, \fesclya, EW \lya, $v_{sep}$, \orat, $\beta_{1200}$, NUV $r_{50}$, and \sigsfr\ exhibit some of the strongest, most significant correlations with \fesclyc. Trends in \fesclyc\ with \lya\ properties like \fesclya, EW \lya, and $v_{sep}$ suggest line-of-sight optical depth plays a key role in identifying LCEs and inferring \fesclyc. Other diagnostics, including \orat, $r_{50}$, \sigsfr, and $\beta_{1200}$, also demonstrate strong, significant correlations with \fesclyc, indicating concentrated star-formation, young stellar populations, and high ionization play pivotal roles in LyC escape.}

From these diagnostics, we have obtained deeper insight into the nature of LCEs and the conditions that may best facilitate LyC escape. LCEs tend to have compact ($r_{50}<0.5$ kpc) regions of intense star formation (\sigsfr$\ga10$ M$_\odot$ yr$^{-1}$ kpc$^{-2}$), indicating stellar feedback may be crucial for LyC escape.

The distribution of \fesclyc\ over other properties like ionization parameter (measured by \orat), burst age (measured by H$\beta$ EW), stellar mass, and metallicity suggests a broad range in LCE galaxy properties that persists across many diagnostics. { This diversity suggests that different physical conditions can lead to LyC escape or that a time delay between star formation and LyC escape can decouple \fesclyc\ from certain galaxy properties. With that being said, galaxies with high \orat, high H$\beta$ EW, and low $\beta$ are more likely to be strong LCEs (\fesclyc$\ga0.05$), suggesting that galaxies like the Green Peas are the most substantial LyC leakers.}

{To better understand how certain properties can affect one another and the inferred \fesclyc, we juxtapose \fesclyc\ diagnostics. \citet{2013ApJ...777...39Z} proposed a comparison of $\beta_{1200}$ with EW(H$\beta$). We find that, while not following the predictions from \citet{2013ApJ...777...39Z}, LCEs with higher \fesclyc\ tend to have higher EW(H$\beta$) ($>100$ \AA) and steeper UV continua ($<-2$). Moreover, LCEs distinguish themselves as having either higher EW(H$\beta$) and steeper UV continua or lower EW(H$\beta$) ($<100$ \AA) and more shallow UV continua ($>-2$).

We also compare \orat\ and EW(H$\beta$), finding the two are strongly correlated. Through comparisons with mass and metallicity, we find higher \orat\ ($>5$), higher \fesclyc\ LCEs have low masses and low metallicities while the lower \orat\ ($<5$), lower \fesclyc\ LCEs have higher masses and higher metallicities. These results point to a distinct difference in LyC escape mechanism, as discussed above. The oxygen ionization structure indicated by \orat\ and $O_{31}$ \citep[e.g.,][]{2015A&A...576A..83S} demonstrates that LCEs cannot be readily described by a simple density-bounded, isotropic escape scenario. Even so, all LCEs display concentrated star formation, indicating that locally intense star formation is necessary for LyC escape regardless of mechanism.}

For further insight into LCEs, we compare our results with cosmological and galaxy evolution simulation predictions. Consistent with some simulations \citep[e.g.,][]{2017MNRAS.470..224T, 2020arXiv200715012S}, we demonstrate that compact, UV-faint, low-mass galaxies are far more likely to be significant LyC leakers than other galaxies. While we do find that concentrated star formation is a significant indicator of LyC escape, simulations that predict this relationship suggest LCEs ought to have higher stellar masses than we find in our sample \citep[e.g.,][]{2017MNRAS.468.2176S,2020ApJ...889..161N}.

{ While we demonstrate that \lya\ is the best \fesclyc\ indicator at low redshift and that our LCEs are comparable to LyC-leaking LAEs at $z\sim2-3$, the increasing IGM opacity at $z\ga4$ renders \lya, at best, difficult to detect from galaxies at the epoch of reionization. From a holistic interpretation of other indirect diagnostics, a galaxy with high EW \lya, high \fesclya, high \orat, high EW(H$\beta$), low $\beta_{1200}$, low NUV $r_{50}$, and high \sigsfr\ is likely an LCE. Thus, observing rest-frame properties like \orat, EW(H$\beta$), and \sigsfr\ with \emph{JWST} is critical to identifying and understanding the galaxies responsible for cosmic reionization.}

\begin{acknowledgments}
Support for this work was provided by NASA through grant number \emph{HST}-GO-15626 from the Space Telescope Science Institute. Additional work was based on observations made with the NASA/ESA Hubble Space Telescope, obtained from the data archive at the Space Telescope Science Institute from \emph{HST} proposals 13744, 14635, 15341, and 15639. STScI is operated by the Association of Universities for Research in Astronomy, Inc. under NASA contract NAS 5-26555.

Funding for the Sloan Digital Sky
Survey IV has been provided by the
Alfred P. Sloan Foundation, the U.S.
Department of Energy Office of
Science, and the Participating
Institutions. SDSS-IV acknowledges support and
resources from the Center for High
Performance Computing  at the
University of Utah. The SDSS
website is \url{www.sdss.org}.
SDSS-IV is managed by the
Astrophysical Research Consortium
for the Participating Institutions
of the SDSS Collaboration.

RA acknowledges support from ANID Fondecyt Regular 1202007.

\end{acknowledgments}

\software{{\sc\ astropy\ } \citep{astropy:2013,astropy:2018}, {\sc\ matplotlib\ } \citep{matplotlib}, {\sc\ numpy\ } \citep{numpy}, {\sc\ scipy\ } \citep{scipy} }

\newpage
\bibliographystyle{aasjournal}
\bibliography{biblio}

\end{document}

%% file: authors.tex
\correspondingauthor{Sophia Flury}
\email{sflury@umass.edu}

\author[0000-0002-0159-2613]{Sophia R. Flury}
\affiliation{Department of Astronomy, University of Massachusetts Amherst, Amherst, MA 01002, United States}

\author{Anne E. Jaskot}
\affiliation{Department of Astronomy, Williams College, Williamstown, MA 01267, United States}

\author{Harry C. Ferguson}
\affiliation{Space Telescope Science Institute, 3700 San Martin Drive
Baltimore, MD 21218, United States}

\author[0000-0003-0960-3580]{G\'abor Worseck}
\affiliation{Institut f\"ur Physik und Astronomie, Universit\"at Potsdam, Karl-Liebknecht-Str. 24/25, D-14476 Potsdam, Germany}

\author{Kirill Makan}
\affiliation{Institut f\"ur Physik und Astronomie, Universit\"at Potsdam, Karl-Liebknecht-Str. 24/25, D-14476 Potsdam, Germany}

\author{John Chisholm}
\affiliation{Department of Astronomy, University of Texas at Austin, Austin, TX 78712, United States}

\author{Alberto Saldana-Lopez}
\affiliation{Observatoire de Gen\`eve, Universit\'e de Gen\`eve, 51 Ch. des Mailletes, 1290 Versoix, Switzerland}

\author{Daniel Schaerer}
\affiliation{Observatoire de Gen\`eve, Universit\'e de Gen\`eve, 51 Ch. des Mailletes, 1290 Versoix, Switzerland}

\author[0000-0003-0503-4667]{Stephan R. McCandliss}
\affiliation{Center for Astrophysical Sciences\\ Department of Physics and Astronomy, Johns Hopkins University, 3400 North Charles Street, Baltimore, MD 21218, USA}

\author{Bingjie Wang}
\affiliation{Department of Physics and Astronomy, Johns Hopkins University, Baltimore, MD 21218, United States}

\author{N. M. Ford}
\affiliation{Department of Astronomy, Williams College, Williamstown, MA 01267, United States}

\author{M. S. Oey}
\affiliation{Department of Astronomy, University of Michigan, Ann Arbor, MI 48109, United States}

\author{Timothy Heckman}
\affiliation{Department of Physics and Astronomy, Johns Hopkins University, Baltimore, MD 21218, United States}

\author{Zhiyuan Ji}
\affiliation{Department of Astronomy, University of Massachusetts Amherst, Amherst, MA 01002, United States}

\author{Mauro Giavalisco}
\affiliation{Department of Astronomy, University of Massachusetts Amherst, Amherst, MA 01002, United States}

\author[0000-0001-5758-1000]{Ricardo Amor\'in}
\affil{Instituto de Investigaci\'on Multidisciplinar en Ciencia y Tecnolog\'ia, Universidad de La Serena, Ra\'ul Bitr\'an 1305, La Serena, Chile}
\affiliation{Departamento de Astronom\'ia, Universidad de La Serena, Av. Juan Cisternas 1200 Norte,  La Serena, Chile}

\author{Hakim Atek}
\affiliation{Institut d’astrophysique de Paris, CNRS UMR7095, Sorbonne Universit\'e, 98bis Boulevard Arago, F-75014 Paris, France}

\author{Jeremy Blaizot}
\affiliation{Univ Lyon, Univ Lyon1, Ens de Lyon, CNRS, Centre de Recherche Astrophysique de Lyon UMR5574, F-69230, Saint-Genis-Laval, France}

\author{Sanchayeeta Borthakur}
\affiliation{School of Earth \& Space Exploration, Arizona State University, Tempe, AZ 85287, United States}

\author{Cody Carr}
\affiliation{Minnesota Institute for Astrophysics, School of Physics and
Astronomy, University of Minnesota, 316 Church Str SE, Minneapolis, MN 55455, United States}

\author{Marco Castellano}
\affiliation{INAF, Osservatorio Astronomico di Roma, via Frascati 33, I-00078 Monteporzio Catone, Italy}

\author{Stephane De Barros}
\affiliation{Observatoire de Gen\`eve, Universit\'e de Gen\`eve, 51 Ch. des Mailletes, 1290 Versoix, Switzerland}

\author{Mark Dickinson}
\affiliation{National Optical-Infrared Astronomy Research Laboratory, Tucson, AZ, United States}

\author{Steven L. Finkelstein}
\affiliation{Department of Astronomy, The University of Texas at Austin, Austin, TX, United States}

\author{Brian Fleming}
\affiliation{Laboratory for Atmospheric and Space Physics, Boulder, Colorado, United States}

\author{Fabio Fontanot}
\affiliation{INAF–Osservatorio Astronomico di Trieste, Via G.B. Tiepolo, 11, I-34143 Trieste, Italy}

\author{Thibault Garel}
\affiliation{Observatoire de Gen\`eve, Universit\'e de Gen\`eve, 51 Ch. des Mailletes, 1290 Versoix, Switzerland}

\author{Andrea Grazian}
\affiliation{INAF-Osservatorio Astronomico di Padova, Vicolo dell’Osservatorio 5, I-35122, Padova, Italy}

\author{Matthew Hayes}
\affiliation{The Oskar Klein Centre, Department of Astronomy, Stockholm University, AlbaNova, SE-10691 Stockholm, Sweden}

\author{Alaina Henry}
\affiliation{Space Telescope Science Institute, 3700 San Martin Drive
Baltimore, MD 21218, United States}


\author{Valentin Mauerhofer}
\affiliation{Observatoire de Gen\`eve, Universit\'e de Gen\`eve, 51 Ch. des Mailletes, 1290 Versoix, Switzerland}
\affiliation{Univ. Lyon, Univ. Lyon 1, ENS de Lyon, CNRS, Centre de Recherche Astrophysique de Lyon UMR5574, 69230 Saint-Genis-Laval, France}

\author{Genoveva Micheva}
\affiliation{Leibniz-Institute for Astrophysics Potsdam, An der Sternwarte 16, 14482 Potsdam, Germany}


\author{Goran Ostlin}
\affiliation{The Oskar Klein Centre, Department of Astronomy, Stockholm University, AlbaNova, SE-10691 Stockholm, Sweden}

\author{Casey Papovich}
\affiliation{George P. and Cynthia Woods Mitchell Institute for Fundamental Physics and Astronomy, Department of Physics and Astronomy,
Texas A\&M University, College Station, TX, United States}

\author{Laura Pentericci}
\affiliation{INAF, Osservatorio Astronomico di Roma, via Frascati 33, I-00078 Monteporzio Catone, Italy}

\author{Swara Ravindranath}
\affiliation{Space Telescope Science Institute, 3700 San Martin Drive
Baltimore, MD 21218, United States}

\author{Joakim Rosdahl}
\affiliation{Univ Lyon, Univ Lyon1, Ens de Lyon, CNRS, Centre de Recherche Astrophysique de Lyon UMR5574, F-69230, Saint-Genis-Laval, France}

\author{Michael Rutkowski}
\affiliation{Department of Physics and Astronomy, Minnesota State University, Mankato, MN, 56001, United States}

\author{Paola Santini}
\affiliation{INAF, Osservatorio Astronomico di Roma, via Frascati 33, I-00078 Monteporzio Catone, Italy}

\author{Claudia Scarlata}
\affiliation{Minnesota Institute for Astrophysics, School of Physics and
Astronomy, University of Minnesota, 316 Church str SE, Minneapolis, MN 55455, United States}

\author{Harry Teplitz}
\affiliation{Infrared Processing and analysis Center, California Institute of Technology, Pasadena, CA 91125, United States}

\author{Trinh Thuan}
\affiliation{Astronomy Department, University of Virginia, Charlottesville, VA 22904, United States}

\author{Maxime Trebitsch}
\affiliation{Astronomy, Kapteyn Astronomical Institute, Landleven 12, 9747 AD Groningen, The Netherlands}

\author{Eros Vanzella}
\affiliation{INAF, Osservatorio Astronomico di Bologna, via Gobetti 93/3 I-40129 Bologna, Italy}

\author{Anne Verhamme}
\affiliation{Observatoire de Gen\`eve, Universit\'e de Gen\`eve, 51 Ch. des Mailletes, 1290 Versoix, Switzerland}
\affiliation{Univ. Lyon, Univ. Lyon 1, ENS de Lyon, CNRS, Centre de Recherche Astrophysique de Lyon UMR5574, 69230 Saint-Genis-Laval, France}


%% file: kendall_tau.tex
\begin{deluxetable*}{l | r c r | r c r | r c r}[!ht]
\caption{Kendall $\tau$ correlation coefficients for proposed LyC \fesclyc\ diagnostics for the combined LzLCS and \pubsamp\ sample, accounting for upper limits in \fesclycrel\ and \fesclyc\ following \citet{1996MNRAS.278..919A}. $p$ values indicate the false-positive probability that the correlation is real. $\sigma$ values, given by ${\rm probit}(1-p)$, indicate the significance of the correlation, i.e., the number of standard deviations separating the correlation statistic from that of the null hypothesis. The characteristic $1\sigma$ uncertainties in $\tau$ estimated by bootstrapping values of the \fesclyc\ metrics and indirect diagnostics are $\pm0.05$. { Each correlation is assessed using all 89 objects in the combined LzLCS sample except $v_{sep}$, which is only measured for 7 LzLCS and 20 published objects.} \label{tab:corrcoef}}
\tablehead{\colhead{Diagnostic} & \multicolumn{3}{c}{\fesclycrel} & \multicolumn{3}{c}{\fesclyc(H$\beta$)} & \multicolumn{3}{c}{\fesclyc(UV)} \\
\colhead{} & \colhead{$\tau$} & \colhead{$p$} & \colhead{$\sigma$} & \colhead{$\tau$} & \colhead{$p$} & \colhead{$\sigma$} & \colhead{$\tau$} & \colhead{$p$} & \colhead{$\sigma$}
}
\startdata
$f~_{esc}^{Ly\alpha}$ & $ 0.292$ & $5.186\times10^{ -5 }$ & $3.882$ & $ 0.343$ & $1.942\times10^{ -6 }$ & $4.618$ & $ 0.324$ & $6.774\times10^{ -6 }$ & $4.351$ \\
EW(Ly$\alpha$) & $ 0.320$ & $8.687\times10^{ -6 }$ & $4.296$ & $ 0.234$ & $1.141\times10^{ -3 }$ & $3.051$ & $ 0.342$ & $2.011\times10^{ -6 }$ & $4.610$ \\
$v_{sep}$ & $-0.493$ & $3.103\times10^{ -4 }$ & $3.422$ & $-0.422$ & $2.033\times10^{ -3 }$ & $2.873$ & $-0.530$ & $1.055\times10^{ -4 }$ & $3.705$ \\
$\log_{10}O_{31}$ & $-0.149$ & $0.039$ & $1.761$ & $-0.144$ & $0.045$ & $1.693$ & $-0.151$ & $0.036$ & $1.796$ \\
$\log_{10}$[O I]/H$\beta$ & $-0.148$ & $0.041$ & $1.745$ & $-0.145$ & $0.044$ & $1.709$ & $-0.145$ & $0.044$ & $1.705$ \\
$\log_{10}O_{32}$ & $ 0.290$ & $5.678\times10^{ -5 }$ & $3.860$ & $ 0.198$ & $6.024\times10^{ -3 }$ & $2.511$ & $ 0.347$ & $1.438\times10^{ -6 }$ & $4.679$ \\
EW(H$\beta$) & $ 0.223$ & $1.953\times10^{ -3 }$ & $2.886$ & $ 0.109$ & $0.132$ & $1.117$ & $ 0.283$ & $8.366\times10^{ -5 }$ & $3.764$ \\
M$_{1500,obs}$ & $ 0.045$ & $0.533$ & $0.000$ & $-0.013$ & $0.857$ & $0.000$ & $ 0.098$ & $0.174$ & $0.940$ \\
M$_{1500,int}$ & $ 0.228$ & $1.591\times10^{ -3 }$ & $2.950$ & $ 0.157$ & $0.029$ & $1.895$ & $ 0.320$ & $8.978\times10^{ -6 }$ & $4.289$ \\
$\beta_{1200}$ & $-0.221$ & $2.200\times10^{ -3 }$ & $2.848$ & $-0.261$ & $2.966\times10^{ -4 }$ & $3.435$ & $-0.283$ & $8.366\times10^{ -5 }$ & $3.764$ \\
$\log_{10}M\star$ & $-0.089$ & $0.216$ & $0.785$ & $-0.074$ & $0.307$ & $0.503$ & $-0.167$ & $0.021$ & $2.040$ \\
COS NUV $r_{50}$ & $-0.388$ & $7.179\times10^{ -8 }$ & $5.261$ & $-0.301$ & $2.938\times10^{ -5 }$ & $4.018$ & $-0.382$ & $1.193\times10^{ -7 }$ & $5.166$ \\
$\rm\log_{10}\Sigma_{SFR,H\beta}$ & $ 0.368$ & $3.884\times10^{ -7 }$ & $4.941$ & $ 0.264$ & $2.650\times10^{ -4 }$ & $3.465$ & $ 0.325$ & $7.099\times10^{ -6 }$ & $4.341$ \\
$\rm\log_{10}\Sigma_{SFR,F_{\lambda1100}}$ & $ 0.070$ & $0.334$ & $0.429$ & $ 0.068$ & $0.347$ & $0.394$ & $-0.035$ & $0.632$ & $0.000$ \\
$\rm\log_{10}{sSFR}$ & $ 0.110$ & $0.128$ & $1.138$ & $ 0.043$ & $0.554$ & $0.000$ & $ 0.181$ & $0.012$ & $2.254$ \\
$\rm\log_{10}\Sigma_{sSFR,H\beta}$ & $ 0.290$ & $6.320\times10^{ -5 }$ & $3.833$ & $ 0.208$ & $4.167\times10^{ -3 }$ & $2.638$ & $ 0.346$ & $1.859\times10^{ -6 }$ & $4.627$ \\
$\rm12+\log_{10}\left(\frac{O}{H}\right)$ & $-0.187$ & $9.484\times10^{ -3 }$ & $2.346$ & $-0.130$ & $0.070$ & $1.475$ & $-0.211$ & $3.420\times10^{ -3 }$ & $2.705$ \\
\enddata
\end{deluxetable*}